\documentclass[10pt,a4paper]{article} \pdfoutput=1 

\usepackage{microtype}

\usepackage{amsmath}
\usepackage{amsthm}
\usepackage{amssymb}
\numberwithin{equation}{section}

\usepackage[dvipsnames]{xcolor}

\usepackage{tabularx}

\usepackage[showseconds=false]{datetime2}

\theoremstyle{plain}
\newtheorem{proposition}{Proposition}[section]

\theoremstyle{definition}

\newtheorem{example}[proposition]{Example}
\newtheorem{remark}[proposition]{Remark}

\newcommand{\R}{\mathbf{R}}
\newcommand{\C}{\mathbf{C}}

\renewcommand{\epsilon}{\varepsilon}
\renewcommand{\phi}{\varphi}

\newcommand{\floor}[1]{\lfloor #1 \rfloor}
\newcommand{\abs}[1]{\left\lvert #1 \right\rvert}

\newcommand{\twin}{\widetilde}

\newcommand{\circleS}{S^1}

\DeclareMathOperator{\sgn}{sgn}
\DeclareMathOperator*{\res}{res}

\DeclareMathOperator{\ad}{ad}
\DeclareMathOperator{\diff}{diff}
\DeclareMathOperator{\Diff}{Diff}

\usepackage{newtxtext,newtxmath}

\usepackage{graphicx}

\usepackage[margin=10pt,font=small,labelfont=bf,labelsep=period]{caption}

\usepackage[nottoc]{tocbibind}

\usepackage{titlesec}
\titleformat*{\section}{\large\bfseries}
\titleformat*{\subsection}{\normalsize\bfseries}

\usepackage[twocolumn, columnsep=7mm, left=19mm, right=14mm, top=25mm, bottom=30mm]{geometry}

\usepackage[colorlinks,linkcolor=blue,citecolor=blue,urlcolor=blue]{hyperref}

\newcommand{\makefrontmatter}{%
  \hypersetup{%
    pdfauthor={Hans Lundmark, Jacek Szmigielski},
    pdftitle={\ourtitle},
  }

  \begin{center}
    {\LARGE \ourtitle}

    \vspace{8mm}
  
    \begin{minipage}{13.5cm}
      \begin{minipage}[t]{4.5cm}
        \large
        \kern0pt
        Hans Lundmark
      \end{minipage}
      \begin{minipage}[t]{9cm}
        \kern0pt
        Department of Mathematics \\
        Linköping University \\
        SE-581\,83 Linköping, Sweden \\
        hans.lundmark@liu.se 
      \end{minipage}
      
      \vspace{5mm}
      
      \begin{minipage}[t]{4.5cm}
        \large
        \kern0pt
        Jacek Szmigielski
      \end{minipage}
      \begin{minipage}[t]{9cm}
        \kern0pt
        Department of Mathematics and Statistics \& \\
        \hspace*{1em}Centre for Quantum Topology and Its Applications (quanTA) \\
        University of Saskatchewan \\
        106 Wiggins Road, Saskatoon, Saskatchewan, S7N\,5E6, Canada \\
        szmigiel@math.usask.ca
      \end{minipage}
    \end{minipage}

    \vspace{6mm}

    {\normalsize \ourdate}
  \end{center}

  \begin{quote}
    \small \textbf{Abstract.} \ourabstract{}
  \end{quote}
  
  \vspace{4mm}
  
}

\usepackage[utf8]{inputenc}
\usepackage[T1]{fontenc}

\newcommand{\ourtitle}{A view of the peakon world through the lens of approximation theory}

\newcommand{\ourabstract}{%
  Peakons (peaked solitons) are particular solutions
  admitted by certain nonlinear PDEs, most famously the Camassa--Holm
  shallow water wave equation.
  These solutions take the form of a train of peak-shaped waves,
  interacting in a particle-like fashion.
  In this article we give an overview of the mathematics of peakons,
  with particular emphasis on the connections to classical problems
  in analysis, such as Padé approximation, mixed Hermite--Padé approximation,
  multi-point Padé approximation,
  continued fractions of Stieltjes type and (bi)orthogonal polynomials.
  The exposition follows the chronological development of our understanding,
  exploring the peakon solutions of the Camassa--Holm, Degasperis--Procesi, Novikov,
  Geng--Xue and modified Camassa--Holm (FORQ) equations.
  All of these paradigm examples are integrable systems arising from the compatibility
  condition of a Lax pair,
  and a recurring theme in the context of peakons is the need to properly
  interpret these Lax pairs in the sense of Schwartz's theory of distributions.
  We trace out the path leading from distributional Lax pairs to explicit formulas
  for peakon solutions via a variety of approximation-theoretic problems,
  and we illustrate the peakon dynamics with graphics.  
}

\newcommand{\ourdate}{July 5, 2022}

\begin{document}

\twocolumn[
\begin{@twocolumnfalse}
\makefrontmatter{}
\end{@twocolumnfalse}
]

\tableofcontents{}

\section{Introduction}
\label{sec:intro}

During the last couple of decades, we have had the pleasure of taking
part in the development of the mathematics of \emph{peakons},
peak-shaped solitons that first appeared as solutions to the
Camassa--Holm shallow water wave equation, and later in many other
related PDEs.
This article is an attempt to give a coherent presentation of selected parts
of this work and the mathematical context where it belongs.
Our aim is to explain in an accessible manner how to
derive explicit formulas for peakon solutions, via Lax pairs, inverse
eigenvalue problems and approximation theory,
and also to illustrate how the study of peakons has inspired
interesting new developments in these areas.
Along the way, we will touch upon some other aspects
of Camassa--Holm-type equations, and give pointers to relevant literature,
but the subject is enormous, and we make no claims to completeness.

Our story thus begins with the highly influential and frequently cited paper of Camassa and Holm~\cite{camassa-holm:1993:CH-orginal-paper} in which the equation
\begin{equation}
  \label{eq:intro-CH-kappa-expanded}
  u_t + 2 \kappa u_x - u_{xxt} + 3 u u_x = 2  u_x u_{xx} + u u_{xxx}
\end{equation}
was proposed as an integrable model of one-dimensional dispersive waves in shallow water,
$u(x,t)$ being the fluid velocity in the $x$ direction.
Here $\kappa$ is a positive physical constant,
but the PDE that we will consider here
and refer to as the Camassa--Holm (CH) equation
is the limiting case with $\kappa = 0$,
\begin{equation}
  \label{eq:intro-CH-expanded}
  u_t - u_{xxt} + 3 u u_x = 2  u_x u_{xx} + u u_{xxx}
  .
\end{equation}
The substitution $u(x,t) = U(x + \kappa t, t) - \kappa$ in~\eqref{eq:intro-CH-kappa-expanded}
leads to equation~\eqref{eq:intro-CH-expanded} for the function~$U$,
so in that sense \eqref{eq:intro-CH-kappa-expanded} and~\eqref{eq:intro-CH-expanded}
are equivalent.
However, if we want to study solutions on the whole real line $x \in \R$ with finite $H^1$-norm
$\int_\R (u^2 + u_x^2) \, dx$
(which is natural, since that is a conserved quantity),
then the cases $\kappa = 0$ and $\kappa \neq 0$ are different,
since the transformation shifts the zero-level
of the solution and thus maps $u(\cdot,t) \in H^1(\R)$ to $U(\cdot,t) \notin H^1(\R)$.

The CH equation~\eqref{eq:intro-CH-expanded} may be written as
\begin{equation}
  \label{eq:intro-CH}
  m_t + (um)_x + u_x m = 0
  ,\qquad
  m = u - u_{xx}
  ,
\end{equation}
or alternatively
\begin{equation}
  \label{eq:intro-CH-b-form}
  m_t + m_x u + 2 m u_x = 0
  ,\qquad
  m = u - u_{xx}
  .
\end{equation}
The adjective \emph{integrable} above refers to
properties associated with the concept of a
\emph{(completely) integrable system},
in particular the existence of a Lax pair
\begin{subequations}
  \label{eq:intro-CH-lax}
  \begin{align}
    \label{eq:intro-CH-lax-x}
    \bigl( \partial_x^2 - \tfrac14 \bigr) \, \psi &= -\tfrac12 \lambda m \psi
    , \\
    \label{eq:intro-CH-lax-t}
    \psi_t &= \tfrac12  \left( \tfrac{1}{\lambda} + u_x \right) \psi - \left( \tfrac{1}{\lambda} + u \right) \psi_x
    ,
  \end{align}
\end{subequations}
whose compatibility (cross-differentiation) results in~\eqref{eq:intro-CH},
but also a bi-Hamiltonian formulation,
an infinite hierarchy of conservation laws,
multisoliton solutions, and so on.
The existence of the Lax pair allows one to reduce a nonlinear PDE
problem to a system of ODEs, reminiscent of the separation of
variables in basic linear PDE theory. With typical boundary
conditions $\lim_{\abs{x} \to \infty} \psi=0$, the first equation~\eqref{eq:intro-CH-lax-x}
becomes a boundary value problem of Sturm--Liouville type with weight~$m$
and an eigenvalue parameter~$\lambda$. The second equation~\eqref{eq:intro-CH-lax-t}
can be viewed as a deformation equation, and one of the miracles of the
subject is that the deformation is \emph{isospectral}, meaning that it
leaves the Sturm--Liouville spectrum invariant.

\begin{figure}
  \centering
  \includegraphics[width=0.95\linewidth]{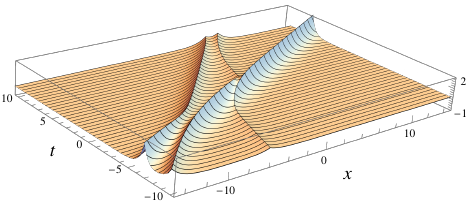}
  \caption{An example of a three-peakon solution of the Camassa--Holm equation
    \eqref{eq:intro-CH}.
    The graph of
    $u(x,t) = \sum_{k=1}^3 m_k(t) \, e^{-\abs{x - x_k(t)}}$
    is plotted for $x \in [-15,15]$ and $t \in [-10,10]$ from the exact solution
    formulas~\eqref{eq:CH-threepeakon-x-m}.
    In this example, all amplitudes $m_k$ are positive,
    so it is a \emph{pure peakon solution} (i.e., there are no \emph{antipeakons} with
    negative~$m_k$).
    See also Figure~\ref{fig:CH-3peakon-positions}.
  }
  \label{fig:CH-pure3peakon-3d}
\end{figure}

As pointed out by Camassa and Holm,
the limiting case~\eqref{eq:intro-CH} is of particular interest since it
admits weak solutions (with finite $H^1$-norm) in the form of peak-shaped travelling waves,
\begin{equation}
  \label{eq:intro-single-peakon}
  u(x,t) = c \, e^{-\abs{x - ct}}
  ,\qquad
  c \in \R
  ,
\end{equation}
known as \emph{peakons} (peaked solitons),
on account of their obviously peaked shape together with the fact
that they can also be combined via superposition to form
\emph{$N$-peakon} or \emph{multipeakon} solutions
of the form
\begin{equation}
  \label{eq:intro-peakons}
  u(x,t) = \sum_{k=1}^N m_k(t) \, e^{-\abs{x - x_k(t)}}
  ,
\end{equation}
which exhibit nonlinear interaction properties
similar to the familiar smooth
multisoliton solutions of the Korteweg--de Vries (KdV) equation
and other integrable PDEs; see Figure~\ref{fig:CH-pure3peakon-3d}.

\begin{figure}
  \centering
  \includegraphics[width=1.0\linewidth]{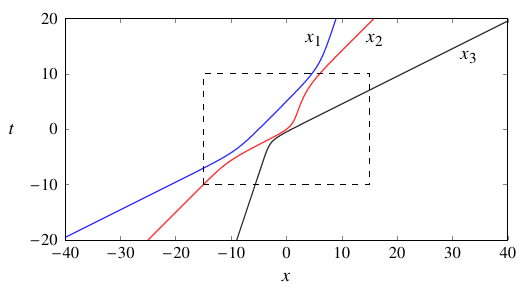}
  \caption{Positions $x = x_k(t)$ of the three individual peakons in the solution
    from Figure~\ref{fig:CH-pure3peakon-3d}, with the dashed rectangle
    indicating the region shown there.
    Note that the ordering $x_1(t) < x_2(t) < x_3(t)$ is preserved for all~$t$,
    and that the peakons asymptotically (as $t \to \pm\infty$)
    move in straight lines in the $(x,t)$-plane,
    like solitary travelling waves.
    The asymptotic velocities are $\{ 2, 1, \tfrac13 \}$ in this example.
    These numbers, which are also the asymptotic values
    of the amplitudes $m_k(t)$,
    are the reciprocals $1/\lambda_k$ of the eigenvalues
    in a certain spectral problem associated with the CH equation,
    as described in Section~\ref{sec:CH},
    and these eigenvalues $\lambda_1 = \tfrac12$,
    $\lambda_2 = 1$ and $\lambda_3 = 3$ appear as parameters
    in the solution formulas~\eqref{eq:CH-threepeakon-x-m}.
  }
  \label{fig:CH-3peakon-positions}
\end{figure}

Individual solitons in a smooth multisoliton solution are in general
not discernible during interactions,
but only when they are well separated from each other.
In contrast, peakons have a well-defined position and amplitude at each instant~$t$.
Indeed, if we define the $k$th peakon in the multi\-peakon solution~\eqref{eq:intro-peakons}
to simply be the $k$th term $m_k(t) \, e^{-\abs{x-x_k(t)}}$,
then we may say that it is located at the position $x=x_k(t)$
where the exponential factor has its peak,
and that its amplitude is~$m_k(t)$;
see Figure~\ref{fig:CH-3peakon-positions}.

The peakon amplitudes $m_k$ may be positive or negative.
(If $m_k$ is zero at some instant, then it is zero for all~$t$ according to
\eqref{eq:intro-CH-peakon-ode-explicit} below, so we may assume $m_k \neq 0$.)
Peakons with negative amplitude
are called \emph{antipeakons}, which leads to a somewhat unfortunate ambiguity where the word
\emph{peakon} may sometimes denote a general term as in the previous paragraph,
and sometimes a peakon with positive amplitude as opposed to an antipeakon.
Henceforth, when we talk about ``peakon solutions'' in general, we will mean multi\-peakon
solutions of the form~\eqref{eq:intro-peakons}, regardless of the signs of the amplitudes.
But a \emph{pure peakon solution} is one where all $m_k > 0$,
a  \emph{pure antipeakon solution} has all $m_k < 0$,
and a \emph{mixed peakon--antipeakon solution} involves amplitudes of both signs.

The function $u$ from \eqref{eq:intro-peakons} is a weak solution
of the CH equation~\eqref{eq:intro-CH}
if and only if the positions $x_k(t)$ and the amplitudes $m_k(t)$ of the individual peakons
satisfy the Hamiltonian system of ODEs
\begin{equation}
  \label{eq:intro-CH-peakon-ode}
  \dot x_k = \frac{\partial H}{\partial m_k}
  ,\qquad
  \dot m_k = -\frac{\partial H}{\partial x_k}
\end{equation}
generated by the Hamiltonian function
\begin{equation}
  \label{eq:intro-CH-hamiltonian}
  H(x_1,\dots,x_N,m_1,\dots,m_N)
  = \frac12 \sum_{i,j=1}^N m_i m_j e^{-\abs{x_i-x_j}}
  ,
\end{equation}
where it is assumed that all $x_k$ are distinct;
usually we label them in increasing order $x_1 < \dots < x_N$.
Explicitly, this system reads
\begin{equation}
  \label{eq:intro-CH-peakon-ode-explicit}
  \begin{aligned}
    \dot x_k &= \sum_{i=1}^N m_i \, e^{-\abs{x_k-x_i}}
    ,\\
    \dot m_k &= m_k \sum_{i=1}^N m_i \sgn(x_k-x_i) \, e^{-\abs{x_k-x_i}}
    ,
  \end{aligned}
\end{equation}
for $1 \le k \le N$, where $\sgn(0)=0$ by definition.
A convenient shorthand notation for this system is obtained by noticing that the
right-hand side of the equation for $\dot x_k$ is obtained
by evaluating the expression $u(x) = \sum_{i=1}^N m_i \, e^{-\abs{x-x_i}}$
at the point $x=x_k$,
and the right-hand side of the equation for $\dot m_k$
equals $-m_k \, u_x(x_k)$, where
\begin{equation}
  u_x(x_k) := \langle u_x \rangle (x_k) = \frac{u_x(x_k^-) + u_x(x_k^+)}{2}
\end{equation}
denotes the arithmetic average of the left and right limits
of the derivative $u_x(x)$ of the same expression $u(x)$ at $x=x_k$.
That is, we may write the system as
\begin{equation}
  \label{eq:intro-CH-peakon-ode-shorthand}
  \dot x_k = u(x_k),
  \qquad
  \dot m_k = -m_k \, u_x(x_k)
  .
\end{equation}
Note in particular that the $k$th peakon,
located at $x = x_k(t)$,
at each instant travels with a velocity $\dot x_k(t)$ equal to
the amplitude $u(x_k(t),t)$ of the composite wave at that location.
Thus, in a pure peakon solution
(such as the one shown in Figure~\ref{fig:CH-pure3peakon-3d})
all peakons travel to the right,
while in a pure antipeakon solution all peakons travel to the left.
In a situation where the peakons start out well separated,
each peakon experiences very little influence
from the exponentially decaying tails of the other peakons,
so $u(x_k) \approx m_k$ and $u_x(x_k) \approx 0$,
and the peakons will all behave nearly like the single-peakon travelling wave~\eqref{eq:intro-single-peakon}:
\begin{equation*}
  \dot x_k(t) \approx m_k(t) \approx \text{constant}
  .
\end{equation*}
But since different peakons may have different velocities,
a faster peakon may catch up with a slower one,
and as they come closer some nonlinear interaction between them will take place.

\begin{figure}
  \centering
  \includegraphics[width=0.95\linewidth]{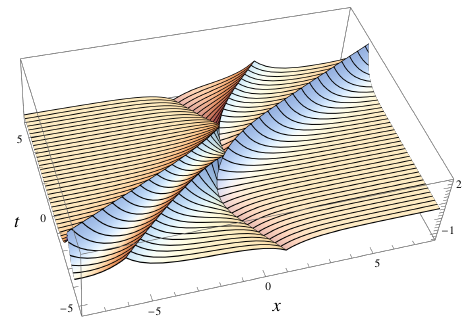}
  \caption{A conservative peakon--antipeakon solution $u(x,t)$
    of the Camassa--Holm equation,
    with two peakons and one antipeakon,
    plotted for $x \in [-8,8]$ and $t \in [-5,5]$.
    See explanations in the text,
    and also Figures~\ref{fig:CH-2peakons-1antipeakon-positions}
    and~\ref{fig:CH-2peakons-1antipeakon-3d-jigsaw}.
  }
  \label{fig:CH-2peakons-1antipeakon-3d}
\end{figure}

\begin{figure}
  \centering
  \includegraphics[width=1.0\linewidth]{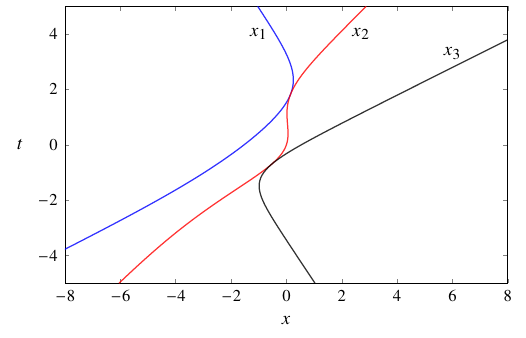}
  \caption{Positions $x = x_k(t)$ of the three individual peakons in the solution shown in
    Figure~\ref{fig:CH-2peakons-1antipeakon-3d}.
    The ordering $x_1(t) < x_2(t) < x_3(t)$ is preserved for all~$t$
    except at the instant of the first collision, where $x_1 < x_2 = x_3$,
    and at the instant of the second collision, where $x_1 = x_2 < x_3$.
    Before the first collision it is~$m_3$ that is negative,
    between the collisions it is~$m_2$,
    and after the second collision it is~$m_1$.
    At a collision where $x_k = x_{k+1}$,
    the amplitudes $m_k$ and $m_{k+1}$ interchange their signs
    by blowing up to $+\infty$ and~$-\infty$ (respectively) as the collision is approached,
    and then ``coming back'' from $-\infty$ and~$+\infty$ afterwards.
    As Figure~\ref{fig:CH-2peakons-1antipeakon-3d} illustrates,
    the solution $u(x,t)$ extends continuously to the instant of collision,
    but the derivative $u_{x}(x,t)$ tends to~$-\infty$ for $x_k(t) < x < x_{k+1}(t)$,
    and then ``comes back'' from $+\infty$ immediately after the collision.
    The peakons asymptotically travel in straight lines as $t \to \pm\infty$,
    in this example with the asymptotic velocities $\{ 2, 1, -\tfrac23 \}$,
    corresponding to the parameter values $\lambda_1 = \tfrac12$,
    $\lambda_2 = 1$ and $\lambda_3 = -\tfrac32$ in the
    solution formulas~\eqref{eq:CH-threepeakon-x-m}.
  }
  \label{fig:CH-2peakons-1antipeakon-positions}
\end{figure}

\begin{figure}
  \centering
  \includegraphics[width=0.95\linewidth]{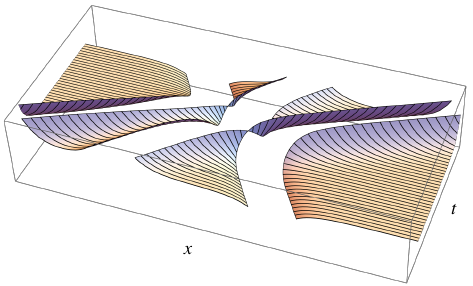}
  \caption{A ``jigsaw puzzle'' version of Figure~\ref{fig:CH-2peakons-1antipeakon-3d},
    where the surface has been cut open along the curves $x = x_k(t)$
    from Figure~\ref{fig:CH-2peakons-1antipeakon-positions}, and the pieces pulled apart and
    shown from a different angle, for better visibility.
  }
  \label{fig:CH-2peakons-1antipeakon-3d-jigsaw}
\end{figure}

In an initially well-separated mixed peakon--antipeakon solution,
the individual (positive) peakons will start out moving to the right
and the individual antipeakons will start out moving to the left,
like travelling waves.
But as a peakon at site~$k$ and an antipeakon at site $k+1$ approach each other,
their dynamics becomes more subtle;
for example, if they are close enough
and $m_k > \abs{m_{k+1}}$,
then $u(x_k)$ and $u(x_{k+1})$ may both be positive,
so that the peakon and the antipeakon will both move to the right.
Despite this, it turns out that what will actually happen is that there will be a
\emph{peakon--antipeakon collision}
at some finite time $t_0$\,:
as $t$ approaches $t_0$ from below,
$x_{k+1}(t) - x_k(t) \to 0$,
$m_k(t) \to \infty$ and $m_{k+1}(t) \to -\infty$,
in such a way that cancellation in the sum~\eqref{eq:intro-peakons} causes
the wave profile $u(x,t)$ to have a continuous limiting shape $u(x,t_0)$.
Moreover, the wave profile becomes ever steeper on the shrinking interval
between the peakon and the antipeakon;
in fact,
$u_x(x,t) \to -\infty$ for $x_k(t) < x < x_{k+1}(t)$
in such a way that the contribution from that interval to the $H^1$-norm of~$u$,
\begin{equation*}
  \int_{x_k(t)}^{x_{k+1}(t)} \bigl( u^2 + u_x^2 \bigr) \, dx
  ,
\end{equation*}
tends to a positive constant as $t \nearrow t_0$.
Much effort has been spent on understanding what happens at such finite-time
blow-ups (both for peakons and more general solutions)
and how to continue the solution into the time region $t \ge t_0$.
Various scenarios are possible, as will be described briefly in
Example~\ref{ex:CH-two-peakons} and in Section~\ref{sec:guide-CH}.
Figures~\ref{fig:CH-2peakons-1antipeakon-3d}, \ref{fig:CH-2peakons-1antipeakon-positions}
and~\ref{fig:CH-2peakons-1antipeakon-3d-jigsaw}
illustrate a so-called \emph{conservative} solution
with two peakons and one antipeakon,
where the $H^1$-norm of the solution drops at the instant of each collision,
but immediately returns to its previous value as the peakon and antipeakon reappear
with their roles reversed -- it is now $m_k$ that is negative and  $m_{k+1}$ that is positive.

The role of peakons in the theory of water waves, with particular emphasis
on variational principles and asymptotic expansions, has been reviewed
in an authoritative work by Holm~\cite{holm:2006:peakons-review}.
Our paper here emphasizes a very different aspect of the mathematics of peakons,
namely the intriguing connections between PDEs admitting peakon solutions on the one hand,
and classical analysis, especially the theory of orthogonal
polynomials and approximation theory, on the other.
This theory, which is still unfolding, has been developed over the
years and at numerous locations, such as
Minneapolis (USA), New Haven (USA), Saskatoon (Canada),
Linköping (Sweden), Montreal (Canada), Shanghai (China)
and Beijing (China).

In Section~\ref{sec:CH} we discuss the Camassa--Holm equation
in more detail, describing the formative ideas that initiated the connection
between approximation theory and peakon solutions of integrable PDEs.
We will show how these tools make it possible to derive explicit
formulas for the general solution of the nonlinear ODEs~\eqref{eq:intro-CH-peakon-ode-explicit}
governing the dynamics of CH peakons,
and to analyze the behaviour of these solutions in great detail,
for example at peakon--antipeakon
collisions~\cite{beals-sattinger-szmigielski:1998:acoustic-scattering-KdV-hierarchy, beals-sattinger-szmigielski:1999:stieltjes, beals-sattinger-szmigielski:2000:moment}.

\begin{figure}
  \centering
  \includegraphics[width=0.95\linewidth]{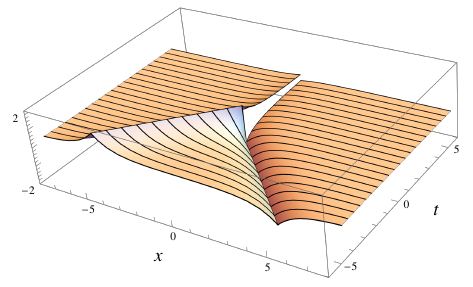}
  \caption{The simplest example of shock formation in the Degasperis--Procesi equation
    \eqref{eq:intro-DP}:
    a peakon and an antipeakon of equal strength collide head-on at $(x,t)=(0,0)$,
    forming a stationary shockpeakon which stays at $x=0$ and decays to zero
    as $t \to \infty$.
    The solution, here plotted for $x \in [-8,8]$ and $t \in [-5,5]$,
    is given by equation~\eqref{eq:intro-DP-symmetric-collision}.
  }
  \label{fig:DP-symmetric-collision}
\end{figure}

We will also discuss some other related PDEs which likewise admit peakon solutions
and have inspired an interesting progression of ideas and techniques.
Historically, the first of these ``post-CH peakon equations''
was the Degasperis--Procesi (DP) equation
\begin{equation}
  \label{eq:intro-DP}
  m_t + (u m)_x + 2 u_x m = 0
  ,\qquad
  m = u - u_{xx}
  ,
\end{equation}
alternatively written as
\begin{equation}
  \label{eq:intro-DP-b-form}
  m_t + m_x u + 3 m u_x = 0
  ,\qquad
  m = u - u_{xx}
\end{equation}
or in expanded form as
\begin{equation}
  \label{eq:intro-DP-expanded}
  u_t - u_{xxt} + 4 u u_x = 3  u_x u_{xx} + u u_{xxx}
  .
\end{equation}
This PDE was identified by Degasperis and Procesi~\cite{degasperis-procesi:1999:asymptotic-integrability}
as being the only equation besides the KdV and CH equations
(and up to coordinate transformations) within the family
$u_t + c_0 u_x + \gamma u_{xxx} - \alpha^2 u_{xxt} = (c_1 u^2 + c_2 u_x^2 + c_3 u u_{xx})_x$
to satisfy asymptotic integrability conditions up to the third order.
Later on, various other integrability tests
\cite{mikhailov-novikov:2002:perturbative,
  hone-wang:2003:prolongation-algebras,
  ivanov:2005:integrability-class-nonlinear-dispersive-wave-equations}
have also identified the CH ($b=2$)
and DP ($b=3$) equations as the only integrable cases in the ``$b$-family''
\begin{equation}
  \label{eq:intro-b-family}
  m_t + m_x u + b m u_x = 0
  ,\qquad
  m = u - u_{xx}
  .
\end{equation}
A few years after the discovery of the DP equation,
Degasperis, Holm and Hone~\cite{degasperis-holm-hone:2002:new-integrable-equation-DP}
showed that it indeed possesses a Lax pair and other attributes of integrability,
as well as peakon solutions of the same form~\eqref{eq:intro-peakons} as the CH equation,
but with a slightly different set of ODEs governing the dynamics of the peakons, namely
\begin{equation}
  \label{eq:intro-DP-peakon-ode}
  \dot x_k = u(x_k)
  ,\qquad
  \dot m_k = -2 m_k \, u_x(x_k)
  .
\end{equation}
Despite being similar in appearance to the CH equation,
the DP equation has quite a different underlying integrability structure,
and its peakon solutions are connected to
approximation theory in a novel and remarkable way,
as we will explain in Section~\ref{sec:DP},
via the concepts of \emph{the discrete cubic string},
\emph{mixed Hermite--Padé approximations}
and \emph{Cauchy biorthogonal polynomials}~\cite{lundmark-szmigielski:2003:DPshort,lundmark-szmigielski:2005:DPlong,bertola-gekhtman-szmigielski:2009:cubicstring,bertola-gekhtman-szmigielski:2010:cauchy}.
Another major difference is that the DP equation admits weak solutions that need not
even be continuous~\cite{coclite-karlsen:2006:DPwellposedness, coclite-karlsen:2007:DPuniqueness},
and in fact peakon--antipeakon collisions lead to the formation
of so-called \emph{shock\-peakons}~\cite{lundmark:2007:DP-shockpeakons}
with jump singularities in~$u$ rather than in~$u_x$.
The simplest case is the antisymmetric one,
where a peakon and an antipeakon of equal strength collide:
\begin{equation}
  \label{eq:intro-DP-symmetric-collision}
  u(x,t) =
  \begin{cases}
    \dfrac{e^{-\abs{x-t}} - e^{-\abs{x+t}}}{1-e^{2t}}
    ,&
    t < 0
    ,\\[1em]
    \dfrac{- \sgn(x) \, e^{-\abs{x}}}{1+t}
    ,&
    t \ge 0
    .
  \end{cases}
\end{equation}
This is illustrated in Figure~\ref{fig:DP-symmetric-collision}.

\begin{figure}
  \centering
  \includegraphics[width=0.95\linewidth]{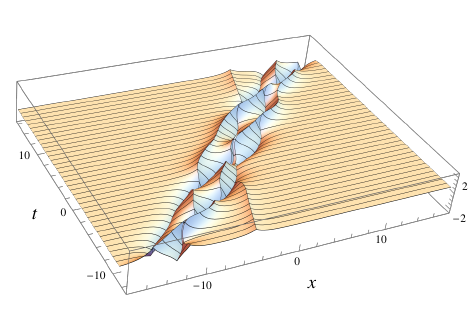}
    \caption{A solution of the Novikov equation \eqref{eq:intro-Novikov}
      where a single peakon interacts with a ``breather-like''
      cluster consisting of two peakons and two antipeakons.
      The graph of $u(x,t)$
      is plotted for $x \in [-18,18]$ and $t \in [-12,15]$ from the exact solution
      formulas~\eqref{eq:Novikov-n-peakon-solution} with $N=5$.
      See also Figure~\ref{fig:novikov-4cluster+single}.
    }
  \label{fig:novikov-4cluster+single-3d}
\end{figure}

\begin{figure}
  \centering \includegraphics[width=1.0\linewidth]{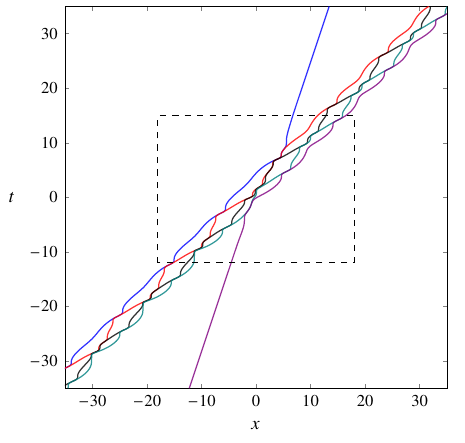}
  \caption{Positions $x = x_k(t)$ of the five individual peakons in the solution
    from Figure~\ref{fig:novikov-4cluster+single-3d}, with the dashed rectangle
    indicating the region shown there.
    The rightmost peakon (at~$x_5$) travels alone before the interaction,
    and then joins the cluster, while the leftmost peakon (at~$x_1$) leaves it.
    Note that the pattern of oscillations within the cluster after the interaction
    is not the same as it was before.
    The reciprocal eigenvalues $1/\lambda_k$ in the associated spectral problem
    are $1 \pm i$ and $1 \pm \tfrac13 i$ (with the common real part giving the
    asymptotic velocity of the cluster, and the imaginary parts giving the
    angular frequencies of the oscillations within it)
    and $\tfrac13$ (giving the asymptotic velocity of the lonely peakon).
  }
  \label{fig:novikov-4cluster+single}
\end{figure}

\begin{figure}
  \centering \includegraphics[width=1.0\linewidth]{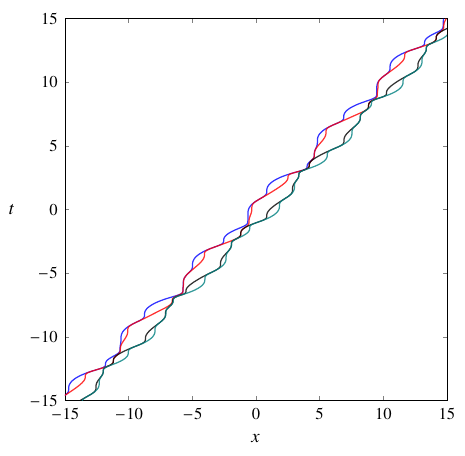}
  \caption{Positions $x = x_k(t)$, $1 \le k \le 4$, for a solution of the Novikov equation
    consisting of a cluster of two peakons and two antipeakons
    displaying quasiperiodic oscillations with two incommensurable frequencies.
    The reciprocal eigenvalues $1/\lambda_k$ in the associated spectral problem
    are $1 \pm i$ and $1 \pm \sqrt5 \, i$.}
  \label{fig:novikov-4cluster-quasiperiodic}
\end{figure}

\begin{figure}
  \centering \includegraphics[width=1.0\linewidth]{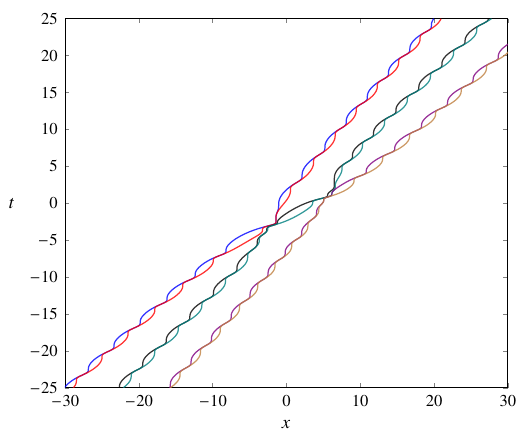}
  \caption{Positions $x = x_k(t)$, $1 \le k \le 6$, for a solution of the Novikov equation
    containing three peakon--antipeakon pairs which
    separate from each other at a logarithmic rate as $t \to \pm\infty$,
    although they all have the same asymptotic velocity.
    This case requires more complicated solution formulas
    than~\eqref{eq:Novikov-n-peakon-solution},
    since the eigenvalues in the associated spectral problem are non-simple
    (there are triple eigenvalues at $1/(1 \pm i)$);
    see Remark~\ref{rem:Novikov-peakon-antipeakon}.}
  \label{fig:novikov-conjugate-triple-roots}
\end{figure}

Later came the Novikov equation,
\begin{equation}
  \label{eq:intro-Novikov}
  m_t + \bigl( (u m)_x + 2 u_x m \bigr) \, u = 0
  ,\qquad
  m = u - u_{xx},
\end{equation}
which differs in appearance from the DP equation only by the extra factor~$u$,
so that the nonlinearity is cubic, as opposed to quadratic for CH and~DP.
It was singled out by Novikov~\cite{novikov:2009:generalizations-of-CH}
using a perturbative symmetry approach,
with Hone and Wang~\cite{hone-wang:2008:cubic-nonlinearity}
providing a Lax pair for it and initiating the study of its peakon solutions,
which are governed by the ODEs
\begin{equation}
  \label{eq:intro-Novikov-peakon-ode}
  \dot x_k = u(x_k)^2
  ,\qquad
  \dot m_k = - m_k \, u(x_k)\, u_x(x_k)
  ,
\end{equation}
which look like the CH peakon ODEs~\eqref{eq:intro-CH-peakon-ode}
except for the extra factor $u(x_k)$ in the equations for $x_k$ and~$m_k$.
Because of the square, $\dot x_k$ cannot be negative, so peakons and antipeakons alike
move to the right. Despite this, peakon--antipeakon collisions do occur,
with $u$ remaining continuous as in the CH case,
rather than developing a shock as in the DP case.
However, mixed peakon--antipeakon solutions of Novikov's equation
display a much greater variety of possible behaviours than
those of the CH equation;
see Figures
\ref{fig:novikov-4cluster+single-3d},
\ref{fig:novikov-4cluster+single},
\ref{fig:novikov-4cluster-quasiperiodic}
and~\ref{fig:novikov-conjugate-triple-roots}
for some examples,
and Remark~\ref{rem:Novikov-peakon-antipeakon} for more information.
As will be explained in Section~\ref{sec:Novikov},
the integrability of Novikov's equation is related to something called
\emph{the dual cubic string}, making it possible to reuse results
from the study of the DP equation in quite a striking way~\cite{hone-lundmark-szmigielski:2009:novikov}.

The next peakon equation that we will discuss,
in Section~\ref{sec:GX},
is the Geng--Xue (GX) equation~\cite{geng-xue:2009:GX-peakon-equation-cubic-nonlinearity},
\begin{equation}
  \label{eq:intro-GX}
  \begin{gathered}
    m_t + \bigl( (u m)_x + 2 u_x m \bigr) \, v = 0
    , \\
    n_t + \bigl( (v n)_x + 2 v_x n \bigr) \, u = 0
    , \\
    m = u - u_{xx}
    ,\quad
    n = v - v_{xx}
    ,
  \end{gathered}
\end{equation}
an integrable two-component system found by generalizing the Lax pair
for the Novikov equation.
(It is sometimes called the two-component Novikov equation,
but beware that there are also other systems going by that name.)
The study of GX peakons is interesting in that it is the first case
which involves the setup of Cauchy biorthogonal polynomials in its full
generality, with two independent spectral measures coming from two
different Lax pairs~\cite{lundmark-szmigielski:2016:GX-inverse-problem,lundmark-szmigielski:2017:GX-dynamics-interlacing}.
A curious detail is that the Lax pairs do not in general provide sufficiently many
constants of motion for solving the peakon ODEs --
the explicit integration hinges on the existence of additional constants of motion
not encoded in the spectral measures~\cite{shuaib-lundmark:2019:GX-noninterlacing}.
Already in the pure peakon case,
peakon solutions of the GX equation show a very rich and complicated behaviour
compared to CH or~DP.
We will describe this briefly in Section~\ref{sec:GX},
but due to the multitude of phenomena and cases that can occur,
we refer to the works cited above for illustrated examples.
One new feature that can be mentioned already here is
that the peakon amplitudes in general grow or decay exponentially as $t \to \pm\infty$,
rather than approaching constant values,
and their logarithms display phase shifts similar to the ones seen for the positions.

Our last example,
to be treated in Section~\ref{sec:mCH},
is the modified Camassa--Holm (mCH) equation,
also known as the FORQ equation,
\begin{equation}
  \label{eq:intro-mCH}
  m_t + \bigl( (u^2-u_x^2) m \bigr)_x = 0
  , \qquad
  m = u - u_{xx}
  .
\end{equation}
This PDE has quite a convoluted history, which
will be discussed briefly in Section~\ref{sec:guide-mCH}.
When dealing with its peakon solutions,
the concept of a \emph{distributional Lax pair}
comes to the forefront.
In this case, when $m=u-u_{xx}$ is a discrete measure (see Section~\ref{sec:CH}), the Lax pair
contains certain problematic terms which involve multiplying a Dirac delta with a discontinuous
function that jumps precisely where the delta is supported,
and in order to preserve the Lax integrability
one is forced to pick a particular interpretation of these terms.
An interesting phenomenon occurring for the mCH equation
(as opposed to the other equations discussed so far)
is that the peakon ODEs obtained in that way are \emph{different} from the peakon ODEs
obtained by defining a general concept of weak solution for the PDE in question,
and requiring the ansatz~\eqref{eq:intro-peakons} to satisfy this definition.
From our point of view here, it is the Lax integrable version of the mCH $N$-peakon ODEs
that are of interest, where explicit solution formulas can be obtained by solving an inverse
spectral problem whose core is formed by certain multi-point Padé approximations.

Finally, Section~\ref{sec:guide} contains various comments and remarks that did not fit
into the narrative of the main text, and we also provide plenty of additional references there.
The literature on peakons,
not to mention other aspects of the Camassa--Holm equation and its relatives,
is very extensive, so it is impossible to give
justice to each and every contribution to the theory,
but we hope that this will at least provide some guidance for the reader who wishes to explore the topic further.
For the sake of readability, the number of references has been kept to a minimum in the other sections.

\section{The Camassa--Holm equation and the eigenvalue problem for a vibrating string}
\label{sec:CH}

Much of the material reviewed in this section is classical mathematics,
whose application to peakons is due to
Beals, Sattinger and Szmigielski~\cite{beals-sattinger-szmigielski:1998:acoustic-scattering-KdV-hierarchy,beals-sattinger-szmigielski:1999:stieltjes,beals-sattinger-szmigielski:2000:moment}.
Further references will be given in Section~\ref{sec:guide-CH}.

The first step in getting familiar with peakon solutions is to understand what happens to
the quantity $m = u - u_{xx}$ in the Camassa--Holm equation~\eqref{eq:intro-CH}
when $u$ is given by the multipeakon ansatz~\eqref{eq:intro-peakons}.
To begin with, consider for simplicity the case
\begin{equation*}
  u = e^{-\abs{x}}
  =
  \begin{cases}
    e^x, & x \le 0
    ,\\
    e^{-x}, & x \ge 0
    .
  \end{cases}
\end{equation*}
Then the first derivative $u_x$ is undefined at $x=0$, since the left and right derivatives are unequal there,
and for $x \neq 0$ we have
\begin{equation*}
  u_x =
  \begin{cases}
    e^x, & x < 0
    ,\\
    -e^{-x}, & x > 0
    .
  \end{cases}
\end{equation*}
The classical derivative of this is of course
\begin{equation*}
  u_{xx} =
  \begin{cases}
    e^x, & x < 0
    ,\\
    e^{-x}, & x > 0
    ,
  \end{cases}
\end{equation*}
which agrees with the original function~$u = e^{-\abs{x}}$ except for being undefined at $x=0$.
However, here we are instead going to take the derivative in the sense of distributions,
so that the jump of size $-2$ at $x=0$ in the first derivative $u_x$
gives rise to a Dirac delta term $-2 \delta(x)$ in the second derivative.
Keeping the same notation $u_{xx}$,
to avoid a proliferation of different symbols for the derivative,
we may thus write the distributional second derivative as
\begin{equation*}
  u_{xx} = -2 \delta(x) +
  \begin{cases}
  e^x, & x < 0
  ,\\
  e^{-x}, & x > 0
  ,
  \end{cases}
\end{equation*}
which in the sense of distributions is the same thing as
the more convenient expression
\begin{equation*}
  u_{xx} = -2 \delta(x) + e^{-\abs{x}}
  .
\end{equation*}
As a consequence, we obtain
\begin{equation*}
  m = u - u_{xx} = 2 \delta(x)
  .
\end{equation*}
Hence, by linearity, if $u(x,t)$ is given by~\eqref{eq:intro-peakons} then the quantity $m = u - u_{xx}$
becomes a linear combination of Dirac deltas,
\begin{equation}
  \label{eq:CH-peakon-m}
  m(x,t) = 2 \sum^N_{k=1} m_k(t) \, \delta \bigl( x-x_k(t) \bigr)
  ,
\end{equation}
i.e., for each fixed~$t$, we have a discrete signed measure $m(\cdot,t)$ on the real line~$\R$.
Now recall that a Dirac delta distribution at $x=a$ may be multiplied by a continuous function~$f$,
according to the rule $f(x) \, \delta(x-a) = f(a) \, \delta(x-a)$.
With this in mind, we realize that the other equation in \eqref{eq:intro-CH},
$m_t + (um)_x + u_x m = 0$,
is not really well-defined as it stands,
since it involves multiplying the discrete measure~$m$
by the function $u_x$ whose value is undefined precisely at the points $x_k$ where $m$ is supported.

One way of resolving this problem is to assign a value to $u_x$ at $x=x_k(t)$, so that
we may define ``$u_x(x) \, \delta(x-x_k) = u_x(x_k) \, \delta(x-x_k)$'',
or more explicitly
\begin{equation*}
  u_x(x,t) \, m(x,t) = 2 \sum^N_{k=1} u_x(x_k(t),t) \, m_k(t) \, \delta \bigl( x-x_k(t) \bigr)
  .
\end{equation*}
It turns out that the correct choice,
which ensures that the same manipulations that for smooth functions
lead to from the Lax equations~\eqref{eq:intro-CH-lax} to the PDE~\eqref{eq:intro-CH-expanded}
are also valid in the discrete case,
is to take $u_x(x_k)$ equal to the average $\tfrac12 \bigl( u_x(x_k^+) + u_x(x_k^-) \bigr)$,
just as in the shorthand notation used in equation~\eqref{eq:intro-CH-peakon-ode-shorthand}.
A short computation, identifying coefficients of $\delta(x-x_k)$ and $\delta'(x-x_k)$,
then shows that the equation $m_t + (um)_x + u_x m = 0$
is indeed satisfied in this regularized distributional sense
if and only if the peakon ODEs~\eqref{eq:intro-CH-peakon-ode-explicit} hold.

As we will see, this type of difficulty presented by \emph{ill-defined} terms is typical of
PDEs admitting peakon solutions.
The essential question is that of uniqueness of regularizing such expressions.
Let us briefly describe a general strategy for addressing this question,
using the term $u_x m$ as an example.
The starting point is the relation $m=u-u_{xx}$. Since $m$ is a
measure, $u$ must be at least continuous. Hence $m-u$ is a measure.
This implies that $u_{xx}$ is a measure, so $u_x$ is a function of
bounded variation (BV).
Such functions, being the difference of monotone functions,
have one-sided limits from the right and from the left at every point,
which allows us to define a product of a BV function~$f$ and the measure~$m$ in a natural way.
Denote the left and right limit of~$f$ at~$x$ by $f^-(x)$ and $f^+(x)$, respectively,
let $\alpha$ and~$\beta$ be two real numbers such that $\alpha +\beta=1$,
and define
\begin{equation}
  \label{eq:CH-multiplication-f-times-m}
  f m = (\alpha f^- + \beta f^+) \, m
  .
\end{equation}
This reduces to the normal multiplication rule at all points of continuity of~$f$,
whereas at the countably many points where $f$ is not continuous
the multiplier of~$m$ is a fixed linear combination of the left and right limits.
It turns out that for peakon equations the choice of $\alpha$ and $\beta$ is dictated by the Lax pairs, if we want to preserve Lax integrability.
In the case of the CH equation the unique choice dictated by Lax integrability is
$\alpha = \beta = \frac12$,
resulting in the arithmetic average of the left and right limits of $u_x$ indicated earlier.
We will revisit this issue in Sections \ref{sec:Novikov} and~\ref{sec:mCH}.

Another approach to making sense of peakon solutions is to rewrite the CH equation as
\begin{equation}
  \begin{split}
    0
    &= (u-u_{xx})_t + 3 uu_x - 2 u_x u_{xx} - uu_{xxx}
    \\
    &= (1-\partial_x^2) \Bigl( u_t + \bigl( \tfrac12 u^2 \bigr)_x \Bigr)
    + \bigl( u^2 + \tfrac12 u_x^2 \bigr)_x
  \end{split}
\end{equation}
and apply the inverse of the differential operator $(1-\partial_x^2)$,
which for solutions vanishing as $\abs{x} \to \infty$
is given by convolution with the function $\tfrac12 e^{-\abs{x}}$.
This gives
\begin{equation}
  \label{eq:CH-eulerian}
  u_t + \partial_x \biggl(
  \tfrac12 u^2 + \tfrac12 e^{-\abs{x}} * \bigl( u^2+\tfrac12 u_x^2 \bigr)
  \biggr)
  = 0
  ,
\end{equation}
and weak solutions are then defined as functions (vanishing at infinity)
which satisfy this equation in a more usual weak sense
(multiply by a test function from a suitable class, integrate by parts, etc.).
With such an approach one finds again, although the calculations are now more involved,
that the peakon ansatz~\eqref{eq:intro-peakons} is a weak solution if and only if
the quantities $x_k$ and~$m_k$ satisfy the system of ODEs~\eqref{eq:intro-CH-peakon-ode-explicit}.

We remark that one may of course also study periodic weak solutions,
in particular periodic peakon solutions,
and then the inverse of $1-\partial_x^2$ will be different,
but we will not consider that case here
(see however Beals et~al. \cite{beals-sattinger-szmigielski:2002:CF-flows-periodic-peakons-2-peakon-2002, beals-sattinger-szmigielski:2005:CF-flows-periodic-peakons-N-peakon-2005}).

Any connection to orthogonal polynomials is totally hidden at this point.
To start revealing that connection,
we make a Liouville transformation, i.e., a change of dependent and independent variables
with the purpose of eliminating the constant term $-\tfrac14$ in the differential operator
$\partial_x^2 - \tfrac14$
appearing in the first Lax equation~\eqref{eq:intro-CH-lax-x}.
Since the time-dependence only enters when considering
the other Lax equation~\eqref{eq:intro-CH-lax-t},
which we will not do for a while yet,
let us for now consider $t$ to be fixed, and omit it in the notation,
so that~\eqref{eq:intro-CH-lax-x} reads
\begin{equation}
  \label{eq:CH-lax-x-again}
  \bigl( \partial_x^2 - \tfrac14 \bigr) \, \psi(x) = -\tfrac12 \lambda \, m(x) \, \psi(x)
  ,\qquad
  x \in \R
  .
\end{equation}
Now let
\begin{equation}
  \label{eq:CH-liouville-trf}
  y = \tanh(x/2)
  ,\qquad
  \psi(x) = \frac{\phi(y)}{\sqrt{1-y^2}}
  .
\end{equation}
For smooth functions it is easily verified using the chain rule that
the Liouville transformation \eqref{eq:CH-liouville-trf} turns
the ODE \eqref{eq:CH-lax-x-again} into
\begin{equation}
  \label{eq:CH-gstring}
  \partial_y^2 \phi(y) = -\lambda \, g(y) \, \phi(y)
  ,\qquad
  -1 < y < 1
  ,
\end{equation}
where
\begin{equation}
  \label{eq:CH-g-m}
  \tfrac12 (1-y^2)^2 g(y) = m(x)
  .
\end{equation}
Note that $\partial_x^2 - \tfrac14$ has become just $\partial_y^2$,
so that the term $-\frac14$ has been eliminated, as promised.

Equation \eqref{eq:CH-gstring}, when considered together with Dirichlet
boundary conditions $\phi(\pm 1)=0$,
is nothing but the classical eigenvalue problem for the vibrational modes of
a string attached at both ends,
like a guitar or violin string,
but with mass density varying from point to point, as described by the function $g(y)$
(which in this physical situation is positive).
Such an \emph{inhomogeneous string} is modelled by the linear wave equation
$g(y) \, \partial^2 w / \partial \tau^2 = \partial^2 w / \partial y^2$
for the deflection $w(y,\tau)$,
and \eqref{eq:CH-gstring} arises when separating the variables as $w(y,\tau) = \phi(y) \, T(\tau)$,
together with the harmonic oscillator ODE $T_{\tau\tau} = -\lambda T$ for the time-dependent part.
The boundary value problem \eqref{eq:CH-gstring} with $\phi(\pm 1)=0$
has a nontrivial solution only for certain positive values of~$\lambda$,
whose square roots are the eigenfrequencies of the string.
The eigen\-oscillations are sinusoidal with respect to the time variable~$\tau$,
but the corresponding spatial eigenfunctions $\phi(y)$ are in general \emph{not} sinusoidal;
the sinusoidal eigenfunctions that most of us are perhaps used to seeing are
an exceptional case that happens for a \emph{homogeneous} string (when $g(y)$ is constant).

We will come back to the CH equation towards the end of this section,
and in particular explain why the particular boundary conditions $\phi(\pm 1)=0$
are relevant,
but first we are going to further explore the Dirichlet eigenvalue problem for the string,
with particular emphasis on the \emph{discrete} case which arises when
considering peakon solutions of the form~\eqref{eq:intro-peakons},
\begin{equation*}
  u(x) = \sum_{k=1}^N m_k \, e^{-\abs{x-x_k}}
  .
\end{equation*}
Then we do not have a smooth function $m(x)$ in~\eqref{eq:CH-lax-x-again}
but instead a discrete measure of the form~\eqref{eq:CH-peakon-m},
\begin{equation*}
  m(x) = 2 \sum_{k=1}^N m_k \, \delta(x-x_k)
  .
\end{equation*}
In this case, we transform the Dirac deltas according to the rule
\begin{equation}
  \label{eq:CH-dirac-transform-rule}
  \delta(x-x_k) = \frac{\delta(y-y_k)}{\frac{dx}{dy}(y_k)}
  = \tfrac12 (1-y_k^2) \, \delta(y-y_k)
  ,
\end{equation}
to obtain \eqref{eq:CH-gstring} with a discrete measure~$g$ on the interval $(-1,1)$,
namely
\begin{subequations}
  \label{eq:CH-peakon-gk-yk}
\begin{equation}
  \label{eq:CH-peakon-g}
  g(y) = \sum_{k=1}^N g_k \, \delta(y-y_k)
  ,\qquad
  g_k = \frac{2 m_k}{1-y_k^2}
  ,
\end{equation}
where (of course)
\begin{equation}
  y_k = \tanh(x_k/2)
  .
\end{equation}
\end{subequations}
This situation corresponds to a \emph{discrete string},
an idealized object consisting of point masses of weight~$g_k$ at the positions~$y_k$,
connected by weightless string.
As we will explain later in this section,
it is through this discrete string,
hiding inside the CH Lax pair~\eqref{eq:intro-CH-lax},
that orthogonal polynomials enter the picture.

But before we come to that, let us give an alternative (and very explicit) way of verifying
the relation \eqref{eq:CH-peakon-g} between the original discrete measure~$m$ and the transformed discrete
measure~$g$,
since this sheds some light on how the ODEs \eqref{eq:CH-lax-x-again} and \eqref{eq:CH-gstring}
work in the discrete case, and also on why the particular change of variables \eqref{eq:CH-liouville-trf}
does the trick of removing the term $-\tfrac14$.
Equation~\eqref{eq:CH-lax-x-again} tells us that the quantity
$\bigl( \partial_x^2 - \tfrac14 \bigr) \, \psi(x)$
must be zero in the intervals where $m$ is zero,
i.e., away from the points $x = x_k$.
Assuming as usual that $x_1 < \dots < x_N$,
this means that $\partial_x^2 \psi(x) = \tfrac14 \psi(x)$
in each of the $N+1$ intervals
\begin{equation*}
  (-\infty,x_1)
  ,\quad
  (x_1,x_2)
  ,\quad
  \ldots
  ,\quad
  (x_{N-1},x_N)
  ,\quad
  (x_N,\infty)
  .
\end{equation*}
If we define $x_0 = -\infty$ and $x_{N+1} = +\infty$ for notational convenience,
the conclusion is that $\psi(x)$ must take the piecewise defined form
\begin{equation}
  \label{eq:CH-psi-piecewise}
  \psi(x) = A_k \, e^{x/2} + B_k \, e^{-x/2}
  ,\qquad
  x_k < x < x_{k+1}
  ,
\end{equation}
for $0 \le k \le N$.
Moreover, $\psi$ should be continuous at each~$x_k$ (for $1 \le k \le N$),
so that the product $m(x) \, \psi(x)$ on the right-hand side
of~\eqref{eq:CH-lax-x-again} makes sense,
and the first derivative $\partial_x \psi$ should have a jump at each~$x_k$,
of size $-\tfrac12 \lambda \cdot 2 m_k \cdot \psi(x_k)$,
so that the second derivative $\partial_x^2 \psi$ on the left-hand side gives rise to
Dirac deltas matching those appearing on the right-hand side.
A bit of calculation shows that these requirements are equivalent to the jump conditions
\begin{equation}
  \label{eq:CH-jump-relation-psi}
  \begin{pmatrix}
    A_k \\ B_k
  \end{pmatrix}
  =
  \left[
    \begin{pmatrix} 1 & 0 \\ 0 & 1 \end{pmatrix}
    - \lambda \, m_k
    \begin{pmatrix} 1 \\ -e^{x_k} \end{pmatrix}
    \bigl( 1, e^{-x_k} \bigr)
  \right]
  \begin{pmatrix}
    A_{k-1} \\ B_{k-1}
  \end{pmatrix}
  ,
\end{equation}
for $1 \le k \le N$,
relating the constants $(A_k, B_k)$ in one interval to the constants $(A_{k-1},B_{k-1})$
in the preceding one.
Thus, for a given value of~$\lambda$,
the solution $\psi(x)$ of equation~\eqref{eq:CH-lax-x-again}
is completely determined by the constants $(A_0,B_0)$,
which may be arbitrary, so that the solution space is two-dimensional
(as expected, since the ODE is of second order).
Now, with
\begin{equation*}
  y = \tanh(x/2)
  = \frac{e^{x/2} - e^{-x/2}}{e^{x/2} + e^{-x/2}}
  = \frac{e^x-1}{e^x+1}
\end{equation*}
as in \eqref{eq:CH-liouville-trf},
we have
\begin{equation*}
  1+y = \frac{2 e^x}{e^x+1}
  , \qquad
  1-y = \frac{2}{e^x+1}
  ,
\end{equation*}
so that an expression of the form
$\psi(x) = A \, e^{x/2} + B \, e^{-x/2}$,
in the kernel of the operator $\partial_x^2 - \tfrac14$,
can be expressed in terms of~$y$ as
\begin{equation*}
  \begin{split}
    \psi(x) &= A \, e^{x/2} + B \, e^{-x/2}
    \\ &
    = \frac{e^x+1}{2e^{x/2}} \left( A \, \frac{2e^x}{e^x+1} + B \, \frac{2}{e^x+1} \right)
    \\ &
    = \dfrac{1}{\sqrt{(1+y)(1-y)}}
    \biggl( A \, (1+y) + B \, (1-y) \biggr)
    \\ &
    = \dfrac{1}{\sqrt{1-y^2}} \, \phi(y)
    ,
  \end{split}
\end{equation*}
where $\phi(y) = A \, (1+y) + B \, (1-y) = (A-B) y + (A+B)$ is a first-degree polynomial,
and hence in the kernel of the operator~$\partial_y^2$.
Consequently, the continuous piecewise hyperbolic function $\psi(x)$ (for $x \in \R$)
given by \eqref{eq:CH-psi-piecewise}
is mapped by \eqref{eq:CH-liouville-trf}
to the continuous piecewise linear function $\phi(y)$ (for $-1 < y < 1$) given by
\begin{equation}
  \label{eq:CH-phi-piecewise}
  \begin{split}
    \phi(y) &
    = A_k \, (1+y) + B_k \, (1-y)
    \\ &
    =  (A_k - B_k) \, y + (A_k + B_k)
    ,\qquad
    y_k < y < y_{k+1}
    .
  \end{split}
\end{equation}
(Since we have defined $x_0 = -\infty$ and $x_{N+1} = +\infty$,
the relation $y_k = \tanh(x_k/2)$ gives $y_0 = -1$ and $y_{N+1} = +1$.)
Clearly we have $\partial_y^2 \phi = 0$ in each interval $(y_k,y_{k+1})$.
Moreover, by multiplying the jump conditions~\eqref{eq:CH-jump-relation-psi} from the left
by the row vector $(1,-1)$, we see that the piecewise constant slope $\partial_y \phi$,
which equals $A_k - B_k$ for $y \in (y_k,y_{k+1})$,
satisfies
\begin{equation*}
  \begin{split}
    &
    (A_k - B_k) - (A_{k-1} - B_{k-1})
    \\ &
    = - \lambda m_k \bigl( 1 + e^{x_k} \bigr) \bigl( A_{k-1} + B_{k-1} e^{-x_k} \bigr)
    \\ &
    = - \lambda m_k \, \frac{e^{x_k}+1}{e^{x_k/2}} \, \bigl( A_{k-1} e^{x_k/2} + B_{k-1} e^{-x_k/2} \bigr)
    \\ &
    = - \lambda m_k \, \frac{e^{x_k}+1}{e^{x_k/2}} \, \psi(x_k)
    \\ &
    = - \lambda m_k \, \frac{2}{\sqrt{1-y_k^2}} \, \frac{\phi(y_k)}{\sqrt{1-y_k^2}}
    \\ &
    = - \lambda \, \underbrace{\frac{2 m_k}{1-y_k^2}}_{=g_k} \, \phi(y_k)
    ,
  \end{split}
\end{equation*}
i.e., it jumps by $-\lambda \, g_k \, \phi(y_k)$ at $y=y_k$,
where $g_k = 2 m_k / (1-y_k^2)$.
So $\phi(y)$ does indeed satisfy the distributional ODE
\eqref{eq:CH-gstring} with the transformed measure~\eqref{eq:CH-peakon-g},
as claimed.

For a given value of $\lambda$,
the solution \eqref{eq:CH-phi-piecewise}
of the discrete string equation $\phi_{yy} = -\lambda g \phi$
is uniquely determined by the constants $A_0$ and~$B_0$
in the leftmost subinterval $(-1,y_1)$ of the interval $(-1,1)$,
or equivalently by the initial values $\phi(-1) = 2B_0$ and $\phi_y(-1) = A_0 - B_0$
at the left endpoint $y=-1$, if we extend $\phi$ to the closed interval $[-1,1]$.
In order to study the eigenvalue problem with Dirichlet boundary conditions,
we can think of it as a shooting problem, where we start with $\phi(-1) = 0$
at the left endpoint
and try to ``aim'' by determining $\lambda$ so that we hit $\phi(1)=0$
when we reach the right endpoint.
We can normalize by letting $\phi_y(-1)=1$,
since eigenfunctions are only determined up to a constant factor anyway.
These choices correspond to $(A_0,B_0)=(1,0)$,
and we will denote this particular solution by $\phi(y;\lambda)$.
By letting
\begin{equation}
  \Phi(y) = \begin{pmatrix} \phi(y) \\ \phi_y(y) \end{pmatrix}
  ,
\end{equation}
the second-order string equation $\phi_{yy} = -\lambda g \phi$
can be written as a system of two first-order equations,
or equivalently a $2 \times 2$ matrix equation,
\begin{equation}
  \label{eq:CH-string-equation-matrix-form}
  \partial_y \Phi(y) =
  \begin{pmatrix}
    0 & 1 \\
    - \lambda g(y) & 0
  \end{pmatrix}
  \Phi(y)
  .
\end{equation}
Then $\Phi(y;\lambda) = \bigl( \phi(y;\lambda), \phi_y(y;\lambda) \bigr)^T$
is the unique solution starting with $\Phi(-1) = (0,1)^T$.
Since $\phi(y) = \phi(y_k) + \phi_y(y_k^+) \, (y-y_k)$
for $y_k \le y \le y_{k+1}$, we have
\begin{equation}
  \label{eq:CH-jump-matrix-Lk}
  \Phi(y_{k+1}^-;\lambda)
  = L_k \, \Phi(y_{k}^+;\lambda)
  ,\quad
  L_k =
  \begin{pmatrix}
    1 & l_k \\
    0 & 1
  \end{pmatrix}
  ,
\end{equation}
where
\begin{equation}
  \label{eq:CH-lk}
  l_k = y_{k+1} - y_k
  ,\quad
  0 \le k \le N
  .
\end{equation}
The jump condition for $\phi_y$ at~$y_k$ becomes
\begin{equation}
  \label{eq:CH-jump-matrix-Gk}
  \Phi(y_{k}^+;\lambda)
  = G_k(\lambda) \, \Phi(y_{k}^-;\lambda)
  ,\quad
  G_k(\lambda) =
  \begin{pmatrix}
    1 & 0 \\
    -\lambda \, g_k & 1
  \end{pmatrix}
  .
\end{equation}
Combining these relations, we can work our way to the right endpoint $y=1$:
\begin{equation}
  \label{eq:CH-string-right-endpoint-matrix-product}
  \Phi(1;\lambda)
  = L_N \, G_N(\lambda) \, L_{N-1} \, G_{N-1}(\lambda) \dotsm L_1 \, G_1(\lambda) \, L_0
  \begin{pmatrix} 0 \\ 1 \end{pmatrix}
  .
\end{equation}
It is not difficult to verify from this that both components of $\Phi(1;\lambda)$
are polynomials in~$\lambda$ of degree~$N$:
\begin{equation}
  \begin{aligned}
    \label{eq:CH-string-right-endpoint-polynomials}
    \phi(1;\lambda) &= 2 + \cdots + (-\lambda)^{N} g_1 g_2 \dotsm g_N \, l_0 l_1 \dotsm l_{N-1} l_N
    ,\\
    \phi_y(1;\lambda) &= 1 + \cdots + (-\lambda)^{N} g_1 g_2 \dotsm g_N \, l_0 l_1 \dotsm l_{N-1}
    .
  \end{aligned}
\end{equation}
For $\lambda$ to be an eigenvalue of the discrete string with Dirichlet boundary
conditions $\phi(\pm 1)=0$, the function $\phi(y;\lambda)$
must hit zero at the right endpoint $y=1$
(recall that it's already zero at the left endpoint $y=-1$, by definition).
In other words, the eigenvalues are precisely the roots of the $N$th-degree polynomial $\phi(1;\lambda)$.
It can be shown~\cite{beals-sattinger-szmigielski:2000:moment}
that these eigenvalues $\lambda = \lambda_k$ ($1 \le k \le N$)
are real and simple,
and that there are as many positive eigenvalues~$\lambda_k$
as there are positive weights~$g_k$.
Hence, since the eigenvalues are obviously nonzero due to $\phi(1;0) = 2 \neq 0$,
there are also as many negative eigenvalues~$\lambda_k$
as there are negative weights~$g_k$.
(For a physical string, the weights are positive, but in order to deal with antipeakons
we need to allow negative weights as well.)

The eigenfunctions
\begin{equation*}
  \phi_k(y) = \phi(y;\lambda_k)
\end{equation*}
are orthogonal in the $L^2$-space on the interval $[-1,1]$ with weight~$g$,
i.e.,
\begin{equation}
  \begin{split}
    0
    = \langle \phi_i, \phi_j \rangle
    &
    = \int_{-1}^1 \phi_i(y) \, \phi_j(y) \, dg(y)
    \\ &
    = \sum_{k=1}^N g_k \, \phi_i(y_k) \, \phi_j(y_k)
  \end{split}
\end{equation}
for $i \neq j$.
This follows easily from the usual computation where $\partial_y^2 \phi_i = -\lambda_i g \phi_i$
is multiplied by $\phi_j$ and subtracted from the corresponding expression with $i$ and~$j$ swapped,
and then integrated over $[-1,1]$.
Note that the $L^2$-space is $N$-dimensional in this discrete case;
since the elements are actually not functions $\phi(y)$ but only
equivalence classes up to equality almost everywhere with respect to the discrete measure~$g$,
they can be represented by the $N$-tuples $\bigl( \phi(y_1),\dots, \phi(y_N) \bigr)$.

Next we define the so-called \emph{Weyl function} of the discrete string:
\begin{equation}
  \label{eq:CH-weyl-function-def}
  W(\lambda) = \frac{\phi_y(1;\lambda)}{\phi(1;\lambda)}
  .
\end{equation}
Clearly, this is a rational function with simple poles at the eigenvalues
$\lambda_1$, \ldots,~$\lambda_N$.
It turns out to be somewhat more convenient to work with the modified Weyl function
$\omega(\lambda) = W(\lambda)/\lambda$,
so that $\omega(\lambda) = O(1/\lambda)$ as $\lambda \to \infty$.
This modified Weyl function has an additional simple pole
at $\lambda=\lambda_0=0$ with residue $W(0)=1/2=a_0$;
denoting the residues at the other poles by~$a_k$,
the partial fractions decomposition of~$\omega$ is
\begin{equation}
  \label{eq:CH-weyl-function-parfrac}
  \omega(\lambda)
  = \frac{W(\lambda)}{\lambda}
  = \frac{1/2}{\lambda} + \sum_{k=1}^N \frac{a_k}{\lambda - \lambda_k}
  = \sum_{k=0}^N \frac{a_k}{\lambda - \lambda_k}
  .
\end{equation}
This sum can be written as an integral
\begin{equation}
  \label{eq:CH-weyl-function-stieltjes-transform}
  \omega(\lambda)
  = \frac{W(\lambda)}{\lambda}
  = \int \frac{d\alpha(z)}{\lambda-z}
\end{equation}
with respect to the discrete measure
\begin{equation}
  \label{eq:CH-peakon-spectral-measure}
  \alpha(\lambda)
  = \tfrac12 \delta(\lambda) + \sum_{k=1}^N a_k \, \delta(\lambda-\lambda_k)
  = \sum_{k=0}^N a_k \, \delta(\lambda-\lambda_k)
  ,
\end{equation}
called the \emph{spectral measure} of the discrete string,
and an integral of the form \eqref{eq:CH-weyl-function-stieltjes-transform} is known as
the \emph{Stieltjes  transform} (or Cauchy transform) of the measure~$\alpha$.

It should be clear that the spectral measure $\alpha(\lambda)$
is uniquely determined by the measure~$g(y)$,
i.e., by the positions~$y_k$ and the weights~$g_k$ in the discrete string,
since the Weyl function $W(\lambda)$ can be explicitly computed from these numbers
via~\eqref{eq:CH-string-right-endpoint-matrix-product}.
The eigenvalues~$\lambda_k$ and the residues~$a_k$
are only defined up to a permutation of the indices,
but if we pick some definite way of ordering the eigenvalues,
say in increasing order $\lambda_1 < \dots < \lambda_N$,
then the spectral data $\{ \lambda_k, a_k \}_{k=1}^N$ are uniquely determined
by the string data $\{ y_k, g_k \}_{k=1}^N$.

The residues $\{ a_k \}_{k=1}^N$ are always positive,
regardless of the signs of the weights~$g_k$.
This can be proved as follows.
To begin with, $\phi = \phi(y;\lambda)$ satisfies
\begin{equation*}
  \begin{split}
    (\phi_y \, \phi_\lambda - \phi \, \phi_{\lambda y})_y
    &
    = \phi_{yy} \, \phi_\lambda - \phi \, \phi_{\lambda yy}
    \\ &
    = \phi_{yy} \, \phi_\lambda + \phi \, (-\phi_{yy})_{\lambda}
    \\ &
    = (-\lambda g \phi) \, \phi_\lambda + \phi \, (\lambda g \phi)_\lambda
    \\ &
    = - \lambda g \phi \, \phi_\lambda + \phi \, ( g \phi + \lambda g \phi_\lambda )
    \\ &
    = g \phi^2
    ,
  \end{split}
\end{equation*}
and thus, if we multiply by~$\lambda$,
\begin{equation}
  \label{eq:CH-ak-proof-0}
  \begin{split}
    \lambda (\phi_y \, \phi_\lambda - \phi \, \phi_{\lambda y})_y
    &
    = \lambda g \phi^2
    = (\lambda g \phi) \, \phi
    \\ &
    = - \phi_{yy} \, \phi
    \\ &
    = \phi_y^2 - (\phi \, \phi_y)_y
    .
  \end{split}
\end{equation}
We have $\phi(-1;\lambda)=0$ for all~$\lambda$ by definition, which implies that $\phi_\lambda(-1;\lambda)=0$
for all~$\lambda$ too.
If we now integrate~\eqref{eq:CH-ak-proof-0} over $y \in [-1,1]$,
and then evaluate at $\lambda=\lambda_k$ in order to use $\phi(1;\lambda_k)=0$,
the only thing which survives is therefore
\begin{equation}
  \label{eq:CH-ak-proof-1}
  \lambda_k \, \phi_y(1;\lambda_k) \, \phi_\lambda(1;\lambda_k)
  = \int_{-1}^1 \phi_y(y;\lambda_k)^2 \, dy
  .
\end{equation}
The right-hand side is clearly positive, so $\phi_\lambda(1;\lambda_k) \neq 0$,
which incidentally proves our earlier claim that the eigen\-values are simple,
and we can use the method of differentiating the denominator to obtain the residue at a simple pole.
Together with~\eqref{eq:CH-ak-proof-1}, this gives
\begin{equation}
  \label{eq:CH-ak-proof-2}
  \begin{split}
    a_k
    & = \res_{\lambda=\lambda_k} \frac{\phi_y(1;\lambda)}{\lambda \, \phi(1;\lambda)}
    = \left[ \frac{\phi_y(1;\lambda)}{\partial_\lambda \bigl( \lambda \, \phi(1;\lambda) \bigr) } \right]_{\lambda=\lambda_k}
    \\ &
    = \frac{\phi_y(1;\lambda_k)}{\lambda_k \, \phi_\lambda(1;\lambda_k)}
    = \frac{\phi_y(1;\lambda_k)^2}{\lambda_k \, \phi_\lambda(1;\lambda_k) \, \phi_y(1;\lambda_k)}
    \\ &
    = \frac{\phi_y(1;\lambda_k)^2}{ \int_{-1}^1 \phi_y(y;\lambda_k)^2 \, dy}
    > 0
    ,
  \end{split}
\end{equation}
as claimed.
The determination of the spectral data $\{ \lambda_k, a_k \}_{k=1}^N$
from the string parameters $\{ y_k, g_k \}_{k=1}^N$
is referred to as the \emph{(forward) spectral problem} for the discrete string,
and it is natural to investigate the \emph{inverse spectral problem}: can the
string parameters be reconstructed from the spectral data?
As Krein found out in the 1950s, the answer is yes, since the problem
is equivalent to already solved problems about continued fractions and Padé approximation.
To get a feeling for the kind of relations involved,
let us first take a down-to-earth look at the forward and inverse problems in the case $N=2$.

\begin{example}
  \label{ex:CH-string-two-masses}
  Consider a discrete string consisting of
  two point masses of nonzero weight $g_1$ and~$g_2$, respectively,
  at the two points $y_1 < y_2$ in the interval $(-1,1)$.
  From~\eqref{eq:CH-string-right-endpoint-matrix-product} we have
  \begin{equation*}
    \begin{split}
      \begin{pmatrix} \phi(1;\lambda) \\ \phi_y(1;\lambda) \end{pmatrix}
      &
      = L_2 \, G_2(\lambda) \cdot L_1 \, G_1(\lambda) \cdot L_0
      \begin{pmatrix} 0 \\ 1 \end{pmatrix}
      \\ &
      =
      \begin{pmatrix} 1 - \lambda g_2 l_2 & l_2 \\ -\lambda g_2 & 1 \end{pmatrix}
      \begin{pmatrix} 1 - \lambda g_1 l_1 & l_1 \\ -\lambda g_1 & 1 \end{pmatrix}
      \begin{pmatrix} l_0 \\ 1 \end{pmatrix}
      \\ &
      =
      \begin{pmatrix}
        2 - C \lambda + D \lambda^2
        \\
        1 - A \lambda + B \lambda^2
      \end{pmatrix}
      ,
    \end{split}
  \end{equation*}
  where we use the (temporary) abbreviations
  \begin{equation}
    \begin{aligned}
      A &= g_1 l_0 + g_2 (l_0 + l_1)
      ,\\
      B &= g_1 g_2 l_0 l_1
      ,\\
      C &= g_1 l_0 (l_1 + l_2) + g_2 (l_0 + l_1) l_2
      ,\\
      D &= g_1 g_2 l_0 l_1 l_2
      ,
    \end{aligned}
  \end{equation}
  with the positive interval lengths
  $l_0 = y_1 - (-1)$, $l_1 = y_2 - y_1$ and $l_2 = 1 - y_2$;
  note that $l_0 + l_1 + l_2 = 2$ is the length of the whole
  interval $[-1,1]$.
  Thus the modified Weyl function is
  \begin{equation*}
    \begin{split}
      \frac{W(\lambda)}{\lambda}
      &
      = \frac{\phi_y(1;\lambda)}{\lambda \, \phi(1;\lambda)}
      = \frac{1 - A \lambda + B \lambda^2}{\lambda \, (2 - C \lambda + D \lambda^2)}
      \\ &
      = \frac{1/2}{\lambda}
        + \frac{(\tfrac12 C - A) + (B - \tfrac12 D) \lambda}{2 - C \lambda + D \lambda^2}
      \\ &
      = \frac{1/2}{\lambda}
        + \frac{\frac{2B-D}{2D} \, \lambda + \frac{C-2A}{2D}}{\lambda^2 - \frac{C}{D} \, \lambda + \frac{2}{D}}
      ,
    \end{split}
  \end{equation*}
  which we compare to the expression from \eqref{eq:CH-weyl-function-parfrac},
  \begin{equation*}
    \begin{split}
      \frac{W(\lambda)}{\lambda}
      &
      = \frac{1/2}{\lambda} + \frac{a_1}{\lambda - \lambda_1} + \frac{a_2}{\lambda - \lambda_2}
      \\ &
      = \frac{1/2}{\lambda}
        + \frac{(a_1 + a_2) \lambda - (\lambda_2 a_1 + \lambda_1 a_2)}{\lambda^2 - (\lambda_1 + \lambda_2) \lambda + \lambda_1 \lambda_2}
      ,
    \end{split}
  \end{equation*}
  to obtain
  \begin{equation}
    \label{eq:CH-two-masses-forward-inverse-problem}
    \begin{aligned}
      \lambda_1 + \lambda_2 &= C/D
      ,&
      a_1 + a_2 &= (2B-D)/2D
      ,\\
      \lambda_1 \lambda_2 &= 2/D
      ,&
      \lambda_2 a_1 + \lambda_1 a_2 &= (2A-C)/2D
      .
    \end{aligned}
  \end{equation}

  For the forward spectral problem, we compute $A$, $B$, $C$ and~$D$ from
  $\{ y_1, y_2, g_1, g_2 \}$,
  solve the quadratic equation
  $\phi(1;\lambda) = 2 - C \lambda + D \lambda^2 = 0$
  to find
  \begin{equation*}
    \lambda_{1,2} = \tfrac{C}{2D} \pm \sqrt{\bigl( \tfrac{C}{2D} \bigr)^2 - \tfrac{2}{D}}
    ,
  \end{equation*}
  and then a pair of linear equations to find $a_1$ and~$a_2$.
  If $g_1$ and~$g_2$ are of opposite sign, then $\lambda_1$ and~$\lambda_2$
  are real and of opposite sign, since $D<0$.
  If $g_1$ and~$g_2$ have the same sign, the quantity under the square root is still positive,
  since it can be written as
  \begin{equation*}
    \frac{\bigl[ g_1 l_0 (2-l_0) - g_2 l_2 (2-l_2) \bigr]^2 + 4 g_1 g_2 (l_0 l_2)^2}{(2D)^{2}}
    ;
  \end{equation*}
  it is also less than $(C/2D)^2$ since $D>0$,
  so $\lambda_1$ and $\lambda_2$ are real and of the same sign (equal to the sign of $g_1$ and~$g_2$).
  These observations verify, in this particular case,
  the general claim that we made earlier
  about the sign pattern of the eigenvalues.
  And as we proved above, $a_1$ and~$a_2$ must be positive.

  For the inverse spectral problem,
  \eqref{eq:CH-two-masses-forward-inverse-problem} is equivalent to
  \begin{equation}
    \label{eq:CH-two-masses-inverse-problem-naively}
    \begin{aligned}
      g_1 l_0 + g_2 (2-l_2) = A &= \frac{1 + 2 a_1}{\lambda_1} + \frac{1 + 2 a_2}{\lambda_2}
      ,\\
      g_1 g_2 l_0 (2-l_0-l_2) = B & = \frac{1 + 2(a_1+a_2)}{\lambda_1 \lambda_2}
      ,\\
      g_1 l_0 (2-l_0) + g_2 (2-l_2) \, l_2 = C & = 2 \left( \frac{1}{\lambda_1} + \frac{1}{\lambda_2} \right)
      ,\\
      g_1 g_2 l_0 (2 - l_0 - l_2) \, l_2 = D & = \frac{2}{\lambda_1 \lambda_2}
      ,
    \end{aligned}
  \end{equation}
  where the right-hand sides are known through the given quantities
  (nonzero distinct real numbers $\lambda_1$ and~$\lambda_2$,
  and positive real numbers $a_1$ and~$a_2$),
  and we seek the four unknowns $g_1$, $g_2$, $l_0 = 1+y_1$
  and $l_2 = 1-y_2$.
  From $D/B$ we immediately find
  $l_2 = 1/(\tfrac12 + a_1 + a_2)$,
  and then $B / (C - l_2 A)$ gives us~$g_2$.
  With these quantities known, we may compute $g_1 l_0 = A - g_2 (2-l_2)$,
  then $l_0 = 2 - l_2 - B /( g_2 \cdot (g_1 l_0) )$,
  and finally $g_1 = (g_1 l_0) / l_0$,
  with the following outcome:
  \begin{equation}
    \label{eq:CH-two-masses-inverse-problem-solution}
    \begin{aligned}
      y_1 &= 1 - \frac{\lambda_1^2 a_1 + \lambda_2^2 a_2}{ \tfrac12 ( \lambda_1^2 a_1 + \lambda_2^2 a_2 ) + (\lambda_1-\lambda_2)^2 a_1 a_2}
      ,\\
      y_2 &= 1 - \frac{1}{\tfrac12 + a_1 + a_2}
      ,\\
      g_1 &= \frac{ \bigr( \tfrac12 ( \lambda_1^2 a_1 + \lambda_2^2 a_2 ) + (\lambda_1 - \lambda_2)^2 a_1 a_2 \bigr)^2}{(\lambda_1 a_1 + \lambda_2 a_2) (\lambda_1 - \lambda_2)^2 \lambda_1 \lambda_2 a_1 a_2}
      ,\\
      g_2 &= \frac{(\tfrac12 + a_1 + a_2)^2}{\lambda_1 a_1 + \lambda_2 a_2}
      .
    \end{aligned}
  \end{equation}
  Since $a_{1,2} > 0$ we see that $y_{1,2} \in (-1,1)$,
  and moreover
  \begin{equation}
    \label{eq:CH-two-masses-inverse-problem-l1}
    \begin{split}
      l_1 &
      = y_2 - y_1
      \\ &
      = \frac{(\lambda_1 a_1 + \lambda_2 a_2)^2}{(\tfrac12 + a_1 + a_2) \, \bigl( \tfrac12 (\lambda_1^2 a_1 + \lambda_2^2 a_2) + (\lambda_1-\lambda_2)^2 a_1 a_2 \bigr)}
      ,
    \end{split}
  \end{equation}
  so that $y_1 \le y_2$ always, with equality iff $\lambda_1 a_1 + \lambda_2 a_2 = 0$.
  The conclusion is that discrete strings with $g_{1,2} > 0$ are in one-to-one correspondence with
  tuples $(\lambda_1,\lambda_2,a_1,a_2)$ such that $0 < \lambda_1 < \lambda_2$ and $a_{1,2}>0$,
  that discrete strings with $g_{1,2}< 0$ are in one-to-one correspondence with
  tuples such that $\lambda_1 < \lambda_2 < 0$ and $a_{1,2}>0$,
  and that discrete strings with $g_1$ and $g_2$ of opposite sign are in one-to-one correspondence with
  tuples such that $\lambda_1 < 0 < \lambda_2$, $a_{1,2}>0$ and $\lambda_1 a_1 + \lambda_2 a_2 \ne 0$.
\end{example}

Next we will describe the solution of the inverse spectral problem for arbitrary~$N$.
This will explain the structure apparent in the
formulas~\eqref{eq:CH-two-masses-inverse-problem-solution},
something that our ``brute force'' calculations in the example above
gave no clues about.

Fix an integer~$r$ with $1 \le r \le N$ and let
\begin{equation}
  \label{eq:CH-matrix-product-even}
  X(\lambda) = L_N \, G_N(\lambda) \dotsm L_{N+1-r} \, G_{N+1-r}(\lambda)
\end{equation}
be the product of the leftmost $2r$ factors in~\eqref{eq:CH-string-right-endpoint-matrix-product}.
Note that each factor has determinant~$1$, and hence $\det\bigl( X(\lambda) \bigr) = 1$ as well.
The entries in the first column of~$X(\lambda)$,
let us call them $Q(\lambda) = X_{11}(\lambda)$ and $P(\lambda) = X_{21}(\lambda)$,
are polynomials in~$\lambda$ of degree~$r$,
whose constant terms come from the first column in the matrix product
\begin{equation*}
  \begin{split}
    &
    L_N \, G_N(0) \dotsm L_{N+1-r} \, G_{N+1-r}(0)
    \\ &
    =
    \begin{pmatrix} 1 & l_N \\ 0 & 1 \end{pmatrix}
    \dotsm
    \begin{pmatrix} 1 & l_{N+1-r} \\ 0 & 1 \end{pmatrix}
    = \begin{pmatrix} 1 & l_N + \dotsb + l_{N+1-r} \\ 0 & 1 \end{pmatrix}
    ,
  \end{split}
\end{equation*}
revealing that $Q(0)=1$ and $P(0)=0$,
while the highest-degree coefficients (of $\lambda^r$) come from the first column in the matrix product
\begin{equation*}
  \begin{split}
    &
    L_N
    \begin{pmatrix} 0 & 0 \\ -g_N & 0 \end{pmatrix}
    \dotsm
    L_{N+1-r}
    \begin{pmatrix} 0 & 0 \\ -g_{N+1-r} & 0 \end{pmatrix}
    \\ &
    = (-g_N) \dotsm (-g_{N+1-r})
    \begin{pmatrix} l_N & 0 \\ 1 & 0 \end{pmatrix}
    \dotsm
    \begin{pmatrix} l_{N+1-r} & 0 \\ 1 & 0 \end{pmatrix}
    \\ &
    = (-1)^r g_N \dotsm g_{N+1-r}
    \begin{pmatrix} l_N \, l_{N-1} \dotsm l_{N+1-r} & 0 \\ \phantom{l_N \,} l_{N-1} \dotsm l_{N+1-r} & 0 \end{pmatrix}
    ,
  \end{split}
\end{equation*}
so that in particular we have
\begin{equation}
  \label{eq:CH-pade-first-Q-highest-coeff}
  \begin{split}
    Q(\lambda)
    = 1 + \cdots
    + (-\lambda)^{r} g_{N} \dotsm g_{N+1-r} \, l_{N} \dotsm l_{N+1-r}
    .
  \end{split}
\end{equation}
We also see from this that the entries $X_{12}(\lambda)$ and $X_{22}(\lambda)$
in the second column have degree at most $r-1$.

On the other hand, the product of the remaining factors
in~\eqref{eq:CH-string-right-endpoint-matrix-product},
\begin{equation*}
  \begin{split}
    & \Phi(y_{N+1-r}^-;\lambda)
    \\ &
    = L_{N-r} \, G_{N-r}(\lambda) \, L_{N-1} \, G_{N-1}(\lambda) \dotsm L_1 \, G_1(\lambda) \, L_0
    \begin{pmatrix} 0 \\ 1 \end{pmatrix}
    ,
  \end{split}
\end{equation*}
is a vector that we may call $(q(\lambda),p(\lambda))^T$,
where $q(\lambda)$ and $p(\lambda)$ can be seen to have degree~$N-r$.
So \eqref{eq:CH-string-right-endpoint-matrix-product} can be written as
\begin{equation*}
  \begin{pmatrix}
    \phi(1;\lambda) \\
    \phi_y(1;\lambda)
  \end{pmatrix}
  =
  \begin{pmatrix}
    X_{11}(\lambda) & X_{12}(\lambda) \\
    X_{21}(\lambda) & X_{22}(\lambda)
  \end{pmatrix}
  \begin{pmatrix} q(\lambda) \\ p(\lambda) \end{pmatrix}
  ,
\end{equation*}
where we can divide the components to obtain
\begin{equation*}
  W(\lambda) = \frac{\phi_y(1;\lambda)}{\phi(y;\lambda)}
  = \frac{X_{21}(\lambda) \, q(\lambda) + X_{22}(\lambda) \, p(\lambda)}{X_{11}(\lambda) \, q(\lambda) + X_{12}(\lambda) \, p(\lambda)}
  .
\end{equation*}
Thus, if we for simplicity's sake omit $\lambda$ for a moment, we get
\begin{equation*}
  \begin{split}
    W - \frac{P}{Q}
    &
    = \frac{X_{21} \, q + X_{22} \, p}{X_{11} \, q + X_{12} \, p}
    - \frac{X_{21}}{X_{11}}
    \\ &
    = \frac{X_{11} (X_{21} \, q + X_{22} \, p) - X_{21} (X_{11} \, q + X_{12} \, p)}{(X_{11} \, q + X_{12} \, p) \, X_{11}}
    \\ &
    = \frac{0 \, q + \det(X) \, p}{(X_{11} \, q + X_{12} \, p) \, X_{11}}
    \\ &
    = \frac{p}{(Q \, q + X_{12} \, p) \, Q}
    = O \left( \frac{1}{\lambda^{2r}} \right)
    ,
  \end{split}
\end{equation*}
as $\lambda \to \infty$ (considering the degrees given above).
After multiplication by~$Q$, this gives
the \emph{Padé approximation} condition
\begin{equation*}
  W(\lambda) \, Q(\lambda)
  = P(\lambda)
  + O \left( \frac{1}{\lambda^{r}} \right)
  ,
\end{equation*}
as $\lambda \to \infty$,
which expresses how well the rational Weyl function~$W$
is approximated by $P/Q$, another rational function involving polynomials of lower degrees.
Since $P(0)=0$, we can divide by~$\lambda$ and express this condition
in terms of the modified Weyl function $\omega(\lambda) = W(\lambda)/\lambda$
and the polynomial $\widetilde P(\lambda) = P(\lambda)/\lambda$:
\begin{equation}
  \label{eq:CH-first-pade-approx}
  \omega(\lambda) \, Q(\lambda)
  = \widetilde P(\lambda)
  + O \left( \frac{1}{\lambda^{r+1}} \right)
  .
\end{equation}
The polynomials $Q(\lambda)$ and $\widetilde P(\lambda)$
(of degree $r$ and $r-1$, respectively,
and with $Q(0)=1$)
are uniquely determined by this condition.
Indeed, using \eqref{eq:CH-weyl-function-stieltjes-transform} we
can expand $\omega(\lambda)$ in powers of $1/\lambda$,
\begin{equation}
  \label{eq:CH-W-power-series}
  \begin{split}
    \omega(\lambda)
    &
    = \frac{W(\lambda)}{\lambda}
    = \int \frac{d\alpha(z)}{\lambda (1-z/\lambda)}
    \\ &
    = \int \left( 1 + \frac{z}{\lambda}
      + \left( \frac{z}{\lambda} \right)^2
      + \left( \frac{z}{\lambda} \right)^3
      + \dotsb \right) \frac{d\alpha(z)}{\lambda}
    \\ &
    = \frac{\alpha_0}{\lambda} + \frac{\alpha_1}{\lambda^2} + \frac{\alpha_2}{\lambda^3} + \dotsb
    ,
  \end{split}
\end{equation}
where
\begin{equation}
  \label{eq:CH-def-moments}
  \alpha_n = \int z^n \, d\alpha(z)
  = \sum_{k=0}^N \lambda_k^n \, a_k
\end{equation}
is the $n$th moment of the spectral measure~\eqref{eq:CH-peakon-spectral-measure},
and if we insert the Laurent series~\eqref{eq:CH-W-power-series}
into~\eqref{eq:CH-first-pade-approx} together with
\begin{equation}
  Q(\lambda) = 1 + q_1 \lambda + q_2 \lambda^2 + \dotsb + q_r \lambda^r
  ,
\end{equation}
and multiply the two expressions on the left-hand side,
the absence of the powers
$1/\lambda$, \ldots, $1/\lambda^r$ on the right-hand side
imposes $r$ linear equations for the $r$ unknown coefficients~$q_i$:
\begin{equation}
  \label{eq:CH-stieltjes-orthogonality-matrix-form}
  \begin{pmatrix}
    \alpha_0 & \alpha_1 & \alpha_2 & \dots & \alpha_r \\
    \alpha_1 & \alpha_2 & \alpha_3 & \dots & \alpha_{r+1} \\
    \vdots & \vdots & \vdots & \vdots & \vdots \\
    \alpha_{r-1} & \alpha_r & \alpha_{r+1} & \dots & \alpha_{2r-1}
  \end{pmatrix}
  \begin{pmatrix}
    1 \\ q_1 \\ q_2 \\ \vdots \\ q_r
  \end{pmatrix}
  =
  \begin{pmatrix}
    0 \\ 0 \\ 0 \\ \vdots \\ 0
  \end{pmatrix}
  .
\end{equation}
We can write this as
\begin{equation*}
  \begin{pmatrix}
    \alpha_1 & \alpha_2 & \dots & \alpha_r \\
    \alpha_2 & \alpha_3 & \dots & \alpha_{r+1} \\
    \vdots & \vdots & \vdots & \vdots \\
    \alpha_r & \alpha_{r+1} & \dots & \alpha_{2r-1}
  \end{pmatrix}
  \begin{pmatrix}
    q_1 \\ q_2 \\ \vdots \\ q_r
  \end{pmatrix}
  =
  -
  \begin{pmatrix}
    \alpha_0 \\ \alpha_1 \\ \vdots \\ \alpha_{r-1}
  \end{pmatrix}
\end{equation*}
and solve for $q_1, \dots, q_r$ using Cramer's rule, to obtain
\begin{equation}
  \label{eq:CH-Qn-even-det}
  Q(\lambda) =
  \frac{
    \begin{vmatrix}
      1 & \lambda & \lambda^2 & \dots & \lambda^r \\
      \alpha_0 & \alpha_1 & \alpha_2 & \dots & \alpha_r \\
      \alpha_1 & \alpha_2 & \alpha_3 & \dots & \alpha_{r+1} \\
      \vdots & \vdots & \vdots & \vdots & \vdots \\
      \alpha_{r-1} & \alpha_r & \alpha_{r+1} & \dots & \alpha_{2r-1}
    \end{vmatrix}
  }{
    \begin{vmatrix}
      \alpha_1 & \alpha_2 & \dots & \alpha_r \\
      \alpha_2 & \alpha_3 & \dots & \alpha_{r+1} \\
      \vdots & \vdots & \vdots & \vdots \\
      \alpha_r & \alpha_{r+1} & \dots & \alpha_{2r-1}
    \end{vmatrix}
  }
  .
\end{equation}
(We will see in Remark~\ref{rem:CH-range-of-spectral-map}
that the determinant in the denominator is typically nonzero,
although exceptional cases may occur when the eigenvalues~$\lambda_k$ do not all have the same sign.)
Once the polynomial $Q(\lambda)$ is known,
$\widetilde P(\lambda) = P(\lambda)/\lambda$ is also determined by~\eqref{eq:CH-first-pade-approx},
as the polynomial part of the Laurent series for $\omega(\lambda) \, Q(\lambda)$.

Let us introduce the notation
\begin{equation}
  \label{eq:CH-def-Hankel-det}
  \Delta_k^n =
  \begin{cases}
    1
    , & k=0
    ,\\
    \det \bigl( \alpha_{n+i+j-2} \bigr)_{i,j=1,\dots,k}
    \, , & k>0
    .
  \end{cases}
\end{equation}
In other words, if
\begin{equation}
  H =
  \begin{pmatrix}
    \alpha_0 & \alpha_1 & \alpha_2 & \alpha_3 & \dots \\
    \alpha_1 & \alpha_2 & \alpha_3 & \alpha_4 & \dots \\
    \alpha_2 & \alpha_3 & \alpha_4 & \alpha_5 & \dots \\
    \alpha_3 & \alpha_4 & \alpha_5 & \alpha_6 & \dots \\
    \vdots & \vdots & \vdots & \vdots & \ddots
  \end{pmatrix}
\end{equation}
is an infinite Hankel matrix
(i.e., constant along anti-diagonals)
containing the moments~$\alpha_k$,
then $\Delta_k^n$ (for $k>0$) is the determinant of a $k\times k$ submatrix
with $\alpha_n$ in the upper left corner.

Then the denominator in~\eqref{eq:CH-Qn-even-det} is~$\Delta_r^1$,
and from the cofactor of $\lambda^r$ in the upper right corner
we see that the highest coefficient is
$q_r = (-1)^r \Delta_r^0 / \Delta_r^1$,
which upon comparison with~\eqref{eq:CH-pade-first-Q-highest-coeff}
yields
\begin{equation}
  \label{eq:CH-string-inverse-first-determinant-quotient}
  g_{N} \dotsm g_{N+1-r} \, l_{N} \dotsm l_{N+1-r}
  = \Delta_r^0 / \Delta_r^1
  .
\end{equation}
Equation~\eqref{eq:CH-string-inverse-first-determinant-quotient}
holds for any $r=1,\dots,N$, and with this information we are halfway
to the solution of the inverse problem.

For the second half, let instead
\begin{equation}
  \label{eq:CH-matrix-product-odd}
  X(\lambda) = L_N \, G_N(\lambda) \dotsm L_{N+1-r} \, G_{N+1-r}(\lambda) \, L_{N-r}
\end{equation}
be the product of the first $2r+1$
factors in~\eqref{eq:CH-string-right-endpoint-matrix-product},
where we have fixed an~$r$ with $0 \le r \le N$,
and consider the entries in the second column, for which we will reuse
the same letters $Q$ and~$P$ again.
All four entries in~$X(\lambda)$, in particular
$Q(\lambda) = X_{12}(\lambda)$ and $P(\lambda) = X_{22}(\lambda)$,
are of degree~$r$, with $P(0)=1$ and
\begin{equation}
  \label{eq:CH-pade-second-Q-highest-lowest-coeff}
  \begin{split}
    Q(\lambda)
    &
    = (l_N + \dotsb + l_{N-r})
    + \cdots
    \\ & \quad
    + (-\lambda)^{r} g_{N} \dotsm g_{N+1-r} \, l_{N} \dotsm l_{N-r}
    .
  \end{split}
\end{equation}
(If $r=0$, this is to be understood as the constant polynomial $Q(\lambda) = l_N$.)
Again, \eqref{eq:CH-string-right-endpoint-matrix-product} becomes
\begin{equation*}
  \begin{pmatrix}
    \phi(1;\lambda) \\
    \phi_y(1;\lambda)
  \end{pmatrix}
  = X(\lambda)
  \begin{pmatrix} q(\lambda) \\ p(\lambda) \end{pmatrix}
  ,
\end{equation*}
but now with
$(q,p)^T = \Phi(y_{N-r}^+;\lambda)$,
so that $q$ and~$p$ have degree $N-r-1$ and $N-r$, respectively,
and we find
\begin{equation*}
  \begin{split}
    W - \frac{P}{Q}
    &
    = \frac{X_{21} \, q + X_{22} \, p}{X_{11} \, q + X_{12} \, p}
    - \frac{X_{22}}{X_{12}}
    \\ &
    = \frac{-\det(X) \, q + 0 \, p}{(X_{11} \, q + X_{12} \, p) \, X_{12}}
    \\ &
    = \frac{-q}{(X_{11} \, q + Q \, p) \, Q}
    = O \left( \frac{1}{\lambda^{2r+1}} \right)
    ,
  \end{split}
\end{equation*}
as $\lambda \to \infty$.
Since $P(0)=1$, we can write $P(\lambda) = 1 + \lambda \, \widetilde P(\lambda)$ with a
polynomial~$\widetilde P$ of degree $r-1$.
We then multiply by $Q$ and divide by~$\lambda$ to get
\begin{equation}
  \label{eq:CH-second-pade-approx}
  \omega(\lambda) \, Q(\lambda)
  = \widetilde P(\lambda)
  + \frac{1}{\lambda}
  + O \left( \frac{1}{\lambda^{r+2}} \right)
  .
\end{equation}
Compared to \eqref{eq:CH-first-pade-approx},
this is a slightly different type of Padé approximation,
but again the polynomials $Q$ and~$\widetilde P$ (of degree $r$ and $r-1$,
respectively) are uniquely determined by this condition.
Indeed, $\widetilde P$ will be determined once $Q$ is, and writing
\begin{equation*}
  Q(\lambda) = q_0 + q_1 \lambda + q_2 \lambda^2 + \dotsb + q_r \lambda^r
  ,
\end{equation*}
comparison of the coefficients of $1/\lambda$, \ldots, $1/\lambda^{r+1}$ on both sides
of~\eqref{eq:CH-second-pade-approx} gives $r+1$ linear equations
for the $r+1$ unknown coefficients~$q_i$:
\begin{equation}
  \begin{pmatrix}
    \alpha_0 & \alpha_1 & \alpha_2 & \dots & \alpha_r \\
    \alpha_1 & \alpha_2 & \alpha_3 & \dots & \alpha_{r+1} \\
    \vdots & \vdots & \vdots & \vdots & \vdots \\
    \alpha_{r-1} & \alpha_r & \alpha_{r+1} & \dots & \alpha_{2r-1} \\
    \alpha_{r} & \alpha_{r+1} & \alpha_{r+2} & \dots & \alpha_{2r}
  \end{pmatrix}
  \begin{pmatrix}
    q_0 \\ q_1 \\ q_2 \\ \vdots \\ q_r
  \end{pmatrix}
  =
  \begin{pmatrix}
    1 \\ 0 \\ \vdots \\ 0 \\ 0
  \end{pmatrix}
  .
\end{equation}
From Cramer's rule,
\begin{equation}
  \label{eq:CH-Qn-odd-det}
  Q(\lambda) =
  \frac{
    \begin{vmatrix}
      1 & \lambda & \lambda^2 & \dots & \lambda^r \\
      \alpha_1 & \alpha_2 & \alpha_3 & \dots & \alpha_{r+1} \\
      \vdots & \vdots & \vdots & \vdots & \vdots \\
      \alpha_{r} & \alpha_{r+1} & \alpha_{r+2} & \dots & \alpha_{2r}
    \end{vmatrix}
  }{
    \begin{vmatrix}
      \alpha_0 & \alpha_1 & \alpha_2 & \dots & \alpha_r \\
      \alpha_1 & \alpha_2 & \alpha_3 & \dots & \alpha_{r+1} \\
      \vdots & \vdots & \vdots & \vdots & \vdots \\
      \alpha_{r} & \alpha_{r+1} & \alpha_{r+2} & \dots & \alpha_{2r}
    \end{vmatrix}
  }
  .
\end{equation}
(In Remark~\ref{rem:CH-range-of-spectral-map}
we will see that the determinant $\Delta_{r+1}^0$
in this denominator is always nonzero.)
In particular, we obtain the highest and lowest coefficients
$q_r = (-1)^r \Delta_r^1 / \Delta_{r+1}^0$
and
$q_0 = \Delta_r^2 / \Delta_{r+1}^0$,
which together with~\eqref{eq:CH-pade-second-Q-highest-lowest-coeff} gives
\begin{equation}
  \label{eq:CH-string-inverse-second-determinant-quotient}
    g_{N} \dotsm g_{N+1-r} \, l_{N} \dotsm l_{N-r}
    = \Delta_r^1 / \Delta_{r+1}^0
\end{equation}
and
\begin{equation}
  \label{eq:CH-string-inverse-third-determinant-quotient}
  l_N + \dotsb + l_{N-r}
  = \Delta_r^2 / \Delta_{r+1}^0
  ,
\end{equation}
for $0 \le r \le N$.

Combining \eqref{eq:CH-string-inverse-first-determinant-quotient}
and \eqref{eq:CH-string-inverse-second-determinant-quotient},
we obtain
\begin{equation*}
  \begin{split}
    &
    l_N = \frac{\Delta_0^1}{\Delta_1^0}
    , \quad
    l_N g_N = \frac{\Delta_1^0}{\Delta_1^1}
    , \quad
    l_N g_N l_{N-1} = \frac{\Delta_1^1}{\Delta_2^0}
    , \\
    &
    l_N g_N l_{N-1} g_{N-1} = \frac{\Delta_2^0}{\Delta_2^1}
  ,
  \end{split}
\end{equation*}
and so on, and we can solve for the unknown quantities $g_k$ and~$l_k$
by looking at the ratios of successive expressions in this
sequence.
Since that means dividing two expressions which are both quotients of
two Hankel determinants, we get the answer in the form of
quotients involving four determinants:
\begin{subequations} \label{eq:CH-formula-gk-lk-yk}
\begin{equation}
  \label{eq:CH-formula-gk}
  g_{N+1-k} = \frac{\bigl( \Delta_{k}^0 \bigr)^2}{\Delta_{k-1}^1 \, \Delta_{k}^1}
  , \quad
  1 \le k \le N
  ,
\end{equation}
\begin{equation}
  \label{eq:CH-formula-lk}
  l_{N-k} = \frac{\bigl( \Delta_{k}^1 \bigr)^2}{\Delta_{k}^0 \, \Delta_{k+1}^0}
  , \quad
  0 \le k \le N
  .
\end{equation}
From \eqref{eq:CH-string-inverse-third-determinant-quotient} we moreover obtain
\begin{equation}
  \label{eq:CH-formula-yk}
  y_{N+1-k} = 1 - \sum_{j=1}^k l_{N+1-j}
  = 1 - \frac{\Delta_{k-1}^2}{\Delta_{k}^0}
  , \quad
  1 \le k \le N
  .
\end{equation}
\end{subequations}
These formulas provide the general solution to the inverse spectral problem
for the discrete string with Dirichlet boundary conditions,
since the given spectral data $\{ \lambda_k, a_k \}_{k=1}^N$
through~\eqref{eq:CH-def-moments}
determine the moments~$\{ \alpha_n \}_{n \ge 0}$,
which in turn through~\eqref{eq:CH-def-Hankel-det}
define the determinants~$\Delta_k^n$,
which in turn through~\eqref{eq:CH-formula-gk-lk-yk}
give the string parameters $\{ y_k, g_k \}_{k=1}^N$.

\begin{example}
  \label{ex:CH-string-two-masses-cont}
  Let us write out the formulas~\eqref{eq:CH-formula-gk-lk-yk} for $N=2$,
  for comparison with the results
  obtained in Example~\ref{ex:CH-string-two-masses}.
  Since $\lambda_0=0$, the constant $a_0=1/2$ only enters in the zeroth moment
  $\alpha_0 = \tfrac12 + a_1 + a_2$;
  the next three moments are
  $\alpha_1 = \lambda_1 a_1 + \lambda_2 a_2$,
  $\alpha_2 = \lambda_1^2 a_1 + \lambda_2^2 a_2$
  and
  $\alpha_3 = \lambda_1^3 a_1 + \lambda_2^3 a_2$.
  Thus,
  \begin{equation}
    \begin{aligned}
      y_1 &= 1 - \frac{\Delta_{1}^2}{\Delta_{2}^0}
      ,\quad &
      g_1 &= \frac{\bigl( \Delta_{2}^0 \bigr)^2}{\Delta_{1}^1 \, \Delta_{2}^1}
      ,\\
      y_2 &= 1 - \frac{\Delta_{0}^2}{\Delta_{1}^0}
      ,&
      g_2 &= \frac{\bigl( \Delta_{1}^0 \bigr)^2}{\Delta_{0}^1 \, \Delta_{1}^1}
      ,
    \end{aligned}
  \end{equation}
  and also
  \begin{equation}
    l_1 = y_2 - y_1
    = \frac{\bigl( \Delta_{1}^1 \bigr)^2}{\Delta_{1}^0 \, \Delta_{2}^0}
    ,
  \end{equation}
  where
  \begin{equation}
    \label{eq:CH-string-two-masses-determinants}
    \begin{aligned}
      \Delta_0^1 &= \Delta_0^2 = 1
      ,\\
      \Delta_1^0 & = \alpha_0 = \tfrac12 + a_1 + a_2
      ,\\
      \Delta_1^1 & = \alpha_1 = \lambda_1 a_1 + \lambda_2 a_2
      ,\\
      \Delta_1^2 & = \alpha_2 = \lambda_1^2 a_1 + \lambda_2^2 a_2
      ,\\
      \Delta_2^0 & =
      \begin{vmatrix} \alpha_0 & \alpha_1 \\ \alpha_1 & \alpha_2 \end{vmatrix}
      = \begin{vmatrix}
        \tfrac12 + a_1 + a_2 & \lambda_1 a_1 + \lambda_2 a_2 \\
        \lambda_1 a_1 + \lambda_2 a_2 & \lambda_1^2 a_1 + \lambda_2^2 a_2
      \end{vmatrix}
      \\ &
      = \tfrac12 (\lambda_1^2 a_1 + \lambda_2^2 a_2) + (\lambda_1 - \lambda_2)^2 a_1 a_2
      ,\\
      \Delta_2^1 & =
      \begin{vmatrix} \alpha_1 & \alpha_2 \\ \alpha_2 & \alpha_3 \end{vmatrix}
      = \begin{vmatrix}
        \lambda_1 a_1 + \lambda_2 a_2 & \lambda_1^2 a_1 + \lambda_2^2 a_2 \\
        \lambda_1^2 a_1 + \lambda_2^2 a_2 & \lambda_1^3 a_1 + \lambda_2^3 a_2
      \end{vmatrix}
      \\ &
      = (\lambda_1 - \lambda_2)^2 \lambda_1 \lambda_2 a_1 a_2
      ,
    \end{aligned}
  \end{equation}
  in agreement with \eqref{eq:CH-two-masses-inverse-problem-solution}
  and~\eqref{eq:CH-two-masses-inverse-problem-l1}.
  From the derivation of \eqref{eq:CH-formula-gk-lk-yk}
  we learn that even if the system
  of equations \eqref{eq:CH-two-masses-inverse-problem-naively}
  is nonlinear in the sought variables $l_k$ and~$g_k$,
  it actually implies \emph{linear} relations (with the moments $\alpha_k$ as coefficients)
  between certain \emph{combinations} of these variables, namely the coefficients in the polynomials
  \begin{equation*}
    \begin{aligned}
      Q_1(\lambda) &= l_2
      ,\\
      Q_2(\lambda) &= 1 - l_2 g_2 \lambda
      ,\\
      Q_3(\lambda) &= (l_2+l_1) - l_2 g_2 l_1 \lambda
      ,\\
      Q_4(\lambda) &= 1 - (l_2 g_2 + l_2 g_1 + l_1 g_1) \lambda + l_2 g_2 l_1 g_1 \lambda^2
      ,\\
      Q_5(\lambda) &= (l_2+l_1+l_0)
      \\ & \quad
      - (l_2 g_2 l_1 + l_2 g_2 l_0 + l_2 g_1 l_0 + l_1 g_1 l_0) \lambda
      \\ & \quad
      + l_2 g_2 l_1 g_1 l_0 \lambda^2
      .
    \end{aligned}
  \end{equation*}
  This is the reason why these coefficients, and hence the sought variables,
  can be expressed in terms of determinants involving the moments~$\alpha_k$.
  (Here the subscript in~$Q_k$ indicates the number of factors, $k=2r$ or $k=2r+1$,
  used in the construction of these polynomials above;
  see \eqref{eq:CH-matrix-product-even} and~\eqref{eq:CH-matrix-product-odd}.)
\end{example}

Note that the solution formulas \eqref{eq:CH-formula-gk-lk-yk} for the inverse spectral
problem are completely explicit,
in contrast to the forward spectral problem, which involves finding the roots~$\lambda_k$
of a polynomial of degree~$N$.
We can make them even more explicit by evaluating the Hankel determinants
using a pretty computation, usually attributed to Heine, which briefly goes as follows.
For $k>0$, write down the determinant $\Delta_k^0 = \det(\alpha_{i+j-2})_{i,j=1,\dots,k}$ with
$\alpha_r = \int z^r \, d\alpha(z)$,
but use a separate dummy variable in each column, say $z_j$ in column~$j$.
From each column, the integral and and a factor $z_j^{j-1}$
can be taken outside the determinant by multilinearity;
what remains is a Vandermonde determinant
$\Delta(z_1,\dots,z_k) = \prod_{i > j} (z_i - z_j)$.
Next, do the same, but with a permutation of the variables, say $z_{\pi(j)}$ in column~$j$,
giving another expression for the same determinant~$\Delta_k^0$.
Averaging these expressions over all permutations $\pi \in S_k$ produces another
Vandermonde determinant factor
from the signs of the permutations and all the factors $z_{\pi(j)}^{j-1}$,
so that
\begin{equation}
  \begin{split}
    \Delta_k^0
    &
    = \frac{1}{k!} \int_{\R^k} \Delta(z_1,\dots,z_k)^2 \, d\alpha(z_1) \dotsm d\alpha(z_k)
    \\ &
    = \int_{z_1 < \dots < z_k} \Delta(z_1,\dots,z_k)^2 \, d\alpha(z_1) \dotsm d\alpha(z_k)
    .
  \end{split}
\end{equation}
With a discrete measure
$\alpha(\lambda) = \sum_{k=0}^N a_k \, \delta(\lambda-\lambda_k)$,
the integral turns into a sum:
\begin{equation}
  \label{eq:CH-delta-zero-heine}
  \Delta_k^0 =
  \sum_{0 \le i_1 < \dots < i_k \le N} \Delta(\lambda_{i_1},\dots,\lambda_{i_k})^2 \, a_{i_1} \dotsm a_{i_k}
  .
\end{equation}
As an example, with $N=2$ this formula gives
\begin{equation*}
  \Delta_2^0
  = (\lambda_0 - \lambda_1)^2 a_0 a_1 + (\lambda_0 - \lambda_2)^2 a_0 a_2 + (\lambda_1 - \lambda_2)^2 a_1 a_2
  ,
\end{equation*}
which in our case, where $\lambda_0=0$ and $a_0=1/2$ by definition,
simplifies to the same expression as in~\eqref{eq:CH-string-two-masses-determinants}:
\begin{equation*}
  \Delta_2^0
  = \tfrac12 ( \lambda_1^2 a_1 + \lambda_2^2 a_2 ) + (\lambda_1 - \lambda_2)^2 a_1 a_2
  .
\end{equation*}
The more general determinant $\Delta_k^n$ is obtained by replacing $a_i$ with $\lambda_i^n a_i$ everywhere
in~\eqref{eq:CH-delta-zero-heine},
since $\alpha_{n+r}$ is the $r$th moment of the measure obtained by modifying $\alpha$
in the same way. To be explicit, the formula is
\begin{equation}
  \label{eq:CH-delta-k-heine}
  \Delta_k^n =
  \sum_{0 \le i_1 < \dots < i_k \le N} \Delta(\lambda_{i_1},\dots,\lambda_{i_k})^2 \, \lambda_{i_1}^n \dotsm \lambda_{i_k}^n \, a_{i_1} \dotsm a_{i_k}
  .
\end{equation}
If $n>1$, then all terms with $i_1=0$ vanish (since $\lambda_0=0$),
so in that case we can sum over $1 \le i_1 < \dots < i_k \le N$ instead.
If $k > N$ (or $k > N+1$ in the case $k=0$), then $\Delta_k^n=0$, since the sum is empty
(there are no increasing $k$-tuples to sum over).

\begin{remark}
  \label{rem:CH-range-of-spectral-map}
  Since all the residues~$a_i$ are positive,
  it is clear from \eqref{eq:CH-delta-k-heine} that all the determinants
  appearing in~\eqref{eq:CH-formula-gk-lk-yk}
  are nonzero provided that all the eigenvalues~$\lambda_k$ (for $1 \le k \le N$)
  have the same sign, while $\Delta_k^1$ (for $1 \le k \le N-1$)
  may be zero if this condition is not met.
  This is relevant because of~\eqref{eq:CH-formula-lk},
  which shows that $l_{N-k} \ge 0$ always, with equality if and only if $\Delta_k^1 = 0$.
  So discrete strings with all weights $g_k$ positive
  are in one-to-one correspondence with spectral data such that
  $0 < \lambda_1 < \dots < \lambda_N$ and all $a_k > 0$,
  strings with all weights negative
  are in one-to-one correspondence with spectral data such that
  $\lambda_1 < \dots < \lambda_N < 0$ and all $a_k > 0$,
  while strings with $p \in \{ 1,\dots,N-1 \}$ negative weights and $N-p$ positive weights
  are in one-to-one correspondence with spectral data such that
  $\lambda_1 < \dots < \lambda_p < 0 < \lambda_{p+1} < \dots < \lambda_N$, all $a_k > 0$,
  and in addition $\Delta_k^1 \neq 0$ for $1 \le k \le N-1$.
\end{remark}

\begin{remark}
  \label{rem:CH-orthpoly}
  For $1 \le r \le N$, let us write $Q_r(\lambda)$
  for the $r$th-degree
  polynomial $Q(\lambda) = X_{11}(\lambda)$
  that we obtained from the product~$X(\lambda)$ with $2r$ factors
  in~\eqref{eq:CH-matrix-product-even}.
  (This was denoted by $Q_{2r}(\lambda)$ in Example~\ref{ex:CH-string-two-masses-cont}.)
  Then, using~\eqref{eq:CH-def-moments},
  equation~\eqref{eq:CH-stieltjes-orthogonality-matrix-form}
  can be written as
  \begin{equation}
    \label{eq:CH-stieltjes-orthogonality-integral-form}
    \int \lambda^j Q_{r}(\lambda) \, d\alpha(\lambda) = 0
    ,\quad
    0 \le j \le r-1
    ,
  \end{equation}
  so that $Q_{r}(\lambda)$ is orthogonal, with respect to the measure~$\alpha$,
  to all polynomials of lower degree.
  With $Q_0(\lambda)=1$ obtained as a special case from the empty product (the identity matrix),
  $\{ Q_{r} \}_{r=0,\dots,N}$ is therefore a family of orthogonal polynomials
  with respect to~$\alpha$.
  (Since $\alpha$ is only supported at $N+1$ points, the $L^2$-space with respect to~$\alpha$
  is $(N+1)$-dimensional, so there can be no more orthogonal polynomials than that.)

  One can verify directly that $Q_{r}$ given by~\eqref{eq:CH-Qn-even-det}
  satisfies the orthogonality condition \eqref{eq:CH-stieltjes-orthogonality-integral-form}, since
  \begin{equation*}
    \begin{split}
      &
      \int \lambda^j Q_{r}(\lambda) \, d\alpha(\lambda)
      \\ &
      =
      \frac{1}{\Delta_r^1}
      \begin{vmatrix}
        \alpha_j & \alpha_{j+1} & \alpha_{j+2} & \dots & \alpha_{j+r}\\
        \alpha_0 & \alpha_1 & \alpha_2 & \dots & \alpha_r \\
        \alpha_1 & \alpha_2 & \alpha_3 & \dots & \alpha_{r+1} \\
        \vdots & \vdots & \vdots & \vdots & \vdots \\
        \alpha_{r-1} & \alpha_r & \alpha_{r+1} & \dots & \alpha_{2r-1}
      \end{vmatrix}
    \end{split}
  \end{equation*}
  vanishes whenever $0 \le j \le r-1$, due to two rows being equal.
\end{remark}

\begin{remark}
  The solution to the inverse problem can also be formulated in terms of
  continued fractions of the type studied in depth by Stieltjes
  in a famous memoir from the end of the
  19th century~\cite{stieltjes:1894:recherches-fractions-continues-1,stieltjes:1895:recherches-fractions-continues-2}.
  If we use $v_k$ (for ``value'')
  to denote the value of the function $\phi(y;\lambda)$ at the point~$y_k$
  and $s_k$ (for ``slope'') to denote the constant value
  of the derivative $\phi_y(y;\lambda)$ in the interval $(y_{k},y_{k+1})$,
  then \eqref{eq:CH-jump-matrix-Lk} and \eqref{eq:CH-jump-matrix-Gk} can be written as
  \begin{equation}
    \label{eq:CH-value-slope-recursion}
    v_{k+1} = v_{k} + s_{k} \, l_{k}
    ,\quad
    s_{k} = s_{k-1} - \lambda g_{k} v_{k}
    .
  \end{equation}
  This recursion implies that the modified Weyl function can be written
  \begin{equation*}
    \begin{split}
      \omega(\lambda)
      &
      = \frac{W(\lambda)}{\lambda}
      = \frac{\phi_y(1;\lambda)}{\lambda \, \phi(1;\lambda)}
      = \frac{s_N}{\lambda \, v_{N+1}}
      = \frac{s_N}{\lambda \, (v_N + s_N \, l_N)}
      \\ &
      = \frac{1}{\lambda \, \left( l_N + \dfrac{v_N}{s_N} \right)}
      = \frac{1}{\lambda \, \left( l_N + \dfrac{v_N}{s_{N-1} - \lambda \, g_N v_N} \right)}
      \\ &
      = \frac{1}{\lambda \, l_N + \dfrac{1}{-g_N + \dfrac{s_{N-1}}{\lambda \, v_N}}}
      ,
    \end{split}
  \end{equation*}
  and continuing in that way produces the following Stieltjes continued fraction expansion:
  \begin{equation}
    \label{eq:CH-W-cfrac}
    \omega(\lambda)
    = \cfrac{1}{\lambda \, l_N +
      \cfrac{1}{-g_N +
        \cfrac{1}{\lambda \, l_{N-1} +
          \cfrac{1}{\raisebox{1.5ex}{$\ddots$} +
            \cfrac{1}{-g_1 +
              \cfrac{1}{\lambda \, l_0}
            }}}}}
    .
  \end{equation}
  Recalling also the expansion~\eqref{eq:CH-W-power-series} of $\omega(\lambda)$
  in powers of $1/\lambda$,
  \begin{equation*}
    \omega(\lambda)
    = \frac{\alpha_0}{\lambda} + \frac{\alpha_1}{\lambda^2} + \frac{\alpha_2}{\lambda_3} + \dotsb
    ,
  \end{equation*}
  we see that $f(\lambda) = -\omega(-\lambda)$ matches the result given by
  formula~(7) in section~11 of Stieltjes's memoir,
  which we basically reproved above using different notation and terminology,
  and which says that if the sequence
  $(\alpha_k)_{k \ge 0}$ is such that all determinants $\Delta_k^0$ and $\Delta_k^1$
  are nonzero, then the Laurent series
  \begin{equation*}
    f(\lambda) = \frac{\alpha_0}{\lambda}-\frac{\alpha_1}{\lambda^2}+\frac{\alpha_2}{\lambda^3}-\dotsb
  \end{equation*}
  can be uniquely developed in a continued fraction
  \begin{equation*}
    \cfrac{1}{c_1\lambda+ \cfrac{1}{c_2+\cfrac{1}{c_3\lambda+\cfrac{1}{c_4 + \dotsb}}}},
  \end{equation*}
  where
  \begin{equation}
    \label{eq:CH-Stieltjes-cn-odd-even}
    c_{2k}=\frac{(\Delta_k^0)^2}{\Delta_{k-1}^1 \Delta_k^1}
    , \qquad
    c_{2k+1}=\frac{(\Delta_k^1)^2}{\Delta_k^0 \Delta_{k+1}^0}
    ,
  \end{equation}
  and moreover
  \begin{equation}
    \label{eq:CH-Stieltjes-cn-sum}
    c_1 + c_3 + \dots + c_{2k+1} = \frac{\Delta_{k}^2}{\Delta_{k+1}^0}
    .
  \end{equation}
  Our discrete case is somewhat degenerate, since the Hankel determinants $\Delta_k^n$
  all vanish when the size~$k$ becomes too large,
  but \eqref{eq:CH-Stieltjes-cn-odd-even} still gives
  all the coefficients in the terminating continued fraction~\eqref{eq:CH-W-cfrac}.
  Actually these formulas predate Stieltjes;
  in an earlier work \cite[p.~185]{stieltjes:1918:complete-works-2}
  he writes that their proof presents no difficulty,
  and refers the reader to texts by Frobenius and Stickelberger
  \cite{frobenius-stickelberger:1880:addition-multiplication-elliptischen-funktionen,frobenius:1881:relationen-zwischen-naherungsbruchen}
  for the details.
\end{remark}

Now we finally return to the Camassa--Holm equation and its peakon solutions.
So far we have only studied the first Lax equation \eqref{eq:intro-CH-lax-x}
for a fixed value of~$t$,
with a discrete measure~$m$ as in \eqref{eq:CH-peakon-m}
when $u(x,t)$ is given by the peakon ansatz~\eqref{eq:intro-peakons}.
After the Liouville transformation \eqref{eq:CH-liouville-trf},
this turned into the string equation \eqref{eq:CH-gstring}
with a discrete measure~$g$ as in~\eqref{eq:CH-peakon-g}.
Now we switch on the time dependence again, so to speak,
and consider the second Lax equation \eqref{eq:intro-CH-lax-t},
repeated here for convenience:
\begin{equation*}
  \psi_t = \tfrac12  \left( \tfrac{1}{\lambda} + u_x \right) \psi - \left( \tfrac{1}{\lambda} + u \right) \psi_x
  .
\end{equation*}
The time evolution of $u(x,t)$ determined by the CH equation,
in the form of the ODEs~\eqref{eq:intro-CH-peakon-ode-explicit}
when we are talking about peakon solutions,
is exactly the condition required for this second Lax equation to be compatible with the first one.
That is, if $\psi(x,t)$ satisfies \eqref{eq:intro-CH-lax-x} at some time~$t$
and for some value of~$\lambda$,
then if $\psi(x,t)$ evolves according to \eqref{eq:intro-CH-lax-t} where $u(x,t)$ is a solution
of the CH equation,
it will remain a solution of \eqref{eq:intro-CH-lax-x} with the same~$\lambda$.
The corresponding statements of course also hold for $\phi(y,t)$, the image of $\psi(x,t)$
under the Liouville transformation.
Moreover, the boundary conditions $\phi(\pm 1,t)=0$ that we imposed on the discrete string
are compatible with this time evolution.
Indeed, the particular solution $\phi(y,t;\lambda)$ which satisfies $\phi=0$ and $\phi_y=1$ at $y=-1$
is the image of a function $\psi(x,t;\lambda)$ which equals $e^{x/2}$ for $x < x_1(t)$ (so that
$\psi_t =0$ and $\psi_x = \tfrac12 \psi$ there),
and since $u(x,t)$ given by \eqref{eq:intro-peakons}
is a (time-dependent) multiple of $e^x$ for $x < x_1(t)$ (so that $u=u_x$ there),
both sides of \eqref{eq:intro-CH-lax-t} are identically zero in the region $x < x_1(t)$.
And in the region $x > x_N(t)$, the preimage~$\psi$ equals $A(t;\lambda) \, e^{x/2} + B(t;\lambda) \, e^{-x/2}$,
where $A=A_N$ and $B=B_N$ in the notation of~\eqref{eq:CH-psi-piecewise},
while $u$ is a multiple of $e^{-x}$, say $u=U(t) \, e^{-x}$ (so that $u_x = -u$),
and hence \eqref{eq:intro-CH-lax-t} becomes
\begin{equation}
  \label{eq:CH-time-dependence-A-B}
  \begin{split}
    &
    A_t \, e^{x/2} + B_t \, e^{-x/2}
    \\ &
    = \tfrac12  \bigl( \tfrac{1}{\lambda} - U \, e^{-x} \bigr) (A \, e^{x/2} + B \, e^{-x/2})
    \\ & \quad
    -  \tfrac12 \bigl( \tfrac{1}{\lambda} + U \, e^{-x} \bigr) (A \, e^{x/2} - B \, e^{-x/2})
    \\ &
    = \bigl( \tfrac{1}{\lambda} B - AU \bigr) \, e^{-x/2}
    ,
  \end{split}
\end{equation}
which implies that $A_t = 0$ and $B_t = \tfrac{1}{\lambda} B - AU$.
Thus the polynomial $A = A(\lambda)$ is actually time-independent, and hence so are its roots,
which by definition are the eigenvalues~$\lambda_k$.
The CH equation therefore induces an \emph{isospectral deformation} of the string with Dirichlet
boundary conditions;
as time passes, the mass distribution of the string changes, but its Dirichlet spectrum remains the same.

Moreover, evaluating $B_t = \tfrac{1}{\lambda} B - AU$ at $\lambda=\lambda_k$ gives
$B_t(\lambda_k) = B(\lambda_k) / \lambda_k$,
which we can use to find the evolution of the residues $a_k(t)$ in
the modified Weyl function
\begin{equation*}
  \frac{W(t;\lambda)}{\lambda}
  = \frac{\phi_y(1,t;\lambda)}{\lambda \, \phi(1,t;\lambda)}
  = \frac{A(\lambda) - B(t;\lambda)}{2\lambda\,A(\lambda)}
  = \frac{1/2}{\lambda} + \sum_{k=1}^N \frac{a_k(t)}{\lambda-\lambda_k}
  .
\end{equation*}
(Here the second equality comes from~\eqref{eq:CH-phi-piecewise}.)
Like this:
\begin{equation*}
  \sum_{k=1}^N \frac{a_k(t)}{\lambda-\lambda_k} = - \frac{B(t;\lambda)}{2 \lambda \, A(\lambda)}
\end{equation*}
gives
\begin{equation*}
  \sum_{k=1}^N \frac{\dot a_k(t)}{\lambda-\lambda_k} = - \frac{B_t(t;\lambda)}{2 \lambda \, A(\lambda)}
  ,
\end{equation*}
so that (since the poles $\lambda_k$ are simple)
\begin{equation*}
  \begin{split}
    \dot a_k(t)
    &
    = \res_{\lambda=\lambda_k} \frac{- B_t(t;\lambda)}{2 \lambda \, A(\lambda)}
    = \left[ \frac{- B_t(t;\lambda)}{\partial_\lambda \bigl( 2 \lambda \, A(\lambda) \bigr)} \right]_{\lambda = \lambda_k}
    = \frac{- B_t(t;\lambda_k)}{2 \lambda_k \, A'(\lambda_k)}
    \\ &
    = \frac{- B(t;\lambda_k) / \lambda_k}{2 \lambda_k \, A'(\lambda_k)}
    = \frac{1}{\lambda_k} \, \res_{\lambda=\lambda_k} \frac{- B(t;\lambda)}{2 \lambda \, A(\lambda)}
    = \frac{1}{\lambda_k} \, a_k(t).
  \end{split}
\end{equation*}
Thus $\dot a_k = a_k/\lambda_k$ for $1 \le  k \le N$,
which immediately gives
\begin{equation}
  \label{eq:CH-a-of-t}
  a_k(t) = a_k(0) \, e^{t/\lambda_k}
  .
\end{equation}

Since we have solved the inverse spectral problem,
the knowledge of the spectral data for all~$t$
tells us the values of the string parameters $y_k(t)$ and~$g_k(t)$ for all~$t$,
through the formulas~\eqref{eq:CH-formula-gk-lk-yk},
and then the inverse of the Liouville transformation \eqref{eq:CH-peakon-gk-yk},
namely
\begin{equation}
  \label{eq:CH-inverse-liouville-trf}
  x_k = \ln\frac{1+y_k}{1-y_k},
  \qquad
  m_k = \tfrac12 (1-y_k^2) \, g_k
  ,
\end{equation}
tells us the values of the peakon parameters $x_k(t)$ and~$m_k(t)$ for all~$t$.

As we will see presently, the contributions related to the extra pole $\lambda=0$ in $W(\lambda)/\lambda$
cancel out in this calculation, so the results can be expressed in terms of the determinants
\begin{equation}
  \label{eq:CH-def-other-Hankel-det}
  \delta_k^n =
  \begin{cases}
    1
    , & k=0
    ,\\
    \det \bigl( \hat\alpha_{n+i+j-2} \bigr)_{i,j=1,\dots,k}
    \, , & k>0
    ,
  \end{cases}
\end{equation}
which are just like the determinants $\Delta_k^n$ from \eqref{eq:CH-def-Hankel-det},
but computed using the moments
\begin{equation}
  \hat\alpha_r = \int z^r \, d\hat\alpha(z) = \sum_{k=1}^N \lambda_k^r a_k
\end{equation}
of the measure
\begin{equation}
  \hat\alpha(\lambda) = \sum_{k=1}^N a_k \, \delta(\lambda - \lambda_k)
\end{equation}
instead of the moments $\alpha_r = \sum_{k=0}^N \lambda_k^r a_k$ that we had before.
Since $\lambda_0=0$ and $a_0=1/2$,
we have
\begin{equation*}
  \alpha_k =
  \begin{cases}
    \hat\alpha_0 + \tfrac12
    ,&
    k = 0
    ,\\
    \hat\alpha_k
    ,&
    k > 0
    ,
  \end{cases}
\end{equation*}
which means that
\begin{equation*}
  \Delta_k^n =
  \begin{cases}
    \delta_k^0 + \tfrac12 \delta_{k-1}^2
    ,&
    n = 0
    ,\\
    \delta_k^n
    ,&
    n > 0
    ,
  \end{cases}
\end{equation*}
so that
\begin{equation*}
  \begin{split}
    \exp x_{N+1-k}
    &
    = \frac{1 + y_{N+1-k}}{1 - y_{N+1-k}}
    = \frac{1 + \left( 1 - \dfrac{\Delta_{k-1}^2}{\Delta_{k}^0} \right)}{1 - \left( 1 - \dfrac{\Delta_{k-1}^2}{\Delta_{k}^0} \right)}
    \\ &
    = \frac{2 \Delta_{k}^0 - \Delta_{k-1}^2}{\Delta_{k-1}^2}
    = \frac{2 ( \delta_k^0 + \tfrac12 \delta_{k-1}^2 ) - \delta_{k-1}^2}{\delta_{k-1}^2}
    = \frac{2 \delta_k^0}{\delta_{k-1}^2}
  \end{split}
\end{equation*}
and
\begin{equation*}
  \begin{split}
    m_{N+1-k}
    &
    = \tfrac12 (1 - y_{N+1-k}^2) \, g_{N+1-k}
    \\ &
    = \tfrac12 (1 + y_{N+1-k}) (1 - y_{N+1-k}) \, g_{N+1-k}
    \\ &
    = \frac12 \left( 2 - \frac{\Delta_{k-1}^2}{\Delta_{k}^0} \right)
    \cdot
    \frac{\Delta_{k-1}^2}{\Delta_{k}^0}
    \cdot
    \frac{\bigl( \Delta_{k}^0 \bigr)^2}{\Delta_{k-1}^1 \, \Delta_{k}^1}
    \\ &
    = \frac{(2 \Delta_{k}^0 - \Delta_{k-1}^2) \, \Delta_{k-1}^2}{2 \, \Delta_{k-1}^1 \, \Delta_{k}^1}
    = \frac{2 \delta_k^0 \, \delta_{k-1}^2}{2 \delta_{k-1}^1 \, \delta_{k}^1}
    = \frac{\delta_k^0 \, \delta_{k-1}^2}{\delta_k^1 \, \delta_{k-1}^1}
    .
  \end{split}
\end{equation*}
Thus, in the end we find that
\begin{equation}
  \label{eq:CH-xk-mk-solution-formulas}
  x_{N+1-k}(t) = \ln\frac{2 \delta_k^0}{\delta_{k-1}^2}
  , \quad
  m_{N+1-k}(t) = \frac{\delta_k^0 \, \delta_{k-1}^2}{\delta_k^1 \, \delta_{k-1}^1}
  ,
\end{equation}
for $1 \le k \le N$.
The formulas \eqref{eq:CH-xk-mk-solution-formulas},
together with the determinant evaluation
\begin{equation}
  \label{eq:CH-delta-k-heine-again}
  \delta_k^n =
  \sum_{1 \le i_1 < \dots < i_k \le N} \Delta(\lambda_{i_1},\dots,\lambda_{i_k})^2 \, \lambda_{i_1}^n \dotsm \lambda_{i_k}^n \, a_{i_1} \dotsm a_{i_k}
\end{equation}
and the time-dependence $a_k(t) = a_k(0) \, e^{t/\lambda_k}$,
provide completely explicit formulas for the general solution of the $N$-peakon ODEs
\eqref{eq:intro-CH-peakon-ode-explicit}
in terms of elementary functions.

\begin{remark}
  To be precise, \eqref{eq:CH-xk-mk-solution-formulas} gives all solutions such that
  all amplitudes $m_k$ are nonzero,
  which is what is needed in order to describe the peakon solutions~\eqref{eq:intro-peakons}.
  From the point of view of the ODEs \eqref{eq:intro-CH-peakon-ode-explicit},
  the most general solution should also take into account the case where some~$m_k$
  may be zero;
  in this case $m_k$ is identically zero, but the corresponding ODE for $x_k$ is still nontrivial.
  The trajectory $x = x_k(t)$ of such a zero-amplitude ``ghostpeakon'', which is influenced by the other
  peakons but does not influence them, can be found from \eqref{eq:CH-xk-mk-solution-formulas}
  through a limiting procedure~\cite{lundmark-shuaib:2019:ghostpeakons}.
  Ghostpeakon trajectories are \emph{characteristic curves}
  associated with the peakon solution $u(x,t)$ containing the ``non-ghost'' peakons,
  i.e., solutions of the ODE $\dot \xi(t) = u(\xi(t),t)$,
  and knowing these curves is of some interest in the study of peakon--antipeakon
  collisions~\cite{grunert-holden:2016:CH-peakon-antipeakon-alpha-dissipative}.
\end{remark}

\begin{remark}
  The factor of~$2$ in the formula for the positions~$x_k$ in \eqref{eq:CH-xk-mk-solution-formulas} can be removed by using determinants
  of moments with respect to the measure $\sum_{k=1}^N b_k \, \delta(\lambda - \lambda_k)$,
  where $b_k = 2 a_k$.
  This is the form used in some of our other works~\cite{lundmark:2007:DP-shockpeakons,lundmark-shuaib:2019:ghostpeakons}.
\end{remark}

\begin{remark}
  \label{rem:CH-inverse-problem-on-the-real-line}
  Since the original spectral problem \eqref{eq:CH-lax-x-again}
  on the real line is equivalent to the string problem \eqref{eq:CH-gstring}
  on the finite interval $[-1,1]$, it is of course not strictly necessary to pass to
  the finite interval.
  For a quick derivation of the multipeakon solution formulas~\eqref{eq:CH-xk-mk-solution-formulas}
  directly in terms of the forward and inverse spectral problem on the real line,
  see Mohajer and Szmigielski~\cite{mohajer-szmigielski:2011:CH-on-real-axis}.
\end{remark}

\begin{example}[The two-peakon solution]
  \label{ex:CH-two-peakons}
  For $N=2$, the equations of motion \eqref{eq:intro-CH-peakon-ode-explicit} are
  \begin{equation}
    \label{eq:CH-twopeakon-ode}
    \begin{aligned}
      \dot x_1 &= m_1 + m_2 \, e^{x_1-x_2}
      , &
      \dot m_1 &= -m_1 m_2 \, e^{x_1-x_2}
      , \\
      \dot x_2 &= m_1 \, e^{x_1-x_2} + m_2
      , &
      \dot m_2 &= m_1 m_2 \, e^{x_1-x_2}
      ,
    \end{aligned}
  \end{equation}
  where we have assumed that $x_1 < x_2$,
  in order to remove the absolute values in the ODEs.
  (If this holds at some initial time, say $t=0$, then it will hold
  at least in some open time interval around $t=0$.)
  We also assume that $m_1$ and $m_2$ are nonzero, so that there really are
  two peakons in the solution.
  These ODEs can be solved directly in terms of the variables $x_1 \pm x_2$ and $m_1 \pm m_2$,
  as was done already in the original Camassa--Holm paper~\cite{camassa-holm:1993:CH-orginal-paper},
  and studies or expositions of this two-peakon solution have been published by many researchers
  \cite{camassa-holm-hyman:1994:CH-new-integrable,
    constantin:1997:CH-soliton-interactions:Expositiones,
    beals-sattinger-szmigielski:2000:moment,
    beals-sattinger-szmigielski:2001:peakon-antipeakon-NEEDS99,
    alber-miller:2001:CH-twopeakon,
    wahlen:2006:CH-peakon-antipeakon-interaction,
    lundmark:2007:DP-shockpeakons,
    parker:2008:CH-peakons-dynamics,
    grunert-holden:2016:CH-peakon-antipeakon-alpha-dissipative,
    lundmark-shuaib:2019:ghostpeakons,
    cieslak-gaczkowski-kubkowski-malogrosz:2017:CH-multipeakons-as-geodesics,
    cieslak-krynski:2021:CH-geometric-aspects-of-two-and-threepeakons}.
  The solution formulas can be written in several equivalent ways,
  but the form coming from \eqref{eq:CH-xk-mk-solution-formulas} is
  \begin{equation}
    \label{eq:CH-twopeakon-x-m}
    \begin{aligned}
      x_1(t) &= \ln \frac{2 (\lambda_1-\lambda_2)^2 a_1 a_2}{\lambda_1^2 a_1 + \lambda_2^2a_2}
      , &
      x_2(t) &= \ln 2 (a_1+a_2)
      , \\
      m_1(t) &= \frac{\lambda_1^2 a_1 + \lambda_2^2 a_2}{\lambda_1 \lambda_2 \left( \lambda_1 a_1 + \lambda_2 a_2\right)}
      , &
      m_2(t) &= \frac{a_1+a_2}{\lambda_1 a_1+\lambda_2 a_2}
      ,
    \end{aligned}
  \end{equation}
  with $a_k = a_k(t) = a_k(0) \, e^{t/\lambda_k}$,
  where the constants
  $\lambda_1$, $\lambda_2$, $a_1(0)$ and $a_2(0)$
  are determined by initial conditions.
  Recall that $a_1$ and~$a_2$ are positive,
  while $\lambda_1$ and~$\lambda_2$ have the same sign pattern as $m_1$ and~$m_2$
  (both positive, both negative, or one of each sign).
  Since
  \begin{equation*}
    \frac{e^{x_2} - e^{x_1}}{2}
    = (a_1 + a_2) - \frac{(\lambda_1-\lambda_2)^2 a_1 a_2}{\lambda_1^2 a_1 + \lambda_2 ^2a_2}
    = \frac{(\lambda_1 a_1 + \lambda_2 a_2)^2}{\lambda_1^2 a_1 + \lambda_2 ^2a_2}
    ,
  \end{equation*}
  we see that $x_1(t) < x_2(t)$ unless $\lambda_1 a_1(t) + \lambda_2 a_2(t) = 0$,
  in which case a collision $x_1=x_2$ takes place;
  this can only happen in the peakon--antipeakon case when $\lambda_1$ and~$\lambda_2$
  have opposite signs,
  and then it happens at a unique time~$t=t_c$, which is greater than~$0$
  if and only if the peakon starts out to the right of the antipeakon,
  i.e., if $m_1(0) > 0 > m_2(0)$.
  Since the factor $\lambda_1 a_1 + \lambda_2 a_2$ appears in the denominator
  of both $m_1$ and~$m_2$ in~\eqref{eq:CH-twopeakon-x-m},
  the individual amplitudes $m_1$ and~$m_2$ blow up at the collision,
  to $+\infty$ and $-\infty$, respectively, but
  their sum $m_1 + m_2$ has the constant value $\tfrac{1}{\lambda_1} + \tfrac{1}{\lambda_2}$,
  and because of this cancellation, the quantities
  \begin{equation*}
    u(x_1(t),t)
    = m_1 + m_2 e^{x_1-x_2}
    = \dot x_1
    = \frac{1}{\lambda_1} + \frac{1}{\lambda_2} - \frac{\lambda_1 a_1 + \lambda_2 a_2}{\lambda_1^2 a_1 + \lambda_2^2 a_2}
  \end{equation*}
  and
  \begin{equation*}
    u(x_2(t),t)
    = m_1 e^{x_1-x_2} + m_2
    = \dot x_
    2 = \frac{1}{\lambda_1} + \frac{1}{\lambda_2} - \frac{\lambda_1 a_1 + \lambda_2 a_2}{\lambda_1 \lambda_2 (a_1 + a_2)}
  \end{equation*}
  stay bounded and converge to the same constant $\tfrac{1}{\lambda_1} + \tfrac{1}{\lambda_2}$.
  Thus, $u(x,t)$ converges to a single-peakon shape
  \begin{equation*}
    u(x,t_c) = \bigl( \tfrac{1}{\lambda_1} + \tfrac{1}{\lambda_2} \bigr) \, e^{-\abs{x-x_1(t_c)}}
  \end{equation*}
  at the time of collision
  (or $u=0$ in the symmetric case $\lambda_2 = -\lambda_1$).
  The energy integral $E(t) = \int_{\R} (u^2 + u_x^2) \, dx$
  is conserved up to the collision, where it drops discontinuously,
  since the contribution $\int_{x_1(t)}^{x_2(t)} u_x^2 \, dx$ tends to a positive constant
  which is not visible in the integral defining~$E(t_c)$.
  In the \emph{conservative} solution, which is the one provided by the formulas
  \eqref{eq:CH-twopeakon-x-m} for all $t \neq t_c$
  (or by \eqref{eq:CH-xk-mk-solution-formulas} in case of general~$N$),
  the lost energy is immediately regained as the peakon and the antipeakon reappear for $t>t_c$,
  but now with $m_1 < 0 < m_2$.
  In the \emph{dissipative} solution, the energy stays at the new lower level,
  and the solution continues as a single-peakon travelling wave with velocity $\tfrac{1}{\lambda_1} + \tfrac{1}{\lambda_2}$
  for $t \ge t_c$
  (or as $u=0$ if $\lambda_2 = -\lambda_1$).
  There is also the intermediate case of an \emph{$\alpha$-dissipative} solution, for any $0 < \alpha < 1$,
  where the peakon and the antipeakon reappear but with only a fraction $1-\alpha$ of the lost energy
  regained.
  There are general definitions of conservative, dissipative and $\alpha$-dissipative
  global weak solutions, and the peakon--antipeakon solutions described above are just special instances;
  the computations needed in order to verify that they actually satisfy these definitions
  are rather involved, as are the definitions themselves (see Section~\ref{sec:guide-CH} for references).
  For pure peakon solutions, the problem of collisions does not arise, and the various solution concepts coincide.

  Precise asymptotics as $t \to \pm\infty$ are easily obtained from the explicit solution
  formulas~\eqref{eq:CH-twopeakon-x-m},
  or \eqref{eq:CH-xk-mk-solution-formulas} in the general case.
  If we label the eigenvalues such that $1/\lambda_1 > 1/\lambda_2$,
  then $a_1(t) = a_1(0) \, e^{t/\lambda_1}$ dominates over
  $a_2(t) = a_2(0) \, e^{t/\lambda_2}$ as $t \to \infty$, and the other way around as $t \to -\infty$.
  For example, as $t \to \infty$ we have
  \begin{equation}
    \begin{split}
      x_2(t)
      &
      = \ln 2 \bigl( a_1(t) + a_2(t) \bigr)
      \\ &
      = \ln 2 a_1(t) + \ln\left( 1 + \frac{a_2(t)}{a_1(t)} \right)
      \\ &
      = \frac{t}{\lambda_1} + \ln 2 a_1(0) + o(1)
    \end{split}
  \end{equation}
  and
  \begin{equation}
    \begin{split}
      x_1(t)
      &
      = \ln \frac{2 (\lambda_1-\lambda_2)^2 a_1(t) \, a_2(t)}{\lambda_1^2 a_1(t) + \lambda_2 ^2a_2(t)}
      \\ &
      = \ln 2 a_2(t)
      + \ln \frac{(\lambda_1-\lambda_2)^2}{\lambda_1^2}
      - \ln \left( 1 + \frac{\lambda_2^2}{\lambda_1^2} \, \frac{a_2(t)}{a_1(t)} \right)
      \\ &
      = \frac{t}{\lambda_2}
      + \ln 2 a_2(0)
      + 2 \ln \abs{ 1 - \frac{\lambda_2}{\lambda_1} }
      + o(1)
      .
    \end{split}
  \end{equation}
  We see that the peakons asymptotically move in straight lines,
  with asymptotic velocities given by the reciprocal eigenvalues $1/\lambda_2$ (for $x_1$)
  and $1/\lambda_1$ (for $x_2$).
  As $t \to -\infty$, similar calculations show that
  \begin{equation}
    x_1(t) = \frac{t}{\lambda_1}
    + \ln 2 a_1(0)
    + 2 \ln \abs{ 1 - \frac{\lambda_1}{\lambda_2} }
    + o(1)
  \end{equation}
  and
  \begin{equation}
    x_2(t) = \frac{t}{\lambda_2} + \ln 2 a_2(0) + o(1)
    ,
  \end{equation}
  so the same asymptotic velocities appear as $t \to -\infty$, but in the opposite order.
  The line followed by the faster peakon ($x_2$) as $t \to \infty$
  is shifted in the $x$ direction,
  compared to the line followed by the faster peakon ($x_1$) as $t \to -\infty$,
  by the amount
  $-2 \ln \abs{ 1 - \frac{\lambda_1}{\lambda_2} }$,
  and similarly the phase shift of the slower peakon is
  $2 \ln \abs{ 1 - \frac{\lambda_2}{\lambda_1} }$.
\end{example}

\begin{example}[The three-peakon solution]
  \label{ex:CH-three-peakons}
  For $N=3$, the peakon ODEs \eqref{eq:intro-CH-peakon-ode-explicit} take the form
  \begin{equation}
    \begin{aligned}
      \dot x_1 &= m_1 + m_2 E_{12} + m_3 E_{13}
      , \\
      \dot x_2 &= m_1 E_{12} + m_2 + m_3 E_{23}
      , \\
      \dot x_3 &= m_1 E_{13} + m_2 E_{23} + m_3
      , \\
      \dot m_1 &= -m_1 m_2 E_{12} - m_1 m_3 E_{13}
      , \\
      \dot m_2 &= m_1 m_2 E_{12} - m_2 m_3 E_{23}
      , \\
      \dot m_3 &= m_1 m_3 E_{13} + m_2 m_3 E_{23}
      ,
    \end{aligned}
  \end{equation}
  if we assume $x_1 < x_2 < x_3$ as usual,
  and write $E_{ij} = e^{x_i-x_j}$ for $i<j$.
  According to~\eqref{eq:CH-xk-mk-solution-formulas}, the solution is
  \begin{subequations} \label{eq:CH-threepeakon-x-m}
  \begin{equation}
    \begin{aligned}
      x_1(t) &= \ln\frac{2 \delta_3^0}{\delta_2^2}
      , &
      x_2(t) &= \ln\frac{2 \delta_2^0}{\delta_1^2}
      , &
      x_3(t) &= \ln 2 \delta_1^0
      , \\
      m_1(t) &= \frac{\delta_3^0 \delta_2^2}{\delta_3^1 \delta_2^1}
      , &
      m_2(t) &= \frac{\delta_2^0 \delta_1^2}{\delta_2^1 \delta_1^1}
      , &
      m_3(t) &= \frac{\delta_1^0}{\delta_1^1}
      ,
    \end{aligned}
  \end{equation}
  with
  \begin{equation}
    \begin{aligned}
      \delta_1^k &= \lambda_1^k \, a_1 + \lambda_2^k \, a_2 + \lambda_3^k \, a_3
      , \\
      \delta_2^k &=
      (\lambda_1-\lambda_2)^2 \lambda_1^k \lambda_2^k \, a_1 a_2
      \\ & \quad
      + (\lambda_1-\lambda_3)^2 \lambda_1^k \lambda_3^k \, a_1 a_3
      \\ & \quad
      + (\lambda_2-\lambda_3)^2 \lambda_2^k \lambda_3^k \, a_2 a_3
      , \\
      \delta_3^k &=
      (\lambda_1-\lambda_2)^2 (\lambda_1-\lambda_3)^2 (\lambda_2-\lambda_3)^2
      \times
      \\ & \quad
      \lambda_1^k \lambda_2^k \lambda_3^k \, a_1 a_2 a_3
      ,
    \end{aligned}
  \end{equation}
  \end{subequations}
  and $a_k = a_k(t) = a_k(0) \, e^{t/\lambda_k}$.
\end{example}

To finish this section, let us illustrate how the theory of orthogonal polynomials
can be used to study peakon--antipeakon collisions in general~\cite{beals-sattinger-szmigielski:2000:moment}.
A collision $x_k(t) = x_{k+1}(t)$ takes place precisely when
$l_k(t) = y_{k+1}(t) - y_k(t)$ becomes zero in the corresponding discrete string,
which is equivalent to the determinant $\Delta_{N-k}^1(t)$ becoming zero;
see Remark~\ref{rem:CH-range-of-spectral-map}.

Like in Remark~\ref{rem:CH-orthpoly}, let us write $Q_r(\lambda)$ for the polynomial $Q(\lambda)$
given by~\eqref{eq:CH-Qn-even-det},
now implicitly depending on~$t$ (meromorphically) since the moments $\alpha_j$ are defined in terms of
$a_k(t) = a_k(0) \, e^{t/\lambda_k}$.
These orthogonal polynomials $\{ Q_r \}_{r=0}^N$
with $Q_r(0)=1$ are not suitable to use here, since the denominator in~\eqref{eq:CH-Qn-even-det}
is~$\Delta^1_{r}(t)$, which may become zero.
Instead, we can consider \emph{orthonormal} polynomials $\{ \widehat Q_r \}_{r=0}^N$,
which are explicitly given by
\begin{equation}
  \label{eq:CH-orthopol}
  \widehat Q_r(\lambda) =
  \frac{
    \begin{vmatrix}
      1 & \lambda & \lambda^2 & \dots & \lambda^r \\
      \alpha_0 & \alpha_1 & \alpha_2 & \dots & \alpha_r \\
      \alpha_1 & \alpha_2 & \alpha_3 & \dots &  \alpha_{r+1} \\
      \vdots & \vdots & \vdots & \vdots & \vdots \\
      \alpha_{r-1} & \alpha_r & \alpha_{r+1} & \dots & \alpha_{2r-1}
    \end{vmatrix}
  }{\bigl( \Delta^0_{r} \Delta^0_{\vphantom{r} \smash{r+1}} \bigr)^{1/2}}
\end{equation}
for $0 \le r \le N$.
(For $r=0$, this means the constant polynomial
$\widehat Q_0(\lambda) = \bigl( \Delta^0_{1} \bigr)^{-1/2} = \bigl( \alpha_0 \bigr)^{-1/2} > 0$.)
Clearly,
\begin{equation}
  \label{eq:CH-Qhat-zero}
  \widehat Q_r(0) = \frac{\Delta^1_r}{\bigl( \Delta^0_{r} \Delta^0_{\vphantom{r} \smash{r+1}} \bigr)^{1/2}}
  ,
\end{equation}
so by \eqref{eq:CH-formula-lk} we can write the lengths~$l_k$ in terms of these polynomials:
\begin{equation}
  \label{eq:CH-l-Qhat}
  l_{N-k} = \frac{\bigl( \Delta_{k}^1 \bigr)^2}{\Delta_{k}^0 \, \Delta_{k+1}^0}
  = \widehat Q_{k}(0)^2
  .
\end{equation}
As is well known, the orthonormal polynomials $\widehat Q_n(\lambda)$ satisfy a second-order recursion relation,
\begin{equation}
  \label{eq:CH-orthpoly-recursion}
  \lambda \, \widehat Q_n(\lambda)
  = c_n \, \widehat Q_{n+1}(\lambda)
  + d_n \, \widehat Q_n(\lambda)
  + c_{n-1} \, \widehat Q_{n-1}(\lambda)
  ,
\end{equation}
for $1 \le n \le N-1$, where
\begin{equation}
  \label{eq:CH-christoffel-darboux-b}
  c_n
  = \frac{\big( \Delta^0_{n} \Delta^0_{\vphantom{n} \smash{n+2}} \big)^{1/2}}{\Delta^0_{n+1}}
  > 0
  ,
  \qquad 0 \le n \le N-1
  .
\end{equation}
This recursion relation implies the Christoffel--Darboux formula
\begin{equation*}
  c_n \, \frac{\widehat Q_{n+1}(\lambda) \, \widehat Q_n(\kappa) - \widehat Q_n(\lambda) \, \widehat Q_{n+1}(\kappa)}
  {\lambda-\kappa}
  =
  \sum_{i=0}^n \widehat Q_i(\lambda) \, \widehat Q_i(\kappa)
  ,
\end{equation*}
where $0 \le n \le N-1$,
which in the limit $\kappa \to \lambda$ takes the form
\begin{equation}
  \label{eq:CH-limiting-darboux}
  c_n \, \biggl( \widehat Q\rlap{$'$}_{n+1}(\lambda) \, \widehat Q_n(\lambda) - \widehat Q_{n+1}(\lambda) \, \widehat Q\rlap{$'$}_n(\lambda) \biggr)
  = \sum_{i=0}^n \widehat Q_i(\lambda)^2
  ,
\end{equation}
where primes denote differentiation.
Since $\widehat Q_0(0)\neq 0$, the right-hand side
is positive when $\lambda=0$, which implies that
(for any given value of~$t$) no two
consecutive $\widehat Q_n(0)$ can vanish, which in turn by~\eqref{eq:CH-l-Qhat}
implies that no two consecutive~$l_k$ can vanish.
This means that Camassa--Holm peakons can only collide in pairs;
there are no triple collisions $x_{k-1} = x_k = x_{k+1}$.

We can also express the weights~$g_k$ in terms of quantities related to the orthonormal polynomials,
using \eqref{eq:CH-formula-gk}, \eqref{eq:CH-Qhat-zero} and~\eqref{eq:CH-christoffel-darboux-b}:
\begin{equation}
  \label{eq:CH-g-Qhat}
  g_{N+1-k} = \frac{\bigl( \Delta_{k}^0 \bigr)^2}{\Delta_{k-1}^1 \, \Delta_{k}^1}
  = \frac{1}{c_{k-1} \, \widehat Q_{k-1}(0) \, \widehat Q_{k}(0)}
  .
\end{equation}
This, together with the three-term recurrence \eqref{eq:CH-orthpoly-recursion} (at $\lambda=0$), gives
\begin{equation}
  \label{eq:CH-g-sum-Qhat}
  \begin{split}
    &
    g_k + g_{k+1}
    = \frac{1}{c_{N-k} \, \widehat Q_{N-k}(0) \, \widehat Q_{N+1-k}(0)}
    \\ & \qquad\qquad
    + \frac{1}{c_{N-k-1} \, \widehat Q_{N-k-1}(0) \, \widehat Q_{N-k}(0)}
    \\ &
    = \frac{- d_{N-k} \, \widehat Q_{N-k}(0)}{c_{N-k} \, \widehat Q_{N+1-k}(0) \cdot \widehat Q_{N-k}(0) \cdot c_{N-k-1} \, \widehat Q_{N-k-1}(0)}
    \\ &
    = \frac{- d_{N-k}}{c_{N-k} \, \widehat Q_{N+1-k}(0) \cdot c_{N-k-1} \, \widehat Q_{N-k-1}(0)}
    .
  \end{split}
\end{equation}
Note that $\widehat Q_{N-k}(0)$ cancels in the last step.
By studying the time evolution of $\widehat Q_n(\lambda)$,
one can prove that each $\Delta^1_n(t)$ has only simple zeros~\cite{beals-sattinger-szmigielski:2000:moment}.
At such a simple zero of $\Delta^1_{N-k}(t)$,
say $t=t_c$,
which is the time of a collision where $l_k(t)$ becomes zero,
we see from \eqref{eq:CH-l-Qhat} that $l_k(t)$ has a double zero,
from \eqref{eq:CH-g-Qhat} that $g_k(t)$ and $g_{k+1}(t)$ both have a simple pole,
and from \eqref{eq:CH-g-sum-Qhat} that $g_k(t) + g_{k+1}(t)$ has a removable singularity
(the problematic factor $\widehat Q_{N-k}(0)$ in the denominator is gone,
and the adjacent $\widehat Q_{N+1-k}(0)$ and $\widehat Q_{N-k-1}(0)$
tend to nonzero limits, since no two consecutive $\widehat Q_n(0)$ can vanish simultaneously).

Translating these results to the real line, it follows that there are constants $K_0 > 0$, $K_1 > 0$
and~$K_2 \in \R$ such that
\begin{equation}
  x_{k+1}(t) - x_k(t) = K_0 \, (t-t_c)^2  + O \Bigl( (t-t_c)^3 \Bigr)
\end{equation}
and
\begin{equation}
  \begin{aligned}
    m_k(t) &= - \frac{K_1}{t-t_c} + K_2 + o(1)
    ,\\
    m_{k+1}(t) &= \frac{K_1}{t-t_c} + K_2 + o(1)
    .
  \end{aligned}
\end{equation}
So the trajectories $x=x_k(t)$ and $x=x_{k+1}(t)$
are tangential at the time of collision (with first-order contact only),
and in the sum \eqref{eq:intro-peakons} defining $u(x,t)$,
the two terms
\begin{equation*}
  m_{k}(t) \, e^{-\abs{x - x_{k}(t)}}
  + m_{k+1}(t) \, e^{-\abs{x - x_{k+1}(t)}}
\end{equation*}
converge to a single peakon
\begin{equation*}
  2 K_2 \, e^{-\abs{x - x_{k}(t_c)}}
\end{equation*}
as $t \to t_c$
(or cancel out completely, if $K_2=0$).

\section{The Degasperis--Procesi equation and the cubic string}
\label{sec:DP}

The Degasperis--Procesi equation~\eqref{eq:intro-DP},
\begin{equation*}
  m_t + (u m)_x + 2 u_x m = 0
  ,\qquad
  m = u - u_{xx}
  ,
\end{equation*}
differs in appearance from the Camassa--Holm equation~\eqref{eq:intro-CH},
\begin{equation*}
  m_t + (u m)_x + u_x m = 0
  ,\qquad
  m = u - u_{xx}
  ,
\end{equation*}
only by the factor~$2$ in front of the term~$u_x m$,
and it admits $N$-peakon solutions of the same form~\eqref{eq:intro-peakons}
as the CH equation,
\begin{equation*}
  u(x,t) = \sum_{k=1}^N m_k(t) \, e^{-\abs{x - x_k(t)}}
  ,
\end{equation*}
but governed by the ODEs
\begin{equation}
  \label{eq:DP-peakon-ode-again}
  \dot x_k = u(x_k)
  ,\qquad
  \dot m_k = -2 m_k u_x(x_k)
  ,
\end{equation}
which differ from the CH peakon ODEs~\eqref{eq:intro-CH-peakon-ode-shorthand}
only by the same factor~$2$ in the equations for $\dot m_1, \dots, \dot m_N$.
Pure peakon solutions of the DP equation are qualitatively similar
to pure peakon solutions of the CH equation,
but this is no longer true for mixed peakon--antipeakon solutions,
and in fact the mathematical structure underlying the integrability of the DP equation
is quite different;
the Lax pair is
\begin{subequations}
  \label{eq:DP-lax}
  \begin{align}
    \label{eq:DP-lax-x}
    (\partial_x^3 - \partial_x) \psi &= -\lambda \, m\psi
    , \\
    \label{eq:DP-lax-t}
    \psi_t &= \left[ \lambda^{-1} (1-\partial_x^2) + u_x - u \partial_x \right] \psi
    ,
  \end{align}
\end{subequations}
where the first equation involves the third-order differential operator
$\partial_x^3 - \partial_x$
rather than the CH second-order operator $\partial_x^2 - \tfrac14$.
If we consider a fixed value of~$t$, and omit $t$ in the notation,
equation~\eqref{eq:DP-lax-x} reads
\begin{equation}
  \label{eq:DP-lax-x-again}
  \bigl( \partial_x^3 - \partial_x \bigr) \, \psi(x) = - \lambda \, m(x) \, \psi(x)
  ,
  \qquad
  x \in \R
  ,
\end{equation}
where the term $-\partial_x$ can be removed~\cite{lundmark-szmigielski:2003:DPshort,lundmark-szmigielski:2005:DPlong}
by the Liouville transformation
\begin{equation}
  \label{eq:DP-liouville-trf}
  y = \tanh(x/2)
  , \qquad
  \psi(x) = \frac{2\,\phi(y)}{1-y^2}
  .
\end{equation}
Indeed,
as can be verified using the chain rule,
this turns \eqref{eq:DP-lax-x-again}
into what we call the \emph{cubic string} equation
\begin{equation}
  \label{eq:DP-cubic-string}
  \partial_y^3 \phi(y) = -\lambda \, g(y) \, \phi(y)
  ,\qquad
  -1<y<1
  ,
\end{equation}
where
\begin{equation}
  \label{eq:DP-gm}
  \left( \frac{1-y^2}{2} \right)^3 g(y) = m(x)
  .
\end{equation}
The terminology ``cubic string''
for the novel third-order equation~\eqref{eq:DP-cubic-string}
comes, of course, from the analogy to the classical second-order string equation
\begin{equation*}
  \partial_y^2 \phi(y) = -\lambda \, g(y) \, \phi(y)
  ,\qquad
  -1<y<1
\end{equation*}
which appeared as equation~\eqref{eq:CH-gstring} in our study of the CH equation.

When
$m(x) = 2 \sum_{k=1}^N m_k \, \delta(x-x_k)$
is a discrete measure of the form~\eqref{eq:CH-peakon-m},
we transform the Dirac deltas
according to the same rule~\eqref{eq:CH-dirac-transform-rule}
as in the CH case,
and obtain \eqref{eq:DP-cubic-string} with the discrete measure
\begin{equation}
  \label{eq:DP-peakon-g}
  g(y) = \sum_{k=1}^N g_k \, \delta(y-y_k)
  ,\qquad
  g_k = \frac{8 m_k}{(1-y_k^2)^2}
  ,
\end{equation}
where $y_k = \tanh(x_k/2)$.
We may also verify this with the following calculation, analogous to the one
for the CH equation in Section~\ref{sec:CH}.
As before, we let $x_0 = -\infty$ and $x_{N+1} = +\infty$,
and accordingly $y_0 = -1$ and $y_{N+1} = +1$.
Equation \eqref{eq:DP-lax-x-again} tells us that $(\partial_x^3 - \partial_x) \psi(x)$
must be zero in the intervals where $m$ is zero,
i.e., away from the points $x_k$, so that
\begin{equation}
  \label{eq:DP-psi-piecewise}
  \psi(x) = A_k \, e^x + B_k + C_k \, e^{-x}
  ,\qquad
  x_k < x < x_{k+1}
  ,
\end{equation}
for $0 \le k \le N$,
and that moreover $\psi$ and $\partial_x \psi$ should be continuous,
while $\partial_x^2 \psi$ must jump by $-\lambda m_k \psi(x_k)$ at $x=x_k$,
leading after some calculation to the jump conditions
\begin{equation}
  \label{eq:DP-jump-relation-psi}
  \begin{pmatrix} A_k \\ B_k \\ C_k \end{pmatrix}
  =
  \left[
    \begin{pmatrix} 1 & 0 & 0 \\ 0 & 1 & 0 \\ 0 & 0 & 1 \end{pmatrix}
    - \lambda m_k
    \begin{pmatrix} e^{-x_k} \\ -2  \\ e^{x_k} \end{pmatrix}
    \bigl( e^{x_k}, 1, e^{-x_k} \bigr)
  \right]
  \begin{pmatrix} A_{k-1}\\ B_{k-1}\\ C_{k-1} \end{pmatrix}
  ,
\end{equation}
for $1 \le k \le N$.
Next, with $y = \tanh(x/2)$, an expression of the form
$\psi(x) = A \, e^x + B + C \, e^{-x}$,
in the kernel of $\partial_x^3 - \partial_x$,
becomes
\begin{equation*}
  \begin{split}
    &
    \psi(x)
    = A \, e^x + B + C \, e^{-x}
    \\ &
    = \frac{(e^x+1)^2}{2e^x}
    \left(
      \frac{A}{2} \left( \frac{2e^x}{e^x+1} \right)^2 +
      \frac{B}{2} \frac{4e^x}{(e^x+1)^2} +
      \frac{C}{2} \left( \frac{2}{e^x+1} \right)^2
    \right)
    \\ &
    = \frac{2}{1-y^2}
    \biggl(
    A \, \frac{(1+y)^2}{2} +
    B \, \frac{(1+y)(1-y)}{2} +
    C \, \frac{(1-y)^2}{2}
    \biggr)
    \\ &
    = \frac{2}{1-y^2} \, \phi(y)
  \end{split}
\end{equation*}
where $\phi(y)$ is a quadratic polynomial,
and hence in the kernel of~$\partial_y^3$.
So the solution $\psi(x)$ of \eqref{eq:DP-lax-x-again},
given by \eqref{eq:DP-psi-piecewise},
turns into a piecewise quadratic function $\phi(y)$ on the interval $-1 < y < 1$,
given by
\begin{equation}
  \label{eq:DP-phi-piecewise}
  \phi(y) =
  A_k \, \frac{(1+y)^2}{2} +
  B_k \, \frac{(1+y)(1-y)}{2} +
  C_k \, \frac{(1-y)^2}{2}
\end{equation}
when $y_k < y < y_{k+1}$, for $0 \le k \le N$.
The function $\phi$ and its first derivative $\partial_y \phi$ are continuous,
and left-multiplying the jump conditions \eqref{eq:DP-jump-relation-psi} by
the row vector $(1,-1,1)$
we find that the second derivative $\partial_y^2 \phi$,
which is piecewise constant ($\partial_y^3 \phi=0$ away from the points~$y_k$)
and equals $A_k - B_k + C_k$ for $y \in (y_k, y_{k+1})$,
satisfies
\begin{equation*}
  \begin{split}
    &
    (A_k - B_k + C_k) - (A_{k-1} - B_{k-1} + C_{k-1})
    \\ &
    = - \lambda m_k (e^{x_k}+2+e^{-x_k}) (A_{k-1} e^{x_k} + B_{k-1} + C_{k-1} e^{-x_k})
    \\ &
    = - \lambda m_k \, \frac{(e^x+1)^2}{e^{x}} \, \psi(x_k)
    \\ &
    = - \lambda m_k \, \frac{4}{1-y_k^2} \, \frac{2 \, \phi(y_k)}{1-y_k^2}
    \\ &
    = - \lambda \underbrace{\frac{8 m_k}{(1-y_k^2)^2}}_{= g_k} \, \phi(y_k)
    ,
  \end{split}
\end{equation*}
i.e., it jumps by $-\lambda \, g_k \, \phi(y_k)$ at $y=y_k$,
where $g_k = 8 m_k / (1-y_k^2)^2$,
so that $\phi(y)$ indeed satisfies the cubic string equation~\eqref{eq:DP-cubic-string}
in the sense of distributions,
with the transformed discrete measure~\eqref{eq:DP-peakon-g}, as claimed.

The boundary values relevant for the study of peakon solutions turn out to be
the Dirichlet-like lopsided conditions
\begin{equation}
  \label{eq:DP-cubic-string-boundary-conditions-y}
  \phi(-1) = \phi_y(-1) = 0
  , \qquad
  \phi(1) = 0
  .
\end{equation}
To study the corresponding eigenvalue problem as a shooting problem,
let $\phi(y;\lambda)$ be the solution of the discrete cubic string equation
with initial values $\phi(-1) = \phi_y(-1) = 0$ and (for normalization)
$\phi_{yy}(-1) = 1$;
these choices correspond to $(A_0,B_0,C_0) = (1,0,0)$,
and $\lambda$ is an eigenvalue if and only if
$\phi(1;\lambda) = 2 A_N(\lambda) = 0$,
i.e., they are the zeros of $A(\lambda) = A_N(\lambda)$,
which by the jump conditions \eqref{eq:DP-jump-relation-psi}
is a polynomial of degree~$N$.

Since the eigenvalue problem is not selfadjoint,
there is perhaps no obvious reason to expect the eigenvalues to be real,
but they are in fact positive and simple provided that all masses $g_k$ are \textbf{positive}
(corresponding to \textbf{pure} $N$-peakon solutions),
since then the problem can be shown~\cite{lundmark-szmigielski:2005:DPlong}
to be \textbf{oscillatory}
in the sense of Gantmacher and Krein~\cite{gantmacher-krein:2002:oscillation-matrices}.
\emph{We assume from now on (unless otherwise mentioned) that this condition holds.}

It is now natural to define \emph{two} Weyl functions, each a rational function
with simple poles at the eigenvalues~$\lambda_k$
(for $1 \le k \le N$):
\begin{equation}\label{eq:DP-weyl}
  W(\lambda) = \frac{\phi_y(1;\lambda)}{\phi(1;\lambda)}
  , \qquad
  Z(\lambda) = \frac{\phi_{yy}(1;\lambda)}{\phi(1;\lambda)}
  .
\end{equation}
The numerator and denominator in these functions have the same degree,
so we divide by $\lambda$ in order to get functions of order
$O(1/\lambda)$ as $\lambda \to \infty$.
This adds a simple pole at $\lambda = \lambda_0 = 0$, with residue $a_0 = W(0) = 1$
and $b_0 = Z(0) = 1/2$, respectively (since $\phi(y;0) = \tfrac12 (1+y)^2$).
In the forward spectral problem, where the discrete measure~$g$ is given,
we determine spectral data consisting of the eigenvalues~$\lambda_k$
together with the remaining residues $a_k$ and~$b_k$
in the partial fraction decompositions of these modified Weyl functions:
\begin{align}
  \label{eq:DP-parfracW}
  \frac{W(\lambda)}{\lambda} &
  = \frac{1}{\lambda} + \sum_{k=1}^N \frac{a_k}{\lambda - \lambda_k}
  = \sum_{k=0}^N \frac{a_k}{\lambda - \lambda_k}
  , \\
  \label{eq:DP-parfracZ}
  \frac{Z(\lambda)}{\lambda} &
  = \frac{1/2}{\lambda} + \sum_{k=1}^N \frac{b_k}{\lambda - \lambda_k}
  = \sum_{k=0}^N \frac{b_k}{\lambda - \lambda_k}
  ,
\end{align}
where
\begin{equation}
  \lambda_0=0
  ,\quad
  a_0=1
  ,\quad
  b_0=1/2
  .
\end{equation}
Under our assumption that all $g_k$ are positive,
it can be shown that all $a_k$ and~$b_k$ are positive as well.

A crucial fact is that the second Weyl function~$Z$ is actually determined by the
first Weyl function~$W$, so that the residues $b_k$ are redundant, and
we can take the spectral data to be just $\{ \lambda_k , a_k \}_{k=1}^N$.
Indeed, with $\eta(y;\lambda) = \phi(y;-\lambda)$ we have
$\phi_{yyy}=-\lambda g \phi$ and $\eta_{yyy}=+\lambda g \eta$, so that
$0 = \eta \phi_{yyy} + \eta_{yyy} \phi = (\eta \phi_{yy} - \eta_{y} \phi_{y} + \eta_{yy} \phi)_{y}$.
Integration over $y \in [-1,1]$ gives
$0 = \eta(1) \, \phi_{yy}(1) - \eta_{y}(1) \, \phi_{y}(1) + \eta_{yy}(1) \, \phi(1)$,
since the boundary conditions \eqref{eq:DP-cubic-string-boundary-conditions-y} make all the
contributions from the left endpoint $y=-1$ vanish.
Division by $\eta(1) \, \phi(1)$ gives
\begin{equation}
  \label{eq:DP-Z-W-symmetry-relation}
  Z(\lambda) - W(-\lambda) \, W(\lambda) + Z(-\lambda) = 0.
\end{equation}
It is clear that this relation determines the even part of~$Z$ in terms of~$W$,
but since we know that~$Z$ has the form~\eqref{eq:DP-parfracZ},
this is actually enough to determine~$Z$ completely.
Indeed, if we divide \eqref{eq:DP-Z-W-symmetry-relation} by~$\lambda$
and take the residue at $\lambda=\lambda_k$,
we get $b_k - W(-\lambda_k) \, a_k + 0 = 0$,
or in other words
\begin{equation}
  \label{eq:DP-bk}
  b_k = \lambda_k a_k \sum_{j=0}^N \frac{a_j}{\lambda_j+\lambda_k}
  \quad (1 \le k \le N)
  ,
\end{equation}
which determines~$Z(\lambda)$ through~\eqref{eq:DP-parfracZ}.
Here we catch our first glimpse of the \emph{Cauchy kernel}
\begin{equation}
  \label{eq:DP-cauchy-kernel}
  K(x,y)=\frac{1}{x+y},
\end{equation}
which plays an important role in the
inverse spectral theory of the cubic string.

Let us define the spectral measure
\begin{equation}
  \label{eq:DP-measure-alpha}
  \alpha(\lambda)
  = \delta(\lambda) + \sum_{k=1}^N a_k \, \delta(\lambda-\lambda_k)
  = \sum_{k=0}^N a_k \, \delta(\lambda-\lambda_k)
  ,
\end{equation}
together with an auxiliary measure
\begin{equation}
  \label{eq:DP-measure-beta}
  \beta(\lambda)
  = \lambda \, \alpha(\lambda)
  = \sum_{k=1}^N \lambda_k \, a_k \, \delta(\lambda-\lambda_k)
  .
\end{equation}
(Note that the multiplication by $\lambda$ kills the term $\delta(\lambda)$
in~\eqref{eq:DP-measure-alpha}, so that we can start the summation from $k=1$ rather than $k=0$.)
Then $W(\lambda)/\lambda$ is a Stieltjes transform
\begin{equation}
  \label{eq:DP-W-stieltjes-transform}
  \frac{W(\lambda)}{\lambda}
  = \int \frac{d\alpha(z)}{\lambda-z}
  ,
\end{equation}
while $Z(\lambda)/\lambda$ can be written as
\begin{equation}
  \label{eq:DP-Z-stieltjes-transform}
  \begin{split}
    \frac{Z(\lambda)}{\lambda}
    &
    = \frac{1/2}{\lambda} + \sum_{k=1}^N \frac{b_k}{\lambda - \lambda_k}
    \\ &
    = \frac{1/2}{\lambda} + \sum_{k=1}^N \sum_{j=0}^N \frac{\lambda_k a_k a_j}{(\lambda_j + \lambda_k) (\lambda - \lambda_k)}
    \\ &
    = \frac{1/2}{\lambda} + \iint \frac{d\beta(z_1) \, d\alpha(z_2)}{(z_1+z_2)(\lambda-z_1)}
    .
  \end{split}
\end{equation}

By letting
$\Phi = (\phi_1,\phi_2,\phi_3)^T = (\phi,\phi_y,\phi_{yy})^T$,
we can write the cubic string equation~\eqref{eq:DP-cubic-string}
with the boundary conditions~\eqref{eq:DP-cubic-string-boundary-conditions-y}
as a $3 \times 3$ matrix equation
\begin{subequations}
  \label{eq:DP-cubic-string-matrix}
\begin{equation}
  \label{eq:DP-cubic-string-matrix-ODE}
  \frac{\partial}{\partial y}
  \begin{pmatrix} \phi_1 \\ \phi_2 \\ \phi_3 \end{pmatrix}
  =
  \begin{pmatrix}
    0 & 1 & 0 \\
    0 & 0 & 1 \\
    -\lambda \, g(y) & 0 & 0
  \end{pmatrix}
  \begin{pmatrix} \phi_1 \\ \phi_2 \\ \phi_3 \end{pmatrix}
\end{equation}
with the boundary conditions
\begin{equation}
  \label{eq:DP-cubic-string-matrix-boundary-conditions-y}
  \phi_1(-1) = \phi_2(-1) = 0
  ,
  \qquad
  \phi_1(1) = 0
  .
\end{equation}
\end{subequations}
Then $\Phi(y;\lambda)$ is the solution starting out with $\Phi(-1;\lambda) = (0,0,1)^T$.
Since
\begin{equation*}
  \phi(y;\lambda) = \phi(y_k;\lambda) + \phi_y(y_k;\lambda) \, (y-y_k) + \tfrac12 \phi_{yy}(y_k^+;\lambda) (y-y_k)^2
\end{equation*}
on the interval $y_k \le y \le y_{k+1}$, by Taylor's formula,
we find at $y_{k+1}$ that
\begin{equation}
  \label{eq:DP-Lk}
  \Phi(y_{k+1}^-;\lambda) = L_k \, \Phi(y_{k}^+;\lambda)
  ,\quad
  L_k =
  \begin{pmatrix}
    1 & l_k & l_k^2/2 \\
    0 & 1 & l_k \\
    0 & 0 & 1
  \end{pmatrix}
  ,
\end{equation}
while the jump condition for $\phi_{yy}$ at $y_k$ becomes
\begin{equation}
  \label{eq:DP-Gk}
  \Phi(y_k^+;\lambda) = G_k(\lambda) \, \Phi(y_k^-;\lambda)
  ,\quad
  G_k(\lambda) =
  \begin{pmatrix}
    1 & 0 & 0 \\
    0 & 1 & 0 \\
    -\lambda\,g_k & 0 & 1
  \end{pmatrix}
  .
\end{equation}
Combining these formulas, we get
\begin{equation}
  \label{eq:DP-cubic-string-right-endpoint-matrix-product}
  \Phi(1;\lambda)
  = L_N \, G_N(\lambda) \, L_{N-1} \, G_{N-1}(\lambda) \dotsm L_1 \, G_1(\lambda) \, L_0
  \begin{pmatrix} 0 \\ 0 \\ 1 \end{pmatrix}
  .
\end{equation}
Considering the similarity with~\eqref{eq:CH-string-right-endpoint-matrix-product},
one may suspect that the entries in the $3\times 3$ matrices
\begin{equation*}
  \begin{split}
    X_{1}(\lambda) &= L_N
    ,\\
    X_{2}(\lambda) &= L_N \, G_N(\lambda)
    ,\\
    X_{3}(\lambda) &= L_N \, G_N(\lambda) \, L_{N-1}
    ,\\
    X_{4}(\lambda) &= L_N \, G_N(\lambda) \, L_{N-1} \, G_{N-1}(\lambda)
    ,\\
    & \;\;\vdots
    \\
    X_{2N+1}(\lambda) &= L_N \, G_N(\lambda) \, L_{N-1} \, G_{N-1}(\lambda) \dotsm L_1 \, G_1(\lambda) \, L_0
  \end{split}
\end{equation*}
could be used for constructing rational approximations to the Weyl functions
$W$ and~$Z$, and so it is indeed.
For example, for each fixed~$r$ with $1 \le r \le N$,
one can show~\cite[Sect.~4.1]{lundmark-szmigielski:2005:DPlong}
that the second column of $X_{2r}(\lambda)$, call it
\begin{equation*}
  (Q(\lambda),P(\lambda),\widehat P(\lambda))^T
\end{equation*}
satisfies
\begin{subequations} \label{eq:DP-WZ-PQ-approx}
\begin{equation}
  \label{eq:DP-WZ-PQ-approx-1}
  W(\lambda) = \frac{P(\lambda)}{Q(\lambda)} + O(\lambda^{1-r})
  ,
  \quad
  Z(\lambda) = \frac{\widehat P(\lambda)}{Q(\lambda)} + O(\lambda^{1-r})
  ,
\end{equation}
\begin{equation}
  \label{eq:DP-WZ-PQ-approx-2}
  Z(-\lambda) Q(\lambda) - W(-\lambda) P(\lambda) + \widehat P(\lambda) = O(\lambda^{-r})
  ,
\end{equation}
\begin{equation}
  \label{eq:DP-WZ-PQ-approx-3}
  P(0) = 1
  , \quad
  \widehat P(0) = 0
  ,
\end{equation}
and
\begin{equation}
  \label{eq:DP-WZ-PQ-approx-4}
  \deg Q(\lambda) = \deg P(\lambda) = \deg \widehat P(\lambda) = r-1
  .
\end{equation}
\end{subequations}
Note from \eqref{eq:DP-WZ-PQ-approx-1} and~\eqref{eq:DP-WZ-PQ-approx-4} that
\begin{equation*}
  W(\lambda) \, Q(\lambda) = P(\lambda) + O(1)
  ,\quad
  Z(\lambda) \, Q(\lambda) = \widehat P(\lambda) + O(1)
  ,
\end{equation*}
so in contrast to the Padé approximations \eqref{eq:CH-first-pade-approx}
and~\eqref{eq:CH-second-pade-approx}
there are no ``missing powers'' on the right-hand sides which immediately impose
conditions on the coefficients of~$Q$.
It is only when $P$ and $\widehat P$ are expressed in terms of~$Q$ through these relations
and inserted into~\eqref{eq:DP-WZ-PQ-approx-2}
(which is an approximate version of~\eqref{eq:DP-Z-W-symmetry-relation})
that we obtain equations for these coefficients.
More specifically, we find that $Q(\lambda)=\sum_{i=0}^{r-1} q_i \lambda^i$
satisfies the linear system
\begin{equation}
  \label{eq:DP-bimoment-linear-equations}
  \begin{pmatrix}
    I_{00} & I_{01} & \dots & I_{0,r-1} \\
    I_{10} & I_{11} & \dots & I_{1,r-1} \\
    \vdots & \vdots & & \vdots \\
    I_{r-1,0} & I_{r-1,1} & \dots & I_{r-1,r-1} \\
  \end{pmatrix}
  \begin{pmatrix}
    q_0 \\ q_1 \\ \vdots \\ q_{r-1}
  \end{pmatrix}
  =
  \begin{pmatrix}
    \alpha_0 \\ \alpha_1 \\ \vdots \\ \alpha_{r-1}
  \end{pmatrix},
\end{equation}
where the vector entries on the right-hand side,
\begin{equation}
  \alpha_j = \int z^j \, d\alpha(z)
  = \sum_{k=0}^N \lambda_k^j a_k
  ,
\end{equation}
are the moments of the spectral measure~\eqref{eq:DP-measure-alpha},
and where the matrix entries on the left-hand side,
\begin{equation}
  I_{ij} = \iint \frac{z_1^{i} \, z_2^j}{z_1+z_2} \, d\beta(z_1) \, d\alpha(z_2)
  = \sum_{k=1}^N \sum_{l=0}^N \frac{\lambda_k^{i+1} \lambda_l^j}{\lambda_k + \lambda_l} \, a_k a_l
  ,
\end{equation}
are the \emph{bimoments} of the measures \eqref{eq:DP-measure-alpha} and~\eqref{eq:DP-measure-beta}
with respect to the Cauchy kernel~\eqref{eq:DP-cauchy-kernel}.
The bimoment matrix in~\eqref{eq:DP-bimoment-linear-equations}
turns out to be nonsingular
(in fact \emph{totally positive}, a much stronger condition meaning that all its minors are positive),
so the \emph{Hermite--Padé approximation problem}~\eqref{eq:DP-WZ-PQ-approx}
uniquely determines the polynomials $Q$, $P$ and~$\widehat P$.

From the definition of $Q(\lambda)$ as the $(1,2)$ entry of~$X_{2r}(\lambda)$,
one can deduce that
\begin{equation*}
  q_0 = \sum_{i=N+1-r}^N \!\!\!\! l_i = 1 - y_{N+1-r}
\end{equation*}
and
\begin{equation*}
  q_{r-1} = l_{N+1-r} \prod_{i=N+2-r}^N \left( \frac{-g_i \, l_i^2}{2} \right)
  .
\end{equation*}
By Cramer's rule,
the linear system~\eqref{eq:DP-bimoment-linear-equations} gives
formulas for these quantities in terms of bimoment determinants,
and hence in terms of the spectral data.
And with this information extracted from the matrices $X_2$, $X_4$, \ldots,~$X_{2N}$
we can solve for all the variables $y_k$ and~$g_k$
in terms of the spectral data,
hence obtaining determinantal formulas for the solution of
the inverse spectral problem for the discrete cubic string.
Formulas analogous to Heine's formula~\eqref{eq:CH-delta-k-heine},
although more complicated, can be used to evaluate the bimoment determinants
explicitly in terms of the spectral data.
Compared to the CH case, where the ratios of determinants obtained from Cramer's rule
were the end of the story, there is one more complication here,
namely that these quotients contain some common factors that need to be cancelled
in order to obtain the solution formulas in their final simplified form.

\begin{remark}
  The factors remaining after the cancellation (expressions such as $U_k$, $V_k$ and $W_k$ appearing
  in the peakon solution formulas \eqref{eq:DP-n-peakon-solution} below)
  have been identified by Chang and
  collaborators~\cite{chang-hu-li-zhao:2018:novikov-peakons-pfaffians-Toda-lattice-BKP,
    chang:2022:hermite-pade-pfaffian-structures-novikov-and-integrable-lattices},
  in the closely related context of Novikov's equation,
  as being not determinants but \emph{Pfaffians} of certain skew-symmetric matrices.
\end{remark}

\begin{remark}
  \label{rem:DP-CBOP}
  To put all these structures into context,
  Bertola, Gekhtman and Szmigielski
  \cite{bertola-gekhtman-szmigielski:2009:cubicstring,
    bertola-gekhtman-szmigielski:2010:cauchy}
  developed a general theory of \emph{Cauchy biorthogonal polynomials} (CBOPs),
  with connections not only to peakons and approximation theory,
  but also to random matrices~\cite{bertola-gekhtman-szmigielski:2009:twomatrix,
    bertola-gekhtman-szmigielski:2013:strong-asymptotics,
    bertola-gekhtman-szmigielski:2014:cauchy-laguerre-twomatrix-meijerG}.
  Their setup involves two measures $\alpha$ and~$\beta$ on the positive real line~$\R_+$,
  with finite moments
  \begin{equation}
    \alpha_k = \int x^k \, d\alpha(x)
    ,\qquad
    \beta_k = \int y^k \, d\beta(y)
    ,
  \end{equation}
  and finite bimoments
  \begin{equation}
    I_{ij} = \iint x^i \, y^j \, K(x,y) \, d\alpha(x) \, d\beta(y)
    .
  \end{equation}
  with respect the measures $\alpha$ and~$\beta$ and some kernel $K(x,y)$ on $\R_+^2$
  which is is totally positive, meaning that
  \begin{equation*}
    \det \Bigl( K(x_i,y_j) \Bigr)_{i,j=1}^m > 0
  \end{equation*}
  whenever $0 < x_1 < \dots < x_m$ and $0 < y_1 < \dots < y_m$.
  Then, assuming that $\alpha$ and $\beta$ are supported at infinitely many points,
  there are polynomials $\{ p_n , q_n \}_{n=0}^{\infty}$,
  with $p_n$ and $q_n$ of degree~$n$,
  satisfying the biorthogonality condition
  \begin{equation}
    \begin{split}
      \langle p_i \mid q_j \rangle
      &:=
      \iint p_i(x) \, q_j(y) \, K(x,y) \, d\alpha(x) \, d\beta(y)
      \\ &
      =
      \begin{cases}
        1, & i=j
        ,\\
        0, & i \neq j
        ,
      \end{cases}
    \end{split}
  \end{equation}
  and these polynomials are uniquely determined if we normalize by requiring
  the highest coefficient of $p_n$ to be positive and equal to
  the highest coefficient of~$q_n$, for each~$n$.
  They have positive simple zeros, and are explicitly given by
  \begin{equation}
    q_n(x) =
    \frac{
      \begin{vmatrix}
        I_{00} & I_{01} & \dots & I_{0,n-1} & 1 \\
        I_{10} & I_{11} & \dots & I_{1,n-1} & x \\
        \vdots & \vdots & \vdots & \vdots & \vdots \\
        I_{n-1,0} & I_{n-1,1} & \dots & I_{n-1,n-1} & x^{n-1} \\
        I_{n0} & I_{n1}  & \dots & I_{n,n-1} & x^n
      \end{vmatrix}
    }{\sqrt{ D_{n} \, D_{n+1} }}
  \end{equation}
  and
  \begin{equation}
    p_n(y) =
    \frac{
      \begin{vmatrix}
        I_{00} & I_{01} & \dots & I_{0,n-1} & I_{0n} \\
        I_{10} & I_{11} & \dots & I_{1,n-1} & I_{1n} \\
        \vdots & \vdots & \vdots & \vdots & \vdots \\
        I_{n-1,0} & I_{n-1,1} & \dots & I_{n-1,n-1} & I_{n-1,n} \\
        1 & y  & \dots & y^{n-1} & y^n
      \end{vmatrix}
    }{\sqrt{ D_{n} \, D_{n+1} }}
    ,
  \end{equation}
  where $D_0 = 1$ and
  \begin{equation}
    D_n =
    \begin{vmatrix}
      I_{00} & I_{01} & \dots & I_{0,n-1} \\
      I_{10} & I_{11} & \dots & I_{1,n-1} \\
      \vdots & \vdots & \vdots & \vdots  \\
      I_{n-1,0} & I_{n-1,1} & \dots & I_{n-1,n-1}
    \end{vmatrix}
    , \quad
    n \ge 1
    .
  \end{equation}
  CBOPs arise when using the Cauchy kernel
  $K(x,y)=\frac{1}{x+y}$,
  whose total positivity follows from the famous formula for the Cauchy determinant,
  \begin{equation*}
    \det \biggl( \frac{1}{x_i + y_j} \biggr)_{i,j=1}^m
    = \frac{\prod_{1 \le i < j \le m} (x_i-x_j) (y_i-y_j)}{\prod_{i,j=1}^m (x_i+y_j)}
    .
  \end{equation*}
  In this case, the zeros of $q_n$ interlace those of~$q_{n+1}$ for all~$n$,
  and likewise for $p_n$ and $p_{n+1}$.
  There are also four-term recurrence relations, Christoffel--Darboux-type identities,
  Hermite--Padé approximation problems whose solution is given in terms of CBOPs,
  and more.
\end{remark}

\begin{remark}
  \label{rem:DP-CBOP-discrete}
  In the context of the discrete cubic string and Degasperis--Procesi peakons,
  there is essentially only one spectral measure~$\alpha$,
  since the measure $\beta$ given by~\eqref{eq:DP-measure-beta}
  depends in a trivial way on $\alpha$ given by~\eqref{eq:DP-measure-alpha},
  and likewise for the dual cubic string and Novikov peakons in Section~\ref{sec:Novikov}
  where actually $\beta=\alpha$,
  but when we come to Geng--Xue peakons in Section~\ref{sec:GX}
  there will be two independent spectral measures $\alpha$ and~$\beta$.
  In all these cases the spectral measures are supported at finitely many points,
  which is a degenerate situation since the determinants $D_n$ will be zero for large~$n$,
  so there are only finitely many CBOPs.
  Also, $\alpha$ in~\eqref{eq:DP-measure-alpha} is not a measure on~$\R_+$ since it has a point mass
  at the origin, but no problems with division by zero arise, since the support of~$\beta$
  lies in~$\R_+$.
\end{remark}

Returning now to peakon solutions of the DP equation,
we switch on the time-dependence again, and consider the second Lax equation~\eqref{eq:DP-lax-t},
\begin{equation*}
  \psi_t = \left[ \lambda^{-1} (1-\partial_x^2) + u_x - u \partial_x \right] \psi
  .
\end{equation*}
The preimage of $\phi(y,t;\lambda)$ under the Liouville transformation~\eqref{eq:DP-liouville-trf},
call it $\psi(x,t;\lambda)$, satisfies $\psi(x,t;\lambda) = e^x$ in the region $x < x_1(t)$ where $u = u_x$,
so both sides of the Lax equation vanish identically there.
And in the region $x > x_N(t)$ we have
$\psi(x,t;\lambda) = A(t;\lambda) \, e^x + B(t;\lambda) + C(t;\lambda) \, e^{-x}$,
where $(A,B,C)=(A_N,B_N,C_N)$ in the notation of~\eqref{eq:DP-psi-piecewise},
while $u=U(t) \, e^{-x} = -u_x$,
so that the Lax equation becomes
\begin{equation*}
  A_t \, e^x + B_t + C_t \, e^{-x}
  = \frac{B}{\lambda} - 2 A U - B U e^{-x}
  ,
\end{equation*}
which implies that $A_t = 0$, $B_t = B/\lambda - 2 AU$ and $C_t = - BU$.
So the polynomial $A = A(\lambda)$ is time-independent, and hence so are its roots,
the eigenvalues~$\lambda_k$.
This shows that the boundary conditions \eqref{eq:DP-cubic-string-boundary-conditions-y}
are consistent with the time evolution induced by the DP equation,
which therefore induces an isospectral deformation of the cubic string.
Evaluating $B_t = B/\lambda - 2 AU$ at $\lambda = \lambda_k$
shows that $B_t(\lambda_k) = B(\lambda_k) / \lambda_k$,
so exactly as for the CH equation in Section~\ref{sec:CH}
it follows that $\partial_t a_k = a_k/\lambda_k$ for $1 \le  k \le N$,
so that
\begin{equation}
  \label{eq:DP:a-of-t}
  a_k(t) = a_k(0) \, e^{t/\lambda_k}
  .
\end{equation}
And also like in the CH case, this means that we have actually solved the DP
peakon ODEs~\eqref{eq:intro-DP-peakon-ode};
we just take the formulas for $y_k$ and~$g_k$ in terms of the spectral data
and map them back to $x_k$ and~$m_k$ using~\eqref{eq:DP-peakon-g},
and let the spectral data evolve in time according to~\eqref{eq:DP:a-of-t}.
To describe the results, we need a bit of notation.
With
\begin{equation}
  \Psi(z_1,\dots,z_k)
  = \prod_{1 \le i < j \le k} \frac{(z_i - z_j)^2}{z_i + z_j}
  ,
\end{equation}
let
\begin{equation}
  \label{eq:DP-def-Uk}
  U_k = \sum_{1 \le i_1 < \dots < i_k \le N} \Psi(\lambda_{i_1},\dots,\lambda_{i_k}) \, a_{i_1} \dotsm a_{i_k}
\end{equation}
for $1 \le k \le N$,
let $U_0 = 1$, and let $U_k=0$ for other values of~$k$.
Let $V_k$ be like $U_k$ except with $\lambda_i a_i$ instead of $a_i$ for all~$i$,
and finally let
\begin{equation}
  \label{eq:DP-def-Wk}
  W_k = \begin{vmatrix} U_k & V_{k-1} \\ U_{k+1} & V_k \end{vmatrix}
  = U_k V_k - U_{k+1} V_{k-1}
\end{equation}
for all~$k$.
In terms of these quantities, the general pure $N$-peakon solution to the DP equation is given by
\begin{equation}
  \label{eq:DP-n-peakon-solution}
  x_{N+1-k} = \log \frac{U_k}{V_{k-1}}
  , \quad
  m_{N+1-k} =
  \frac{(U_k)^2 \, (V_{k-1})^2}{W_k W_{k-1}}
  ,
\end{equation}
for $1 \le k \le N$.

\begin{example}[The two-peakon solution]
  \label{ex:DP-two-peakons}
  As in the CH case, the DP two-peakon solution can be found by
  direct integration using the variables $x_1 \pm x_2$ and $m_1 \pm m_2$,
  and this was done in the original paper by Degasperis, Holm and
  Hone~\cite{degasperis-holm-hone:2002:new-integrable-equation-DP}.
  The governing ODEs are
  \begin{equation}
    \label{eq:DP-twopeakon-ode}
    \begin{aligned}
      \dot x_1 &= m_1 + m_2 \, e^{x_1-x_2}
      , &
      \dot m_1 &= - 2 m_1 m_2 \, e^{x_1-x_2}
      , \\
      \dot x_2 &= m_1 \, e^{x_1-x_2} + m_2
      , &
      \dot m_2 &= 2 m_1 m_2 \, e^{x_1-x_2}
      ,
    \end{aligned}
  \end{equation}
  where we have assumed that $x_1 < x_2$,
  like in the CH case~\eqref{eq:CH-twopeakon-ode},
  in order to remove the absolute values in the ODEs.
  In our notation, the solution (at least in the pure peakon case) takes the form
  \begin{equation}
    \label{eq:DP-twopeakon-solution}
    \begin{split}
      x_1(t) &= \ln\frac{U_2}{V_1}
      = \ln \frac{\frac{(\lambda_1-\lambda_2)^2}{\lambda_1+\lambda_2}a_1 a_2}{\lambda_1 a_1 + \lambda_2 a_2}
      , \\
      x_2(t) &= \ln\frac{U_1}{V_0} = \ln (a_1+a_2)
      , \\
      m_1(t) &= \frac{(U_2)^2 (V_1)^2}{W_2 W_1} = \frac{(\lambda_1 a_1 + \lambda_2 a_2)^2}{\lambda_1 \lambda_2 \left( \lambda_1 a_1^2 + \lambda_2 a_2^2 + \frac{4 \lambda_1 \lambda_2}{\lambda_1+\lambda_2}a_1 a_2 \right)}
      , \\
      m_2(t) &= \frac{(U_1)^2 (V_0)^2}{W_1 W_0} = \frac{(a_1+a_2)^2}{\lambda_1 a_1^2 + \lambda_2 a_2^2 + \frac{4 \lambda_1 \lambda_2}{\lambda_1+\lambda_2}a_1 a_2}
      ,
    \end{split}
  \end{equation}
  where $a_k = a_k(t) = a_k(0) \, e^{t/\lambda_k}$.
  Like in Example~\ref{ex:CH-two-peakons}, we can extract precise asymptotics as $t \to \pm\infty$
  from these formulas simply by looking at dominant terms. For example,
  if we label the eigenvalues such that $1/\lambda_1 > 1/\lambda_2$,
  then as $t \to \infty$ we have
  \begin{equation}
    x_2(t) = \ln\big( a_1(t) + a_2(t) \bigr)
    = \frac{t}{\lambda_1} + \ln a_1(0) + o(1)
  \end{equation}
  in exactly the same way as in Example~\ref{ex:CH-two-peakons}, and
  \begin{equation}
    \begin{split}
      x_1(t)
      &
      = \ln \frac{\frac{(\lambda_1-\lambda_2)^2}{\lambda_1+\lambda_2} a_1(t) \, a_2(t)}{\lambda_1 a_1(t) + \lambda_2 a_2(t)}
      \\ &
      = \ln a_2(t)
      + \ln \frac{(\lambda_1-\lambda_2)^2}{\lambda_1 (\lambda_1 + \lambda_2)}
      - \ln \left( 1 + \frac{\lambda_2}{\lambda_1} \, \frac{a_2(t)}{a_1(t)} \right)
      \\ &
      = \frac{t}{\lambda_2} + \ln a_2(0)
      + \ln \frac{\bigl( 1 - \frac{\lambda_2}{\lambda_1} \bigr)^2}{1 + \frac{\lambda_2}{\lambda_1}}
      + o(1)
      ,
    \end{split}
  \end{equation}
  while as $t \to -\infty$ we instead have
  \begin{equation}
    x_2(t) = \frac{t}{\lambda_2} + \ln a_2(0) + o(1)
  \end{equation}
  and
  \begin{equation}
    \begin{split}
      x_1(t)
      = \frac{t}{\lambda_1} + \ln a_1(0)
      + \ln \frac{\bigl( 1 - \frac{\lambda_1}{\lambda_2} \bigr)^2}{1 + \frac{\lambda_1}{\lambda_2}}
      + o(1)
      .
    \end{split}
  \end{equation}
  Thus the peakons asymptotically move in straight lines,
  with asymptotic velocities given by the reciprocal eigenvalues $1/\lambda_k$,
  as in the CH case, but here the phase shifts of these lines are different:
  \begin{equation*}
    -\ln \frac{\bigl( 1 - \frac{\lambda_1}{\lambda_2} \bigr)^2}{1 + \frac{\lambda_1}{\lambda_2}}
    \quad\text{and}\quad
    \ln \frac{\bigl( 1 - \frac{\lambda_2}{\lambda_1} \bigr)^2}{1 + \frac{\lambda_2}{\lambda_1}}
  \end{equation*}
  for the faster and the slower peakon, respectively.
\end{example}

\begin{example}[The three-peakon solution]
  \label{ex:DP-three-peakons}
  For $N=3$, the relevant quantities~$U_k$ are given by
  $U_{-1} = 0$,
  $U_0 = 1$,
  \begin{equation*}
    \begin{split}
      U_1 &= a_1+a_2+a_3
      ,
      \\
      U_2 &=
      \frac{(\lambda_1-\lambda_2)^2}{\lambda_1+\lambda_2} a_1 a_2
      +\frac{(\lambda_1-\lambda_3)^2}{\lambda_1+\lambda_3} a_1 a_3
      +\frac{(\lambda_2-\lambda_3)^2}{\lambda_2+\lambda_3} a_2 a_3
      ,
      \\
      U_3 &=
      \frac{(\lambda_1-\lambda_2)^2 (\lambda_1-\lambda_3)^2 (\lambda_2-\lambda_3)^2}
      {(\lambda_1+\lambda_2) (\lambda_1+\lambda_3) (\lambda_2+\lambda_3)}
      a_1 a_2 a_3
    \end{split}
  \end{equation*}
  and $U_4 = 0$,
  while $V_k$ is obtained from $U_k$ by replacing each $a_i$ with $\lambda_i a_i$,
  and consequently
  \begin{equation*}
    \begin{split}
      W_0 &= 1
      , \\
      W_1 &= U_1 V_1 - U_2 V_0
      \\ &
      = \lambda_1 a_1^2 + \lambda_2 a_2^2 + \lambda_3 a_3^2
      \\ & \quad
      + \frac{4 \lambda_1 \lambda_2}{\lambda_1+\lambda_2} a_1 a_2
      + \frac{4 \lambda_1 \lambda_3}{\lambda_1+\lambda_3} a_1 a_3
      + \frac{4 \lambda_2 \lambda_3}{\lambda_2+\lambda_3} a_2 a_3
      ,
      \\
      W_2 &= U_2 V_2 - U_3 V_1
      \\ &
      = \frac{(\lambda_1-\lambda_2)^4}{(\lambda_1+\lambda_2)^2} \lambda_1 \lambda_2 (a_1 a_2)^2
      + \frac{(\lambda_1-\lambda_3)^4}{(\lambda_1+\lambda_3)^2} \lambda_1 \lambda_3 (a_1 a_3)^2
      \\ & \quad
      + \frac{(\lambda_2-\lambda_3)^4}{(\lambda_2+\lambda_3)^2} \lambda_2 \lambda_3 (a_2 a_3)^2
      \\ & \quad
      +
      \frac{4 \lambda_1 \lambda_2 \lambda_3 a_1 a_2 a_3}
      {(\lambda_1+\lambda_2)(\lambda_1+\lambda_3)(\lambda_2+\lambda_3)}
      \times
      \\ & \qquad
      \Bigl(
        (\lambda_1-\lambda_2)^2 (\lambda_1-\lambda_3)^2 a_1
        + (\lambda_2-\lambda_1)^2 (\lambda_2-\lambda_3)^2 a_2
      \\
      & \qquad
        + (\lambda_3-\lambda_1)^2 (\lambda_3-\lambda_2)^2 a_3
      \Bigr)
      ,
      \\
      W_3 &= U_3 V_3 = \lambda_1 \lambda_2 \lambda_3 (U_3)^2
      .
    \end{split}
  \end{equation*}
  Letting $a_k = a_k(t) = a_k(0) \, e^{t/\lambda_k}$ in these expressions, the DP $3$-peakon solution
  (at least in the pure peakon case) is
  \begin{equation}
    \label{eq:DP-threepeakon}
    \begin{aligned}
      x_1(t) &= \ln\frac{U_3}{V_2}
      , &
      m_1(t) &= \frac{(U_3)^2 (V_2)^2}{W_3 W_2} = \frac{(V_2)^2}{\lambda_1 \lambda_2 \lambda_3 W_2}
      , \\
      x_2(t) &= \ln\frac{U_2}{V_1}
      , &
      m_2(t) &= \frac{(U_2)^2 (V_1)^2}{W_2 W_1}
      , \\
      x_3(t) &
      = \ln U_1
      , &
      m_3(t) &= \frac{(U_1)^2 (V_0)^2}{W_1 W_0} = \frac{(U_1)^2}{W_1}
      .
    \end{aligned}
  \end{equation}
\end{example}

\begin{remark}
  \label{rem:DP-inverse-problem-on-the-real-line}
  As in the CH case (Remark~\ref{rem:CH-inverse-problem-on-the-real-line}),
  the peakon solution formulas \eqref{eq:DP-n-peakon-solution} can be derived working directly
  with the inverse problem for \eqref{eq:DP-lax-x-again}
  on the real line, bypassing the transformation to the cubic string~\eqref{eq:DP-cubic-string};
  see Mohajer~\cite{mohajer:2017:DP-peakon-inverse-problem}.
\end{remark}

\begin{remark}
  \label{rem:DP-antipeakons}
  In the peakon--antipeakon case, which has been thoroughly studied by Szmigielski and
  Zhou~\cite{szmigielski-zhou:2013:DP-peakon-antipeakon,szmigielski-zhou:2013:DP-colliding-peakons-shock-formation},
  the eigenvalues~$\lambda_k$ need not be positive, or even real,
  nor need they be simple.
  At least for $N=3$, it is known that there can be no eigenvalues on the imaginary axis,
  and that the number of eigenvalues with negative real part equals the number of antipeakons.
  Likewise, the residues~$a_k$ can be negative or complex.
  But if the eigenvalues (as determined by initial data for $x_k$ and~$m_k$ at some time~$t_0$)
  are simple and satisfy the condition that no $\lambda_i + \lambda_j$ is zero,
  then the solution formulas~\eqref{eq:DP-n-peakon-solution}
  still make sense, and they do satisfy the peakon ODEs~\eqref{eq:intro-DP-peakon-ode},
  but only in a time interval around $t_0$ which is free of collisions.
  Let us look at the initial value problem, where we go forward in time,
  and suppose that there is a collision, with $x_k = x_{k+1}$ for some~$k$,
  at some time $t_1 > t_0$ (but not before that).
  Then, as $t \nearrow t_1$,
  the wave profile $u(x,t)$ develops a jump discontinuity at the location of the collision,
  so that it can no longer be described by the peakon ansatz~\eqref{eq:intro-peakons},
  and instead continues for $t \ge t_1$ in the form of a
  \emph{shockpeakon} solution~\cite{lundmark:2007:DP-shockpeakons}.
  See Figure~\ref{fig:DP-symmetric-collision} in the Introduction for the simplest example
  of this phenomenon.
  Multi-shockpeakon solutions have the form
  \begin{equation}
    u(x,t) = \sum_{k=1}^N \Bigl( m_k(t) - s_k(t) \, \sgn\bigl(x - x_k(t) \bigr) \Bigr) \, e^{-\abs{x - x_k(t)}}
    ,
  \end{equation}
  and are governed by a set of $3N$ ODEs for the positions~$x_k$, amplitudes~$m_k$
  and shock strengths~$s_k$.
  Explicit solutions have only been found in some very particular small cases,
  and it is not known whether those ODEs can be considered as integrable in any sense.

  The peakon trajectories
  $x = x_k(t)$ and $x = x_{k+1}(t)$
  always meet transversally at the collision~\cite[Theorem~4.7]{szmigielski-zhou:2013:DP-colliding-peakons-shock-formation},
  rather than tangentially as in the CH case.
  So at least in some time interval beyond the collision,
  the values obtained from the solution formulas~\eqref{eq:DP-n-peakon-solution}
  will be in the wrong order,
  $x_{k+1}(t) < x_k(t)$,
  and this means that they do no longer satisfy the peakon ODEs.
  An example may help to clarify this point:
  in the two-peakon case, the solution formulas
  \eqref{eq:DP-twopeakon-solution}
  still satisfy the \emph{simplified} peakon ODEs~\eqref{eq:DP-twopeakon-ode}
  also in such a time interval after the collision,
  but those ODEs are only equivalent to the \emph{actual} peakon ODEs~\eqref{eq:intro-DP-peakon-ode}
  if the ordering assumption $x_1 \le x_2$ holds, since otherwise it's not
  true that $e^{-\abs{x_1-x_2}} = e^{x_1-x_2}$.
  This fact was the cause of some puzzlement before it was realized that
  the continuation of the solution past the collision could not be
  obtained within the world of peakons, but required the concept of shockpeakons.

  If there are antiresonances $\lambda_i + \lambda_j = 0$
  (like in the antisymmetric peakon--antipeakon collision shown in Figure~\ref{fig:DP-symmetric-collision},
  for example),
  or if some eigenvalues are non-simple,
  then the solution formulas~\eqref{eq:DP-n-peakon-solution} do not apply, and must be replaced
  by modified versions.
  The most general solution formulas for DP peakon--antipeakon solutions
  have not been written down explicitly, as far as we know.
  This could be done by taking suitable limits in~\eqref{eq:DP-n-peakon-solution},
  but it is doubtful whether it would be worth the trouble;
  no really interesting new phenomena would have time to arise in these cases,
  since the solutions are only described by the modified formulas up until the time of the
  first collision anyway.
  In contrast, peakon--antipeakon solutions of the Novikov equation display a remarkable
  variety of new behaviours in the corresponding situation; see Remark~\ref{rem:Novikov-peakon-antipeakon}
  below.
\end{remark}

\section{The Novikov equation and the dual cubic string}
\label{sec:Novikov}

We now turn to the Novikov equation~\eqref{eq:intro-Novikov},
\begin{equation*}
  m_t + \bigl( (u m)_x + 2 u_x m \bigr) \, u = 0
  ,\qquad
  m = u - u_{xx}
  ,
\end{equation*}
with peakon solutions of the form~\eqref{eq:intro-peakons}
as for the CH and DP equations, but
governed by the ODEs~\eqref{eq:intro-Novikov-peakon-ode},
where in particular $\dot x_k = u(x_k)^2 \ge 0$ always,
so that antipeakons also move to the right, instead of to the left.
The results described here
(concerning pure peakon solutions)
were obtained in our paper with Hone~\cite{hone-lundmark-szmigielski:2009:novikov}.

The $3 \times 3$ matrix Lax pair given by Hone and Wang~\cite{hone-wang:2008:cubic-nonlinearity}
reads
(except for an adjustment of the matrix in \eqref{eq:Novikov-lax-t} by $(3 z^2)^{-1}$ times the identity matrix)
\begin{subequations}
  \label{eq:Novikov-lax}
  \begin{equation}
    \label{eq:Novikov-lax-x}
    \frac{\partial}{\partial x}
    \begin{pmatrix} \psi_1 \\ \psi_2 \\ \psi_3 \end{pmatrix} =
    \begin{pmatrix}
      0 & zm & 1 \\
      0 & 0 & zm \\
      1 & 0 & 0
    \end{pmatrix}
    \begin{pmatrix} \psi_1 \\ \psi_2 \\ \psi_3 \end{pmatrix}
    ,
  \end{equation}
  \begin{equation}
    \label{eq:Novikov-lax-t}
    \frac{\partial}{\partial t}
    \begin{pmatrix} \psi_1 \\ \psi_2 \\ \psi_3 \end{pmatrix} =
    \begin{pmatrix}
      -u u_x & \frac{u_x}{z}-u^2 mz & u_x^2 \\
      \frac{u}{z} & - \frac{1}{z^2} & - \frac{u_x}{z} - u^2 mz \\
      -u^2 & \frac{u}{z} & uu_x
    \end{pmatrix}
    \begin{pmatrix} \psi_1 \\ \psi_2 \\ \psi_3 \end{pmatrix}
    ,
  \end{equation}
\end{subequations}
where $z$ is the spectral parameter.
For $z \neq 0$, equation~\eqref{eq:Novikov-lax-x} is equivalent,
under the change of variables
\begin{equation}
  \label{eq:Novikov-liouville-trf}
  \begin{split}
    y &= \tanh x
    ,\\
    \phi_1(y) &= \psi_1(x) \cosh x - \psi_3(x) \sinh x
    ,\\
    \phi_2(y) &= z \, \psi_2(x)
    ,\\
    \phi_3(y) &= z^2 \, \psi_3(x) / \cosh x
    ,\\
    g(y) &= m(x) \, \cosh^3 x
    ,\\
    \lambda &= -z^2
    ,
  \end{split}
\end{equation}
to the matrix equation
\begin{subequations} \label{eq:Novikov-dual-cubic-string}
\begin{equation}
  \label{eq:Novikov-dual-cubic-string-ODE}
  \frac{\partial}{\partial y}
  \begin{pmatrix} \phi_1 \\ \phi_2 \\ \phi_3 \end{pmatrix}
  =
  \begin{pmatrix}
    0 & g(y) & 0 \\
    0 & 0 & g(y) \\
    -\lambda & 0 & 0
  \end{pmatrix}
  \begin{pmatrix} \phi_1 \\ \phi_2 \\ \phi_3 \end{pmatrix}
\end{equation}
on the interval $-1<y<1$.
(As usual we consider a fixed $t$, for the moment, and don't write out the time-dependence.)
The boundary conditions relevant to peakon solutions turn out to be
\begin{equation}
  \label{eq:Novikov-dual-cubic-string-boundary-conditions-y}
  \phi_2(-1) = \phi_3(-1) = 0
  ,
  \qquad
  \phi_3(1) = 0
  .
\end{equation}
\end{subequations}
Note the resemblance to the matrix form~\eqref{eq:DP-cubic-string-matrix}
of the cubic string equation $\phi_{yyy} = -\lambda g \phi$.
This is more than a superficial similarity,
and we will refer to the eigenvalue problem \eqref{eq:Novikov-dual-cubic-string}
as the \emph{dual cubic string}, for the following reason:
for continuous mass distributions $g(y) > 0$,
\eqref{eq:DP-cubic-string-matrix-ODE}
and
\eqref{eq:Novikov-dual-cubic-string-ODE}
are related via the change of variables defined by the differential equation
\begin{equation}
  \label{eq:duality-transformation}
  \frac{d\tilde{y}}{dy} = g(y) = \frac{1}{\tilde{g}(\tilde{y})},
\end{equation}
where $y$ and $g(y)$ refer to
\eqref{eq:DP-cubic-string-matrix-ODE}
and $\tilde{y}$ and $\tilde{g}(\tilde{y})$ to
\eqref{eq:Novikov-dual-cubic-string-ODE}
-- or the other way around!

This duality manifests itself in a very striking way in the discrete case,
where the measure
\begin{equation*}
  m(x) = 2 \sum_1^N m_k \, \delta(x-x_k)
\end{equation*}
is mapped by the Liouville transformation~\eqref{eq:Novikov-liouville-trf} to
\begin{equation}
  \begin{gathered}
    g(y) = \sum_1^N g_k \, \delta(y-y_k)
    , \\
    y_k = \tanh x_k
    ,\quad
    g_k = 2 m_k \cosh x_k
    .
  \end{gathered}
\end{equation}
In order to study this discrete case of the dual cubic
string~\eqref{eq:Novikov-dual-cubic-string} as a shooting problem,
let $\Phi(y;\lambda)$ be the solution starting with $\Phi(-1;\lambda) = (1,0,0)^T$
at the left endpoint $y=-1$,
and successively extend it throughout the interval $[-1,1]$ as dictated
by~\eqref{eq:Novikov-dual-cubic-string-ODE}.
The component $\phi_3(y;\lambda)$ is continuous and piecewise linear,
while $\phi_1(y;\lambda)$ and $\phi_2(y;\lambda)$ are piecewise constant with jumps at the points~$y_k$;
more precisely, the jump condition at $y_k$ is
\begin{equation}
  \label{eq:Novikov-Gk}
  \Phi(y_k^+;\lambda) =
  \begin{pmatrix}
    1 & g_k & \frac12 g_k^2 \\ 0 & 1 & g_k \\ 0 & 0 & 1
  \end{pmatrix}
  \Phi(y_k^-;\lambda)
\end{equation}
(if we interpret the product $\phi_2(y) \, \delta(y-y_k)$ as
$\langle \phi_2(y_k) \rangle \, \delta(y-y_k)$,
which is the correct choice to preserve Lax integrability),
and the passage from $y_k$ to $y_{k+1}$ is described by
\begin{equation}
  \label{eq:Novikov-Lk}
  \Phi(y_{k+1}^-;\lambda) =
  \begin{pmatrix}
    1 & 0 & 0 \\ 0 & 1 & 0 \\ -\lambda l_k & 0 & 1
  \end{pmatrix}
  \Phi(y_k^+;\lambda)
  ,
\end{equation}
where $l_k = y_{k+1} - y_k$ as usual.
Here we see matrices of exactly the same form as $L_k$ and~$G_k(\lambda)$
from \eqref{eq:DP-Lk} and~\eqref{eq:DP-Gk},
except that the roles of the distances~$l_k$ and the masses $g_k$
have been reversed!

Actually, to get a perfect duality
which also includes
the boundary conditions~\eqref{eq:Novikov-dual-cubic-string-boundary-conditions-y}
for the dual cubic string,
we should not use the Dirichlet-like boundary conditions
\eqref{eq:DP-cubic-string-matrix-boundary-conditions-y}
for the ``primal'' cubic string,
but instead consider Neumann-like boundary conditions,
say $\phi_{yyy} = -\lambda g \phi$ for $y\in\R$,
with $\phi_y(-\infty) = \phi_{yy}(-\infty) = 0$ and
$\phi_{yy}(\infty) = 0$.
In the discrete case,
if  the primal Neumann-like cubic string has
point masses at $y_0 < y_1 < \dots < y_N$,
then its $N+1$ weights~$g_k$ and its $N$ finite lengths $l_k = y_{k+1}-y_k$
correspond to the $N+1$ lengths~$l_k$ and $N$ weights~$g_k$ for the discrete dual cubic string
on $[-1,1]$ with the boundary
conditions~\eqref{eq:Novikov-dual-cubic-string-boundary-conditions-y}.

The eigenvalues of~\eqref{eq:Novikov-dual-cubic-string} are the roots
of $\phi_3(1;\lambda)$, which is a polynomial in~$\lambda$ of degree~$N+1$,
with zero constant term. The root $\lambda_0=0$ can be said to be an artifact
introduced by the Liouville transformation, and only the nonzero eigenvalues
are of interest to the inverse problem.
The suitable Weyl functions turn out to be
\begin{equation}
  W(\lambda)
  = - \frac{\phi_2(1;\lambda)}{\phi_3(1;\lambda)}
  = \sum_{k=1}^N \frac{a_k}{\lambda-\lambda_k}
  ,
\end{equation}
where a common factor of~$\lambda$ in $\phi_2(1;\lambda)$ and $\phi_3(1;\lambda)$ cancels,
and
\begin{equation}
  Z(\lambda)
  = - \frac{\phi_1(1;\lambda)}{\phi_3(1;\lambda)}
  = \frac{1/2}{\lambda} + \sum_{k=1}^N \frac{b_k}{\lambda-\lambda_k}
  .
\end{equation}
Provided that all weights $g_k$ are \emph{positive} (the pure peakon case),
the eigenvalues $\lambda_k$ are positive and simple,
and the residues $a_k$ and~$b_k$ are positive (which is the reason for including
a minus sign in the definitions of $W$ and~$Z$).
Note also that both Weyl functions already are of order $O(1/\lambda)$ as $\lambda \to \infty$,
so there is no need to divide by~$\lambda$ as we have done in the CH and DP cases.
They satisfy an identity similar to~\eqref{eq:DP-Z-W-symmetry-relation}, namely
\begin{equation}
  \label{eq:Novikov-Z-W-symmetry-relation}
  Z(\lambda) + W(-\lambda) \, W(\lambda) + Z(-\lambda) = 0,
\end{equation}
which determines
\begin{equation}
  \label{eq:Novikov-bk}
  b_k = a_k \sum_{j=1}^N \frac{a_j}{\lambda_j+\lambda_k}
  \quad (1 \le k \le N)
  .
\end{equation}
Hence, with the spectral measure
\begin{equation}
  \label{eq:Novikov-measure-alpha}
  \alpha(\lambda)
  = \sum_{k=1}^N a_k \, \delta(\lambda-\lambda_k)
  ,
\end{equation}
we can write
\begin{equation}
  \label{eq:Novikov-W-stieltjes-transform}
  W(\lambda)
  = \int \frac{d\alpha(z)}{\lambda-z}
\end{equation}
and
\begin{equation}
  \label{eq:Novikov-Z-stieltjes-transform}
  \begin{split}
    Z(\lambda)
    &
    = \frac{1/2}{\lambda} + \sum_{k=1}^N \frac{b_k}{\lambda - \lambda_k}
    \\ &
    = \frac{1/2}{\lambda} + \sum_{k=1}^N \sum_{j=1}^N \frac{a_k a_j}{(\lambda_j + \lambda_k) (\lambda - \lambda_k)}
    \\ &
    = \frac{1/2}{\lambda} + \iint \frac{d\alpha(z_1) \, d\alpha(z_2)}{(z_1+z_2)(\lambda-z_1)}
    ,
  \end{split}
\end{equation}
so that we are in the CBOP setup with coinciding measures $\beta=\alpha$;
see Remark~\ref{rem:DP-CBOP-discrete}.

The time evolution of the spectral data induced by the Novi\-kov
peakon ODEs~\eqref{eq:intro-Novikov-peakon-ode},
via the second Lax equation~\eqref{eq:Novikov-lax-t},
is the usual one: $a_k(t) = a_k(0) \, e^{t / \lambda_k}$,
with $\lambda_k$ time-independent.

The inverse spectral problem for the Neumann-like cubic string had been
solved~\cite{kohlenberg-lundmark-szmigielski:2007:inverse-spectral-discrete-cubic-string}
before Novikov's equation was even discovered,
so the hard work was already done,
and those results together with the duality quickly provide the
solution of the inverse spectral problem for the dual cubic string as well,
and hence the explicit solution formulas
for the Novikov peakon ODEs~\eqref{eq:intro-Novikov-peakon-ode} in the pure peakon case:
\begin{equation}
  \label{eq:Novikov-n-peakon-solution}
  x_{N+1-k}(t) = \frac12 \ln\frac{Z_k}{W_{k-1}}
  , \qquad
  m_{N+1-k}(t) = \frac{\sqrt{Z_k W_{k-1}}}{U_k U_{k-1}}
\end{equation}
for $1 \le k \le N$,
where $U_k$ and $W_k$ are as in \eqref{eq:DP-def-Uk} and~\eqref{eq:DP-def-Wk},
while $Z_k$ is obtained from~$W_k$ by replacing every $a_i$ with $a_i/\lambda_i$.

\begin{example}[The two-peakon solution]
  \label{ex:Novikov-twopeakon}
  When $N=2$, the Novikov peakon ODEs~\eqref{eq:intro-Novikov-peakon-ode} take the form
  \begin{equation}
    \label{eq:Novikov-twopeakon-ode}
    \begin{aligned}
      \dot x_1 &= \bigl( m_1 + m_2 \, e^{x_1-x_2} \bigr)^2
      , \\
      \dot x_2 &= \bigl( m_1 \, e^{x_1-x_2} + m_2 \bigr)^2
      , \\
      \dot m_1 &= -m_1 m_2 \, e^{x_1-x_2} \, \bigl( m_1 + m_2 \, e^{x_1-x_2} \bigr)
      , \\
      \dot m_2 &= m_1 m_2 \, e^{x_1-x_2} \, \bigl( m_1 \, e^{x_1-x_2} + m_2 \bigr)
      .
    \end{aligned}
  \end{equation}
  Already this case is sufficiently complicated for direct integration
  to be quite a challenge.
  Hone and Wang~\cite{hone-wang:2008:cubic-nonlinearity} wrote down explicit expressions
  for $x_2 - x_1$, $m_2^2-m_1^2$ and $m_1 m_2$,
  but left an unevaluated antiderivative in their ``somewhat more formidable'' formula for $x_1+x_2$,
  merely indicating how it could be explicitly calculated in principle.
  But the general formulas~\eqref{eq:Novikov-n-peakon-solution} provide a completely explicit solution:
  \begin{subequations} \label{eq:Novikov-twopeakon}
  \begin{equation}
    \begin{split}
      x_1(t)
      &
      = \frac12 \ln\frac{Z_2}{W_1}
      = \frac12 \ln \dfrac{ \frac{(\lambda_1-\lambda_2)^4}{(\lambda_1+\lambda_2)^2 \lambda_1 \lambda_2} \, a_1^2 a_2^2}{\lambda_1 \, a_1^2 + \lambda_2 \, a_2^2 + \frac{4 \, \lambda_1 \lambda_2}{\lambda_1+\lambda_2} \, a_1 a_2}
      , \\
      x_2(t)
      &
      = \frac12 \ln\frac{Z_1}{W_0}
      = \frac12 \ln \left( \frac{a_1^2}{\lambda_1} + \frac{a_2^2}{\lambda_2} + \frac{4}{\lambda_1+\lambda_2} \, a_1 a_2 \right)
      ,
    \end{split}
  \end{equation}
  and
  \begin{equation}
    \begin{split}
      m_1(t)
      &
      = \dfrac{\sqrt{Z_2 W_1}}{U_2 U_1}
      \\ &
      = \dfrac{ \left[ \frac{(\lambda_1 - \lambda_2)^4 \, a_1^2 a_2^2}{(\lambda_1 + \lambda_2)^2 \lambda_1 \lambda_2} \left( \lambda_1 \, a_1^2 + \lambda_2 \, a_2^2 + \frac{4 \, \lambda_1 \lambda_2}{\lambda_1+\lambda_2} \, a_1 a_2 \right) \right]^{1/2}}{\frac{(\lambda_1 - \lambda_2)^2 \, a_1 a_2}{\lambda_1 + \lambda_2} \, (a_1+a_2)}
      , \\
      m_2(t)
      &
      = \dfrac{\sqrt{Z_1 W_0}}{U_1 U_0}
      = \dfrac{\left( \frac{a_1^2}{\lambda_1} + \frac{a_2^2}{\lambda_2} + \frac{4}{\lambda_1+\lambda_2} \, a_1 a_2 \right)^{1/2}}{a_1+a_2}
      ,
    \end{split}
  \end{equation}
  \end{subequations}
  where $a_k = a_k(t) = a_k(0) \, e^{t / \lambda_k}$.
  In the pure peakon case, where $\lambda_1$, $\lambda_2$, $a_1$ and~$a_2$ are positive,
  the expression for~$m_1$ can be simplified to
  \begin{equation*}
    m_1(t) = \frac{ \left( \lambda_1 \, a_1^2 + \lambda_2 \, a_2^2 + \frac{4 \, \lambda_1 \lambda_2}{\lambda_1+\lambda_2} \, a_1 a_2 \right)^{1/2}}{\sqrt{\lambda_1 \lambda_2} \, (a_1+a_2)}
    ,
  \end{equation*}
  but the general formula above is needed in order to describe peakon--antipeakon solutions.
\end{example}

\begin{remark}
  \label{rem:Novikov-inverse-problem-on-the-real-line}
  For derivations of the peakon solution formulas \eqref{eq:Novikov-twopeakon} directly on
  the real line (cf. Remarks \ref{rem:CH-inverse-problem-on-the-real-line}
  and~\ref{rem:DP-inverse-problem-on-the-real-line}),
  see Mohajer and Szmigielski~\cite{mohajer-szmigielski:2012:novikov-on-real-axis},
  and also the recent elegant approach by Chang~\cite{chang:2022:hermite-pade-pfaffian-structures-novikov-and-integrable-lattices},
  which emphasizes the role of Pfaffians (rather than determinants) in this context,
  building on earlier work by Chang, Hu, Li and Zhao~\cite{chang-hu-li-zhao:2018:novikov-peakons-pfaffians-Toda-lattice-BKP}.
\end{remark}

\begin{remark}
  \label{rem:Novikov-peakon-antipeakon}
  The peakon--antipeakon case has been studied by Kardell and Lundmark.
  A preliminary version of this work formed part of Kardell's Ph.D. thesis~\cite{kardell:2016:phdthesis};
  the final version is still under preparation.
  Like for the DP equation (Remark~\ref{rem:DP-antipeakons}),
  the spectrum may now be complex and non-simple.
  The eigenvalues~$\lambda_k$ must have
  positive real part in the case $N=2$,
  and nonnegative real part if $N \ge 3$.
  If the eigenvalues are simple and have positive real part,
  then the solution formulas~\eqref{eq:Novikov-n-peakon-solution}
  still make sense, and they do satisfy the peakon ODEs
  and preserve the ordering $x_1 < \dots < x_N$,
  except at isolated instants $t=t_c$ where some $U_{N-k}(t)$ vanishes,
  causing a collision $x_{k}(t_c) = x_{k+1}(t_c)$ where $m_k(t)$ and~$m_{k+1}(t)$ are undefined.
  However, as for the CH equation,
  the wave profile $u(x,t)$ extends continuously to these times,
  and this provides a globally defined (conservative) peakon solution.
  The order of contact of the colliding trajectories is higher than in the CH case, since
  \begin{equation*}
    e^{2 x_{k+1}(t)} - e^{2 x_k(t)} = \frac{U_{N-k}(t)^4}{W_{N-k-1}(t) \, W_{N-k}(t)}
    ,
  \end{equation*}
  where the denominators can be shown to be positive;
  thus, in the typical case where $U_{N-k}(t)$ has a simple zero at $t=t_c$,
  the distance $x_{k+1}(t) - x_k(t)$ will have a zero of multiplicity~$4$,
  but it is also possible to have higher multiples of~$4$.
  With complex eigenvalues, a group of $n$ eigenvalues $\lambda_k$
  such that all $1/\lambda_k$ share the same real part
  will give rise to a cluster of $n$ peakons travelling together, performing
  an intricate dance among themselves, and interacting with other peakons (or peakon clusters).
  These clusters are somewhat reminiscent of the ``breather'' soliton solutions
  occurring in some other integrable PDEs, like the sine-Gordon equation.
  Merely describing the precise \emph{asymptotics} as $t \to \pm \infty$
  of such an $n$-peakon cluster (as part of an $N$-peakon solution with $n<N$)
  requires the \emph{exact} solution formulas for the $n$-peakon ODEs.
  If some eigenvalues are non-simple or lie on the imaginary axis,
  the solution is described by considerably more complicated formulas,
  obtained from~\eqref{eq:Novikov-n-peakon-solution} by taking suitable limits,
  and in these cases there can be peakons (or peakon clusters)
  with the same limiting velocity but still separating
  at a logarithmic rate as $t \to \pm \infty$,
  or ``asymptotic peakon--antipeakon collisions'' where $x_k(t)$ and $x_{k+1}(t)$ tend to the
  same constant value as $t \to \infty$ or $t \to -\infty$.
  A few examples of this very rich world of possible behaviours were
  shown in Figures
  \ref{fig:novikov-4cluster+single-3d},
  \ref{fig:novikov-4cluster+single},
  \ref{fig:novikov-4cluster-quasiperiodic}
  and~\ref{fig:novikov-conjugate-triple-roots}
  in the Introduction.
\end{remark}

\begin{remark}
  \label{rem:Novikov-Himonas-Holliman-Kenig}
  Himonas, Holliman and Kenig~\cite{himonas-holliman-kenig:2018:novikov-2-peakon-illposedness}
  were able to use estimates to (among other things) prove directly from
  the Novikov two-peakon ODEs~\eqref{eq:Novikov-twopeakon-ode}
  that peakon--antipeakon collisions can actually occur,
  despite the fact that the peakon and the antipeakon both move to the right.
  Apparently they were unaware of the existence of an exact formula
  for peakon--antipeakon solutions
  which may have made their life easier.
  However, it must be emphasized that their methods also
  apply to peakon--antipeakon solutions on the circle (i.e., periodic with respect to~$x$),
  for which exact solution formulas are currently \emph{not} known.
\end{remark}

\section{The Geng--Xue equation and its twin Lax pairs}
\label{sec:GX}

The Geng--Xue equation~\eqref{eq:intro-GX},
\begin{equation}
  \label{eq:GX-again}
  \begin{gathered}
    m_t + \bigl( (u m)_x + 2 u_x m \bigr) \, v = 0
    , \\
    n_t + \bigl( (v n)_x + 2 v_x n \bigr) \, u = 0
    , \\
    m = u - u_{xx}
    ,\quad
    n = v - v_{xx}
    ,
  \end{gathered}
\end{equation}
was obtained by Geng and Xue~\cite{geng-xue:2009:GX-peakon-equation-cubic-nonlinearity}
as the compatibility condition of the Lax pair
\begin{subequations}
  \label{eq:GX-laxI}
  \begin{equation}
    \label{eq:GX-laxI-x}
    \frac{\partial}{\partial x}
    \begin{pmatrix} \psi_1 \\ \psi_2 \\ \psi_3 \end{pmatrix} =
    \begin{pmatrix}
      0 & zn & 1 \\
      0 & 0 & zm \\
      1 & 0 & 0
    \end{pmatrix}
    \begin{pmatrix} \psi_1 \\ \psi_2 \\ \psi_3 \end{pmatrix}
    ,
  \end{equation}
  \begin{equation}
    \label{eq:GX-laxI-t}
    \frac{\partial}{\partial t}
    \begin{pmatrix} \psi_1 \\ \psi_2 \\ \psi_3 \end{pmatrix} =
    \begin{pmatrix}
      -v_xu & \frac{v_x}{z}-vunz & v_xu_x \\
      \frac{u}{z} & v_xu - vu_x - \frac{1}{z^2} & - \frac{u_x}{z} - vumz \\
      -vu & \frac{v}{z} & vu_x
    \end{pmatrix}
    \begin{pmatrix} \psi_1 \\ \psi_2 \\ \psi_3 \end{pmatrix}
    ,
  \end{equation}
\end{subequations}
which clearly reduces to the Lax pair~\eqref{eq:Novikov-lax} for Novikov's equation
if $u=v$ (and hence $m=n$).
But because of the symmetry, it also arises as the compatibility condition of
another Lax pair, with $u$ and~$v$ (and hence $m$ and~$n$) interchanged:
\begin{subequations}
  \label{eq:GX-laxII}
  \begin{equation}
    \label{eq:GX-laxII-x}
    \frac{\partial}{\partial x}
    \begin{pmatrix} \twin\psi_1 \\ \twin\psi_2 \\ \twin\psi_3 \end{pmatrix} =
    \begin{pmatrix}
      0 & zm & 1 \\
      0 & 0 & zn \\
      1 & 0 & 0
    \end{pmatrix}
    \begin{pmatrix} \twin\psi_1 \\ \twin\psi_2 \\ \twin\psi_3 \end{pmatrix}
    ,
  \end{equation}
  \begin{equation}
    \label{eq:GX-laxII-t}
    \frac{\partial}{\partial t}
    \begin{pmatrix} \twin\psi_1 \\ \twin\psi_2 \\ \twin\psi_3 \end{pmatrix} =
    \begin{pmatrix}
      -u_xv & \frac{u_x}{z}-uvmz & u_xv_x \\
      \frac{v}{z} & u_xv - uv_x - \frac{1}{z^2} & - \frac{v_x}{z} - uvnz \\
      -uv & \frac{u}{z} & uv_x
    \end{pmatrix}
    \begin{pmatrix} \twin\psi_1 \\ \twin\psi_2 \\ \twin\psi_3 \end{pmatrix}
    .
  \end{equation}
\end{subequations}
This remark may seem pointless at first, but is in fact crucial, since if $u \neq v$ we
will get different spectral data from the two Lax pairs, and we need to combine these data
in order to solve the inverse spectral problem.

The peakon solutions take the form
\begin{equation}
  \label{eq:GX-peakons}
  \begin{split}
    u(x,t) &= \sum_{k=1}^N m_k(t) \, e^{-\abs{x - x_k(t)}}
    , \\
    v(x,t) &= \sum_{k=1}^N n_k(t) \, e^{-\abs{x - x_k(t)}}
    ,
  \end{split}
\end{equation}
where we impose the restriction that for each~$k$ exactly one of $m_k(t)$ and $n_k(t)$
is identically zero, so that the peakons are \emph{non-overlapping}.
In other words, the peakons in the first component~$u$ are located at \emph{different} sites
than the peakons in the second component~$v$.
The reason for this restriction is that it is difficult to make sense of the PDEs~\eqref{eq:intro-GX}
if peakons are allowed to overlap; we are not aware of any definition of weak or distributional solutions
that manages to avoid the serious problems of undefined products arising in that case
(see Section~\ref{sec:guide-GX}).
It may still be possible to obtain integrable ODEs from the Lax pairs
even in the overlapping case, but we will leave that question for future research.
Anyway, one can make sense of non-overlapping peakons as distributional solutions to~\eqref{eq:intro-GX},
and they are governed by the ODEs
\begin{equation}
  \label{eq:GX-peakon-ode}
  \begin{split}
    \dot x_k &= u(x_k) \, v(x_k)
    ,\\
    \dot m_k &= m_k \bigl( u(x_k) \, v_x(x_k) - 2 \, u_x(x_k) \, v(x_k) \bigr)
    ,\\
    \dot n_k &= n_k \bigl( u_x(x_k) \, v(x_k) - 2 \, u(x_k) \, v_x(x_k) \bigr)
    ,
  \end{split}
\end{equation}
for $1 \le k \le N$.
So there are $N$ peakons in total, and they may be distributed among the two components
$u$ and~$v$ in any non-overlapping way, with $N_1$ peakons in~$u$ and $N_2 = N-N_1$ in~$v$.
We will only deal with pure peakon solutions here, since they are sufficiently complicated already,
and since peakon--antipeakon collisions lead to shockpeakon formation (as for the DP equation),
so that we would need to leave the world of peakons.

For no particular reason other than to start somewhere,
we chose to begin our study of these
ODEs~\cite{lundmark-szmigielski:2016:GX-inverse-problem,lundmark-szmigielski:2017:GX-dynamics-interlacing}
with the \emph{interlacing} case, where $N=2K$ is even
and there are $K$ peakons occurring alternatingly
in the two components:
first a peakon in~$u$ at~$x_1$ with amplitude $m_1 > 0$,
then one in~$v$ at~$x_2$ with amplitude $n_2 > 0$, then one in $u$~again, then in~$v$, and so on.
(Starting with $u$ entails no loss of generality, since the equations are symmetric with
respect to swapping $u$ and~$v$.)
As we shall see, this turned out to be a stroke of luck,
since it is only in this case (and in the odd interlacing case with $N=2K+1$ peakons)
that the two Lax pairs provide a sufficient amount of
spectral data for the inverse spectral problem to be uniquely solvable.
So we have two discrete positive measures:
\begin{equation*}
  m(x) = 2 \sum_{i=1}^K m_{2i-1} \, \delta(x - x_{2i-1})
  ,
\end{equation*}
supported at the odd-numbered sites, and
\begin{equation*}
  n(x) = 2 \sum_{i=1}^K n_{2i} \, \delta(x - x_{2i})
  ,
\end{equation*}
supported at the even-numbered sites.
As before, we begin with a Liouville transformation,
similar to~\eqref{eq:Novikov-liouville-trf}
that we used for Novikov's equation:
\begin{equation}
  \label{eq:GX-liouville-trf}
  \begin{split}
    y &= \tanh x,\\
    \phi_1(y) &= \psi_1(x) \cosh x - \psi_3(x) \sinh x,\\
    \phi_2(y) &= z \, \psi_2(x),\\
    \phi_3(y) &= z^2 \, \psi_3(x) / \cosh x,\\
    g(y) &= m(x) \, \cosh^3 x,\\
    h(y) &= n(x) \, \cosh^3 x,\\
    \lambda &= -z^2.
  \end{split}
\end{equation}
Under this transformation (with $z\neq 0$),
equation \eqref{eq:GX-laxI-x} is equivalent to
\begin{subequations}
\label{eq:GX-spectral-problem-y}
\begin{equation}
  \label{eq:GX-dual-cubic-I}
  \frac{\partial}{\partial y}
  \begin{pmatrix} \phi_1 \\ \phi_2 \\ \phi_3 \end{pmatrix}
  =
  \begin{pmatrix}
    0 & h(y) & 0 \\
    0 & 0 & g(y) \\
    -\lambda & 0 & 0
  \end{pmatrix}
  \begin{pmatrix} \phi_1 \\ \phi_2 \\ \phi_3 \end{pmatrix}
\end{equation}
for $-1<y<1$, and the boundary conditions relevant for peakon solutions turn out to be
the same as in the Novikov case,
\begin{equation}
  \label{eq:GX-dual-cubic-boundary-conditions-y}
  \phi_2(-1) = \phi_3(-1) = 0
  ,
  \qquad
  \phi_3(1) = 0
  .
\end{equation}
\end{subequations}
The measures $m(x)$ and~$n(x)$ are transformed into
\begin{equation}
  \label{eq:GX-gh}
  g(y) = \sum_{i=1}^{K} g_{2i-1} \, \delta(y - y_{2i-1})
  , \quad
  h(y) = \sum_{i=1}^{K} h_{2i} \, \delta(y - y_{2i})
  ,
\end{equation}
where
\begin{equation}
  \label{eq:GX-trf-discrete-measures}
  g_{2i-1} = 2 m_{2i-1} \cosh x_{2i-1}
  , \quad
  h_{2i} = 2 n_{2i} \cosh x_{2i}
  .
\end{equation}
The twin Lax equation~\eqref{eq:GX-laxII-x} is of course transformed into
an equation of the same form as~\eqref{eq:GX-dual-cubic-I} but with $g(y)$ and $h(y)$ swapped,
and we impose the same boundary conditions~\eqref{eq:GX-dual-cubic-boundary-conditions-y}
in that case.

Consider first the eigenvalue problem~\eqref{eq:GX-spectral-problem-y}.
We let $\Phi(y;\lambda)$ be the solution to \eqref{eq:GX-dual-cubic-I}
starting with
$\Phi = (\phi_1,\phi_2,\phi_3)^T = (1,0,0)^T$ at the left endpoint $y=-1$,
and successively compute its values as we move to the right in the interval $[-1,1]$.
As in the Novikov case (see~\eqref{eq:Novikov-Lk}),
the passage from $y_k$ to $y_{k+1}$ is described by
\begin{equation}
  \Phi(y_{k+1}^-;\lambda) =
  \begin{pmatrix}
    1 & 0 & 0 \\ 0 & 1 & 0 \\ -\lambda l_k & 0 & 1
  \end{pmatrix}
  \Phi(y_k^+;\lambda)
\end{equation}
where $l_k = y_{k+1} - y_k$.
But instead of \eqref{eq:Novikov-Gk},
we find at~$y_k$ the jump condition
\begin{equation}
  \label{eq:GX-Gi}
  \Phi(y_k^+;\lambda) =
  \begin{pmatrix}
    1 & 0 & 0 \\ 0 & 1 & g_{2i-1} \\ 0 & 0 & 1
  \end{pmatrix}
  \Phi(y_k^-;\lambda)
\end{equation}
if $k=2i-1$ is odd, and
\begin{equation}
  \label{eq:GX-Hi}
  \Phi(y_k^+;\lambda) =
  \begin{pmatrix}
    1 & h_{2i} & 0 \\ 0 & 1 & 0 \\ 0 & 0 & 1
  \end{pmatrix}
  \Phi(y_k^-;\lambda)
\end{equation}
if $k=2i$ is even.
As a consequence of this, the degree of $\Phi(y;\lambda)$ as a polynomial in~$\lambda$
will only increase about half as quickly as in the Novikov case, as we move to the right.
The spectrum is defined by the roots of $\phi_3(1;\lambda)$, which is
a polynomial of degree~$K+1$ (rather than~$2K$) with zero constant term,
so together with the root $\lambda_0=0$
we get $K$ nonzero eigenvalues $\{ \lambda_i \}_{i=1}^K$ from this first spectral problem.
The polynomial $\phi_2(1;\lambda)$ has degree $K$ and zero constant term,
so the Weyl function
$W(\lambda) = - \phi_2(1;\lambda) / \phi_3(1;\lambda)$
has a partial fraction expansion of the form
\begin{equation}
  W(\lambda)
  = - \frac{\phi_2(1;\lambda)}{\phi_3(1;\lambda)}
  = \sum_{i=1}^K \frac{a_i}{\lambda-\lambda_i}
\end{equation}
defining the $K$ residues $\{ a_i \}_{i=1}^K$.
In the pure peakon case, where all $g_{2i-1}$ and~$h_{2i}$ are positive,
it can be shown that the nonzero eigenvalues $\lambda_i$ are positive and simple,
and the residues~$a_i$ are positive.

For the twin spectral problem, with $g(y)$ and $h(y)$ swapped,
things will be similar, except that $g_{2i-1}$ will be moved to the $(1,2)$ position in~\eqref{eq:GX-Gi},
and $h_{2i}$ to the $(2,3)$ position in~\eqref{eq:GX-Hi}.
This has the effect that the first weight~$g_1$ disappears from the calculation entirely,
and the polynomial degrees will be slightly lower.
Denoting the solution in this case by $\widetilde \Phi(y;\lambda)$,
both $\widetilde \phi_2(1;\lambda)$ and $\widetilde \phi_3(1;\lambda)$
are of degree~$K$ with zero constant term,
so we get only $K-1$ nonzero eigenvalues $\{ \mu_j \}_{j=1}^{K-1}$ (positive and simple),
and the Weyl function takes the form
\begin{equation}
  \widetilde W(\lambda)
  = - \frac{\widetilde \phi_2(1;\lambda)}{\widetilde \phi_3(1;\lambda)}
  = - b_\infty + \sum_{j=1}^{K-1} \frac{b_j}{\lambda-\mu_j}
  ,
\end{equation}
defining a (positive) parameter~$b_\infty$
together with the $K-1$ (positive) residues $\{ b_j \}_{j=1}^{K-1}$.

In summary, the two spectral problems provide us with $4K-1$ numbers,
namely
$\{ \lambda_i, a_i \}_{i=1}^K$,
$\{ \mu_j, b_j \}_{j=1}^{K-1}$
and~$b_\infty$.
This doesn't quite match the number of original parameters,
namely the $4K$ quantities $\{ g_{2i-1}, h_{2i} \}_{i=1}^K$ and $\{ y_k \}_{k=1}^{2K}$.
The missing piece of the puzzle is provided by an additional (positive) parameter~$b_\infty^*$
coming from an adjoint spectral problem in a way that we will not describe here;
it is given by the equality
\begin{equation}
  \label{eq:GX-b-infty-b-infty-star-product}
  b_{\infty} b^*_{\infty}
  = \frac{l_1 l_3 \dotsm l_{2K-1}}{l_0 l_2 l_4 \dotsm l_{2K}}
  \times \biggl( \prod_{j=1}^{K-1} \mu_j \biggr) \biggm/ \biggl( \prod_{i=1}^{K} \lambda_i \biggr)
  .
\end{equation}
Including $b_\infty^*$, the number of parameters in the spectral data is $4K$ as well,
and if we impose the ordering conditions $\lambda_1 < \dots < \lambda_K$
and $\mu_1 < \dots < \mu_{K-1}$,
it turns out (after lots of technical work) that
there is a one-to-one correspondence between interlacing positive discrete
measures $g(y)$ and $h(y)$ as in~\eqref{eq:GX-gh}
and spectral data of this kind.
In other words, the inverse spectral problem is uniquely solvable,
and moreover the solution is given explicitly by formulas involving determinants
containing Cauchy bimoments
\begin{equation}
  I_{ij} = \iint \frac{z_1^i z_2^j}{z_1+z_2} \, d\alpha(z_1) \, d\beta(z_2)
\end{equation}
of the two independent spectral measures
\begin{equation}
  \label{eq:spectral-measures}
  \alpha(z) = \sum_{i=1}^K a_i \, \delta(z-\lambda_i)
  , \quad
  \beta(z) = \sum_{j=1}^{K-1} b_j \, \delta(z-\mu_j)
  .
\end{equation}

\begin{remark}
  We see here that the GX twin spectral problems fit into the framework of
  Cauchy biorthogonal polynomials in its more general form
  with two measures; see Remark~\ref{rem:DP-CBOP-discrete}.
\end{remark}

The time-dependence of the spectral data induced by the GX peakon ODEs~\eqref{eq:GX-peakon-ode}
is $a_i(t) = a_i(0) \, e^{t/\lambda_i}$
and $b_j(t) = b_j(0) \, e^{t/\mu_j}$,
with $\lambda_i$, $\mu_j$, $b_\infty$ and $b_\infty^*$ time-independent,
and mapping the solution of the inverse spectral problem back to the real line
gives the general interlacing pure $2K$-peakon solution of the GX equation.
We refer to the original works~\cite{lundmark-szmigielski:2016:GX-inverse-problem,lundmark-szmigielski:2017:GX-dynamics-interlacing}
for the solution formulas (since stating them would require quite a lot of additional notation),
as well as examples with graphics.
Let us just briefly describe the asymptotics of the $K+K$ interlacing pure peakon solutions
as $t \to \pm \infty$ (where $K \ge 2$, since the case $K=1$ is exceptional and somewhat trivial).
With the eigenvalues numbered in increasing order $0 < \lambda_1 < \ldots < \lambda_K$
and $0 < \mu_1 < \dots < \mu_{K-1}$, define $2K-1$ positive numbers
$c_1 > \dots > c_{2K-1}$ by
\begin{equation}
  \label{eq:GX-asymptotic-velocities}
  \begin{aligned}
    c_{2j}
    &=
    \frac12 \left( \frac{1}{\lambda_{j+1}} + \frac{1}{\mu_{j}} \right)
    ,\quad
    j = 1, \dots, K-1
    ,\\[1.5ex]
    c_{2j-1}
    &=
    \begin{cases}
      \displaystyle
      \frac12 \left( \frac{1}{\lambda_{j}} + \frac{1}{\mu_{j}} \right)
      ,&
      j = 1, \dots, K-1
      ,\\[2ex] \displaystyle
      \frac12 \left( \frac{1}{\lambda_{K}} \right)
      ,&
      j = K
      .
    \end{cases}
  \end{aligned}
\end{equation}
Then the peakons asymptotically travel in straight lines, with the limiting velocities
\begin{equation}
  \begin{split}
    &
    (\dot x_1, \dot x_2, \dot x_3, \dots, \dot x_{2K-1}, \dot x_{2K})
    \\
    \sim
    &
    (c_1, c_1, c_2, \dots, c_{2K-2}, c_{2K-1})
    , \quad
    t \to -\infty
    ,
  \end{split}
\end{equation}
and with the same velocities in the opposite order after all interactions have taken place,
\begin{equation}
  \begin{split}
    &
    (\dot x_1, \dot x_2, \dots, \dot x_{2K-2}, \dot x_{2K-1}, \dot x_{2K})
    \\
    \sim
    &
    (c_{2K-1}, c_{2K-2}, \dots, c_2, c_1, c_1)
    , \quad
    t \to \infty
    .
  \end{split}
\end{equation}
Note that the two fastest peakons travel in parallel lines with the same velocity~$c_1$.

Unlike what we have seen for the other PDEs described so far,
the amplitudes $m_{2i-1}$ and~$n_{2i}$ do \emph{not} in general
tend to constants as $t \to \pm\infty$,
but instead grow or decay exponentially.
Thus, the functions $\ln m_{2i-1}(t)$ and $\ln n_{2i}(t)$ are asymptotically linear,
with a set of $2K-1$ slopes
\begin{equation}
  \label{eq:asymptotic-amplitude-slopes}
  \begin{aligned}
    d_{2j}
    &=
    \frac12 \left( \frac{1}{\lambda_{j+1}} - \frac{1}{\mu_{j}} \right)
    ,\quad
    j = 1, \dots, K-1
    ,\\[1.5ex]
    d_{2j-1}
    &=
    \begin{cases}
      \displaystyle
      \frac12 \left( \frac{1}{\lambda_{j}} - \frac{1}{\mu_{j}} \right)
      ,&
      j = 1, \dots, K-1
      ,\\[2ex] \displaystyle
      \frac12 \left( \frac{1}{\lambda_{K}} \right)
      ,&
      j = K
      ,
    \end{cases}
  \end{aligned}
\end{equation}
that appear in the opposite order as $t \to \infty$ compared to when $t \to -\infty$,
and thus with phase shifts analogous to the ones usually only displayed by the \emph{positions}
of the solitons.

The odd interlacing case with $K+1$ peakons in~$u$ and $K$ in~$v$
is slightly different, in a perhaps surprising way~\cite{shuaib-lundmark:2019:GX-noninterlacing}.
Here the two spectral problems contribute $K$ eigenvalues and residues each,
$\{ \lambda_i, a_i \}_{i=1}^K$
and
$\{ \mu_j, b_j \}_{j=1}^K$,
together with the constants $b_\infty$ and~$b_\infty^*$,
for a total of $4K+2 = 2(2K+1) = 2N$ parameters, at is should be.
There are $2K$ asymptotic velocities as $t \to -\infty$,
with the fastest velocity occurring twice (for $x_1$ and~$x_2$);
in order from left to right they are
\begin{equation}
  \begin{split}
    &
    \frac12 \left( \frac{1}{\lambda_1} + \frac{1}{\mu_1} \right)
    > \frac12 \left( \frac{1}{\lambda_2} + \frac{1}{\mu_1} \right)
    > \frac12 \left( \frac{1}{\lambda_2} + \frac{1}{\mu_2} \right)
    \\ &
    > \frac12 \left( \frac{1}{\lambda_3} + \frac{1}{\mu_2} \right)
    > \dots
    > \frac12 \left( \frac{1}{\lambda_K} + \frac{1}{\mu_K} \right)
    > \frac12 \frac{1}{\mu_K}
    .
  \end{split}
\end{equation}
But as $t \to \infty$, there is a partly \emph{different} set of asymptotic velocities,
namely (from right to left, and with the fastest velocity applying to both
$x_{2K}$ and~$x_{2K+1}$)
\begin{equation}
  \begin{split}
    &
    \frac12 \left( \frac{1}{\lambda_1} + \frac{1}{\mu_1} \right)
    > \frac12 \left( \frac{1}{\lambda_1} + \frac{1}{\mu_2} \right)
    > \frac12 \left( \frac{1}{\lambda_2} + \frac{1}{\mu_2} \right)
    \\ &
    > \frac12 \left( \frac{1}{\lambda_2} + \frac{1}{\mu_3} \right)
    > \dots
    > \frac12 \left( \frac{1}{\lambda_K} + \frac{1}{\mu_K} \right)
    > \frac12 \frac{1}{\lambda_K}
    .
  \end{split}
\end{equation}
Thus, the even-numbered peakons (those in~$v$) have the same set of incoming and outgoing velocities
$\frac12 \left( \lambda_i^{-1} + \mu_i^{-1} \right)$,
with explicitly computable phase shifts,
while the odd-numbered peakons (those in~$u$) have different velocities going in and and coming out,
so that it's meaningless to talk about phase shifts for them;
similarly for the logarithms of the amplitudes.

The results described so far only concern interlacing solutions.
When we relax this condition and allow arbitrary peakon configurations,
we run into an interesting issue.
Suppose, for example, that we start out with the first two peakons at $x_{1}$ and~$x_{2}$
belonging to the first component~$u$, and interlace the peakons from there on.
After the transformation to the interval $[-1,1]$ there will thus be positive weights $g_1$ and $g_2$
at $y_1$ and~$y_2$ in the measure $g(y)$.
Then, when computing the solution $\Phi(y;\lambda)$ to \eqref{eq:GX-dual-cubic-I},
we will encounter the following matrix product when going from
$\Phi(y_0^+;\lambda)$ to $\Phi(y_{3}^-;\lambda)$:
\begin{equation*}
  \begin{split}
    &
    \begin{pmatrix} 1 & 0 & 0 \\ 0 & 1 & 0 \\ -\lambda l_{2} & 0 & 1 \end{pmatrix}
    \begin{pmatrix} 1 & 0 & 0 \\ 0 & 1 & g_{2} \\ 0 & 0 & 1 \end{pmatrix}
    \begin{pmatrix} 1 & 0 & 0 \\ 0 & 1 & 0 \\ -\lambda l_{1} & 0 & 1 \end{pmatrix}
    \\ &
    \times
    \begin{pmatrix} 1 & 0 & 0 \\ 0 & 1 & g_{1} \\ 0 & 0 & 1 \end{pmatrix}
    \begin{pmatrix} 1 & 0 & 0 \\ 0 & 1 & 0 \\ -\lambda l_{0} & 0 & 1 \end{pmatrix}
    \\ &
    =
    \begin{pmatrix} 1 & 0 & 0 \\ -\lambda \bigl( (g_{1} + g_{2}) l_{0} + g_{2} l_{1} \bigr) & 1 & g_{1} + g_{2} \\ -\lambda ( l_{0} + l_{1} + l_{2}) & 0 & 1 \end{pmatrix}
    .
  \end{split}
\end{equation*}
But this matrix is the same as that obtained if we replace the two masses at $y_1$ and~$y_2$
with a single mass of weight $\bar{g}_1 = g_1 + g_2$
at the position $\bar{y}_1 = y_0 + \bar{l}_0 = y_3 - \bar{l}_1$ where $\bar{l}_0$ is determined by
\begin{equation*}
  \bar{g}_1 \bar{l}_0 = (g_{1} + g_{2}) l_{0} + g_{2} l_{1}
  ,
\end{equation*}
namely
\begin{equation*}
  \begin{split}
    &
    \begin{pmatrix} 1 & 0 & 0 \\ 0 & 1 & 0 \\ -\lambda \bar{l}_{1} & 0 & 1 \end{pmatrix}
    \begin{pmatrix} 1 & 0 & 0 \\ 0 & 1 & \bar{g}_{1} \\ 0 & 0 & 1 \end{pmatrix}
    \begin{pmatrix} 1 & 0 & 0 \\ 0 & 1 & 0 \\ -\lambda \bar{l}_{0} & 0 & 1 \end{pmatrix}
    \\ &
    =
    \begin{pmatrix} 1 & 0 & 0 \\ -\lambda \bar{g}_{1} \bar{l}_{0} & 1 & \bar{g}_{1} \\ -\lambda ( \bar{l}_{0} + \bar{l}_{1}) & 0 & 1 \end{pmatrix}
    .
  \end{split}
\end{equation*}
The same phenomenon will happen in the twin spectral problem,
so we will get exactly the same spectral data as for an interlacing case with
the weight $\bar{g}_1$ at~$\bar{y}_1$.
This means that the spectral data do not contain enough information to
let us recover the individual positions and weights
$\{ y_1, y_2, g_1, g_2 \}$,
but only the ``effective position and weight'' $\{ \bar{y}_1, \bar{g}_1 \}$
of the pair as a whole.
And the same thing happens for any group of $k$ consecutive peakons in the same
component ($u$ or~$v$) -- the trace that it leaves in the spectral data is the same
as that of a single ``effective weight'' at some ``effective position'',
and thus the individual weights and positions within the group
cannot be resolved from the spectral data alone.

Nevertheless, it is possible to find the explicit peakon solution formulas also in the
non-interlacing cases. This was done by Shuaib and
Lundmark~\cite{shuaib-lundmark:2019:GX-noninterlacing},
as follows.
Given a non-interlacing peakon configuration, insert auxiliary peakons
to make it interlacing.
For this (larger) interlacing $K+K$ configuration, the solution formulas are known.
Now make a suitable substitution in the spectral data appearing in these solution
formulas, such as replacing $\mu_{K-1}$ with $1/\epsilon$,
replacing $\lambda_K$ with some constant times $1/\epsilon$,
and replacing the corresponding residues $b_{K-1}$ and $a_K$ with
constants times some carefully chosen powers of~$\epsilon$.
Then let~$\epsilon \to 0$.
If the substitution is correctly designed, the effect of this will be that the amplitude of
exactly one of the auxiliary peakons tends to zero, while all the other positions and
amplitudes tend to finite limits,
leaving the solution formulas for the configuration where that particular peakon has been removed
(by being turned into a zero-amplitude ``ghostpeakon''~\cite{lundmark-shuaib:2019:ghostpeakons}).
Continuing in this way, the auxiliary peakons can be killed off one by one,
until we reach the solution formulas for the configuration that we started with.
The constants involved in the successive substitutions will appear as
parameters in these final solution formulas,
and those parameters are what determine the individual positions and amplitudes
of the peakons within each group of consecutive peakons in~$u$ or in~$v$,
while the remaining eigenvalues and residues are related to the ``effective position
and amplitude'' of each such group.
Although this is a simple idea in principle,
the actual implementation is technical (to say the least),
with many details to keep track of,
and a myriad of special cases and exceptions.
Merely stating the general non-interlacing solution formulas takes several pages,
despite using all the abbreviated notation that we chose not to describe
when discussing the interlacing case earlier.
Asymptotically, ``singleton'' peakons behave like in the interlacing case,
while in a group of two or more consecutive peakons in the same component,
only one of the peakons in the group (the leftmost or rightmost one, depending on
whether $t \to -\infty$ or $t \to \infty$)
will behave like a singleton at that site would, while all the remaining peakons in the
group instead approach the next peakon to the right or to the left, respectively.
However, this is only true for ``typical'' groups in the middle;
the two leftmost and the two rightmost groups behave slightly differently.
For detailed explanations of the asymptotics of non-interlacing solutions,
see the original paper cited above, where there are plenty of illustrated examples.

\section{The modified Camassa--Holm equation and distributional Lax integrability}
\label{sec:mCH}

There are several equations going by the name \emph{modified Camassa--Holm equation},
but the one that we will consider here is
\begin{equation}
  \label{eq:mCH-again}
  m_t + \bigl( (u^2-u_x^2) \, m \bigr)_x = 0
  ,\qquad
  m = u-u_{xx}
  ,
\end{equation}
or in expanded form
\begin{equation}
  \label{eq:mCH-expanded}
  \begin{split}
    u_t - u_{xxt}
    &
    + 3 u^2 u_x  + u_x^2 u_{xxx} + 2 u_x u_{xx}^2
    \\ &
    - ( u^2 u_{xxx} + u_x^3 + 4 u u_x u_{xx} )
    = 0
    .
  \end{split}
\end{equation}
We will abbreviate it as the mCH equation;
it is also often called the \emph{FORQ equation}, after
Fokas/Fuchssteiner, Olver, Rosenau and Qiao
(see Section~\ref{sec:guide-mCH}).
This PDE bears some similarity to the CH equation, mainly due to the presence
of the relation $m=u-u_{xx}$, but the nonlinear terms are quite
different; in particular the nonlinearity is cubic in~$u$.

To maintain the focus of the exposition we will mostly follow
the articles by Chang and Szmigielski
\cite{chang-szmigielski:2016:mCH-concept-of-peakons, chang-szmigielski:2017:mCH-liouville-integrability-conservative-peakons, chang-szmigielski:2018:mCH-Lax-integrability-peakon-problem},
where the reader can find further details and references;
here we just mention the book by
Baker and Graves-Morris~\cite{baker-gravesmorris:1996:pade-approximants}
for multi-point Padé approximants in general.
As with all equations discussed here, we will focus on the peakon sector of solutions,
so the peakon ansatz \eqref{eq:intro-peakons} for $u(x,t)$ is in force, and consequently $m$ is a discrete
measure as in~\eqref{eq:CH-peakon-m}.
We also assume that all $m_k$ are positive (the pure peakon case),
and that $x_1 < \dots < x_N$.

The Lax pair for \eqref{eq:mCH-again} reads~\cite{schiff:1996:CH-dual-hierarchies-zero-curvature-formulations, qiao:2006:mCH-new-integrable-cuspons-WMpeakons}
\begin{equation}
  \label{eq:mCH-Lax}
  \Psi_x = \tfrac12 U \Psi
  , \qquad
  \Psi_t = \tfrac12 V \Psi
  ,
\end{equation}
where
\begin{align*}
  \Psi & = (\psi_1,\psi_2)^T
  , \\
  U & = \begin{pmatrix} -1 & z m \\ -z m & 1 \end{pmatrix}
  , \\
  V &= \begin{pmatrix} 4z^{-2} + Q & -2z^{-1} (u-u_x)-z m Q \\ 2 z^{-1}(u+u_x) + z m Q & -Q \end{pmatrix}
  , \\[1ex]
  Q &= u^2-u_x^2
  , \quad
  z \in \C
  .
\end{align*}
We note that the $x$-equation of the Lax pair is ill-defined on the
support of the discrete measure~$m$, where neither component
of~$\Psi$ is continuous.
This creates exactly the same problem that we lightly touched upon in the
introduction. This time, however, we will not a priori define what we
mean by a weak or distributional form of the mCH equation~\eqref{eq:mCH-again}.
Instead we will define the distributional Lax pair and let
its compatibility dictate the ``correct'' interpretation of~\eqref{eq:mCH-again},
which for the purposes of this article is the one preserving integrability.

The vector function $\Psi$ has left and right limits at the points~$x_k$ where $m$ is supported,
and is smooth away from them.
This allows one to define the products $\psi_1 m$ and $\psi_2 m$
using the general philosophy described in Section~\ref{sec:CH}
(see equation~\eqref{eq:CH-multiplication-f-times-m}), by postulating that
\begin{equation}
  \label{eq:mCH-multPsi}
  \psi_i m = 2\sum_{k=1}^N \Bigl( \alpha \, \psi_i(x_k^-) + \beta \, \psi_i(x_k^+) \Bigr) \, m_k \, \delta(x-x_k)
\end{equation}
for $i \in \{ 1,2 \}$ and $\alpha+\beta=1$.
We also note that in the $t$-equation the term $mQ$ needs to be defined as well,
since $Q = u^2-u_x^2$ is not continuous at the support of~$m$.
In other words, we need to define
$Q(x) \, \delta(x-x_k) = Q(x_k) \, \delta(x-x_k)$
for some yet to be determined values $Q(x_k)$.
It turns out~\cite[Appendix~A]{chang-szmigielski:2018:mCH-Lax-integrability-peakon-problem}
that distributional compatibility holds,
i.e., $\partial _t \partial_x \Psi = \partial_x \partial_t \Psi$,
provided that $\alpha$ and~$\beta$ in \eqref{eq:mCH-multPsi} are chosen according to
\begin{equation}
  \label{eq:mCH-alpha-beta}
  (\alpha,\beta)=(1,0)
  \quad\text{or}\quad
  (\alpha,\beta)=(0,1)
  .
\end{equation}
Then, in either case, the compatibility further implies the peakon ODEs
\begin{equation}
  \label{eq:mCH-distrcompatibility}
  \dot m_k=0
  , \qquad
  \dot x_k = Q(x_k)
\end{equation}
where
\begin{equation}
  Q(x_k) := \langle Q \rangle (x_k)
  = \tfrac12 \bigl( Q(x_k^+) + Q(x_k^-) \bigr)
  .
\end{equation}
Thus, the initially ill-defined term
$Q(x) \, m = (u^2-u_x^2) \, m$
which appears in the Lax pair, and also in the mCH equation itself,
is to be defined as $\langle Q \rangle (x) \, m$.
In other words, the mCH equation~\eqref{eq:mCH-again},
if it is to be derived from the Lax pair, should be interpreted as
\begin{equation}
  \label{eq:mCH-distr}
  m_t + \bigl( \langle Q \rangle (x) \, m \bigr)_x = 0
  .
\end{equation}
Since $u$ is continuous, the term $u^2 m$ causes no problems,
so the crucial part is that $u_x^2 m$ needs to be interpreted as
\begin{equation}
  \label{eq:mCH-lax-regularization}
  u_x^2 m = \langle u_x^2 \rangle \, m
\end{equation}
in order to obtain Lax integrable peakon ODEs.

Let us compare this with the weak formulation of~\eqref{eq:mCH-again}
used by Gui, Liu, Olver and Qu~\cite{gui-liu-olver-qu:2013:mCH-wavebreaking-peakons}
(and many other works).
In that approach, one eliminates~$m$ by inverting $(1-\partial_x^2)$
via the convolution formula $u = p * m$ where $p(x) = \frac12 e^{-\abs{x}}$,
and then defines the weak solution via double integration with respect to $t$ and~$x$ against
compactly supported smooth test functions.
Applying this definition of weak solutions to the peakon sector
leads to peakon ODEs which amount to a \emph{different} interpretation of the term $u_x^2 m$,
namely~\cite{chang-szmigielski:2018:mCH-Lax-integrability-peakon-problem}
\begin{equation}
  \label{eq:mCH-weak-regularization}
  u_x^2 m = \frac{\langle u_x^2\rangle+2\langle u_x \rangle^2}{3} \, m
  .
\end{equation}
Let us refer to \eqref{eq:mCH-lax-regularization} as the \emph{Lax regularization},
and \eqref{eq:mCH-weak-regularization}
as the \emph{weak regularization}.
To illustrate the differences between the two,
it is sufficient to consider the case $N=2$, where
\begin{equation*}
  m = 2 m_1 \, \delta(x-x_1) + 2 m_2 \, \delta(x-x_2)
  ,
\end{equation*}
and we assume $x_1 < x_2$ to simplify formulas.
Under the Lax regularization, the peakon ODEs are
\begin{equation}\label{eq:mCH-twopeakonL}
  \begin{aligned}
    \dot m_1 &= 0
    ,&
    \dot x_1 &= 2 m_1 m_2 \, e^{x_1-x_2}
    ,\\
    \dot m_2 &= 0
    ,&
    \dot x_2 &= 2 m_1 m_2 \, e^{x_1-x_2}
    ,
  \end{aligned}
\end{equation}
while according to the weak regularization they are
\begin{equation}
  \label{eq:mCH-twopeakonW}
  \begin{aligned}
    \dot m_1 &= 0
    ,&
    \dot x_1 &= \tfrac 23 m_1^2 + 2m_1 m_2 \, e^{x_1-x_2}
    ,\\
    \dot m_2 &= 0
    ,&
    \dot x_2 &= \tfrac23 m_2^2 + 2 m_1 m_2 \, e^{x_1-x_2}
    .
  \end{aligned}
\end{equation}
Although this looks like a minor difference, the consequences are quite striking.
The function
\begin{equation*}
  x \mapsto u(x,t) = m_1(t) \, e^{-\abs{x-x_1(t)}} + m_2 \, e^{-\abs{x-x_2(t)}}
\end{equation*}
belongs to the Sobolev space $H^1(\R)$ for each fixed~$t$,
and a simple computation shows that
\begin{equation}
  \lVert u \rVert^2_{H^1}
  = \int_{\R} (u^2 + u_x^2) \, dx
  = 2 (m_1^2+m_2^2) + 4 m_1m_2 e^{x_1-x_2}
  ,
\end{equation}
so from \eqref{eq:mCH-twopeakonL} and \eqref{eq:mCH-twopeakonW} we find that
\begin{equation}
  \frac{d}{dt} \lVert u \rVert^2_{H^1} = 0
\end{equation}
in the Lax regularization,
but
\begin{equation}
  \frac{d}{dt} \lVert u \rVert^2_{H^1}
  = \tfrac83 m_1 m_2 (m_1^2 - m_2^2) \, e^{x_1 - x_2}
\end{equation}
in the weak regularization.
In the smooth sector,
$H_0 = \lVert u \rVert^2_{H^1}$
is one of the original Hamiltonians of~\eqref{eq:mCH-again},
and hence conserved~\cite{qiao:2006:mCH-new-integrable-cuspons-WMpeakons}.
The computation above shows that $H_0$ remains a constant of motion in the peakon sector
in the Lax regularization,
but not in the weak regularization.
For more about this, see Anco and Kraus~\cite{anco-kraus:2018:hamiltonian-peakons-weak-solution-mCH}.

Now we turn our attention to the approximation aspects of the Lax pair~\eqref{eq:mCH-Lax};
from now on, we will be using the Lax regularization~\eqref{eq:mCH-lax-regularization} exclusively.
First, we note that one can simplify the $x$-equation in \eqref{eq:mCH-Lax} by performing the gauge transformation
\begin{equation}
  \Phi
  = (\phi_1, \phi_2)^T
  = \operatorname{diag} \bigl( z^{-1} e^{x/2}, e^{-x/2} \bigr) \, \Psi
  ,
\end{equation}
which leads to
\begin{equation}
  \label{eq:mCH-xLax}
  \begin{aligned}
    \Phi_x = \begin{pmatrix} 0 & h \\ -\lambda g & 0 \end{pmatrix} \Phi
    , \\
    g(x) = \sum_{k=1}^N g_k \, \delta(x-x_k)
    , \\
    h(x) = \sum_{k=1}^N h_k \, \delta(x-x_k)
    ,
  \end{aligned}
\end{equation}
where
\begin{equation}
  \label{eq:mCH-gk-hk}
  g_k = m_k \, e^{-x_k}
  , \quad
  h_k = m_k \, e^{x_k}
  , \quad
  \lambda = z^2
  .
\end{equation}
For future use, note that $g_k h_k = m_k^2$.
Next, we require that
$\phi_1(-\infty) = \phi_2(\infty) = 0$,
a condition which is compatible with the time evolution induced by the mCH equation.
Thus, we consider the boundary value problem
\begin{equation}
  \label{eq:mCH-xLaxBVP-}
  \Phi_x = \begin{pmatrix} 0 & h \\ -\lambda g & 0 \end{pmatrix} \Phi
  ,\quad
  \phi_1(-\infty) = \phi_2(\infty) = 0
  ,
\end{equation}
where we will interpret the matrix product using the left regularization $(\alpha, \beta) = (1,0)$ in~\eqref{eq:mCH-alpha-beta},
meaning that
\begin{equation}
  \label{eq:mCH-left-regularization}
  \Phi(x) \, \delta(x-x_k) := \Phi(x_k^-) \, \delta(x-x_k)
  .
\end{equation}
(We could express this by saying that we are seeking solutions $\Phi(x)$ that are continuous from the left.)
Writing $\Phi(x;\lambda)$ for the solution starting out with
$\phi_2(-\infty;\lambda) = 1$,
we observe that it is a piecewise constant two-component vector,
and one can rephrase \eqref{eq:mCH-xLax} as a difference equation. Indeed, with
\begin{subequations} \label{eq:mCH-def-q-p}
\begin{equation}
  \begin{pmatrix} q_0(\lambda) \\ p_0(\lambda) \end{pmatrix}
  =
  \begin{pmatrix} 0 \\ 1 \end{pmatrix}
\end{equation}
and
\begin{equation}
  \begin{pmatrix} q_k(\lambda) \\ p_k(\lambda) \end{pmatrix}
  =
  \Phi(x_k^+;\lambda)
  ,\qquad
  1 \le k \le N
  ,
\end{equation}
\end{subequations}
we can translate the jump condition
\begin{equation}
  \Phi(x_k^+;\lambda) - \Phi(x_k^-;\lambda)
  = \begin{pmatrix} 0 & h_k \\ -\lambda g_k & 0 \end{pmatrix} \Phi(x_k^-;\lambda)
\end{equation}
(note that \eqref{eq:mCH-left-regularization} is used here)
into the recurrence
\begin{subequations} \label{eq:mCH-transition}
\begin{equation}
  \begin{pmatrix} q_k(\lambda) \\ p_k(\lambda) \end{pmatrix}
  = T_k(\lambda) \begin{pmatrix} q_{k-1}(\lambda) \\ p_{k-1}(\lambda) \end{pmatrix}
  ,\qquad
  1 \le k \le N
  ,
\end{equation}
where the transition matrix equals
\begin{equation}
  T_k(\lambda) =\begin{pmatrix} 1 & h_k \\ -\lambda g_k & 1 \end{pmatrix}
  .
\end{equation}
\end{subequations}
Thus,
\begin{equation*}
  \begin{pmatrix} q_k(\lambda) \\ p_k(\lambda) \end{pmatrix}
  = T_k(\lambda) \dotsm T_1(\lambda) \begin{pmatrix} 0 \\ 1 \end{pmatrix}
  =: S_k(\lambda) \begin{pmatrix} 0 \\ 1 \end{pmatrix}
  .
\end{equation*}
It is not difficult to derive explicit expressions for
the entries in the matrix $S_k(\lambda)$,
and in particular for $q_k(\lambda)$ and~$p_k(\lambda)$,
in terms of $\{ g_i, h_i \}$.
For example (assuming that $N \ge 4$),
\begin{equation*}
  \begin{split}
    p_4(\lambda)
    & = 1 - \bigl( h_1 g_2 + h_1 g_3 + h_1 g_4 + h_2 g_3 + h_2 g_4 + h_3 g_4 \bigr) \, \lambda
    \\ & \quad
    + h_1 g_2 h_3 g_4 \, \lambda^2
    ,
  \end{split}
\end{equation*}
and in general
\begin{subequations} \label{eq:mCH-pk-qk-explicit}
\begin{equation}
  \label{eq:mCH-pk-explicit}
  p_k(\lambda)
  = 1 + \sum_{r=1}^{\lfloor k/2 \rfloor}
  \Biggl( \,\, \sum_{\substack{I, J \in \binom{[k]}{r} \\[0.5ex] I<J}}
  h_{i_1} g_{j_1} \dotsm h_{i_r} g_{j_r}
  \,\, \Biggr)
  (-\lambda)^r
  ,
\end{equation}
where $\binom{[k]}{r}$ denotes the set of $r$-element subsets of $[k] = \{ 1,\dots,k \}$,
and the notation $I<J$ means that the index sets $I$ and $J$ are ``interlacing'',
\begin{equation*}
  1 \le i_1 < j_1 < \dotsb < i_r < j_r \le k
  .
\end{equation*}
Similarly,
\begin{equation}
  \label{eq:mCH-qk-explicit}
  q_k(\lambda)
  = \sum_{r=0}^{\lfloor (k-1)/2 \rfloor}
  \Biggl( \,\, \sum_{\substack{I \in \binom{[k]}{r+1} \\[0.5ex] J \in \binom{[k]}{r} \\[0.5ex] I<J}}
  h_{i_1} g_{j_1} \dotsm h_{i_r} g_{j_r} h_{i_{r+1}}
  \,\, \Biggr)
  (-\lambda)^r
  .
\end{equation}
\end{subequations}

By analyzing the $t$-member of the Lax pair \eqref{eq:mCH-Lax}
in the asymptotic region $x>x_N$, we arrive at
the evolution equations for $q_N(\lambda)$ and~$p_N(\lambda)$:
\begin{equation}
  \label{eq:mCH-q-p-evol}
    \dot q_N = \frac{2}{\lambda} \Bigl( q_N - L p_N \Bigr)
  , \qquad
  \dot p_N = 0
  ,
\end{equation}
where $L = \sum_{k=1}^N h_k$~\cite{chang-szmigielski:2018:mCH-Lax-integrability-peakon-problem}.
In particular, the polynomial $p_N(\lambda)$ is time-invariant,
and since the boundary condition $\phi_2(\infty;\lambda)=0$ translates into $p_N(\lambda)=0$, we see that
the spectrum of~\eqref{eq:mCH-xLaxBVP-}, i.e., the set of zeros of $p_N(\lambda)$,
is time-invariant too.
No obvious information about the nature of this spectrum is available at this point,
but in fact it is positive and simple,
which can be proved in an indirect and perhaps surprising way by studying the structure of
the Weyl function, which we define as
\begin{equation}
  \label{eq:mCH-Weyl}
  W(\lambda)
  = \frac{\phi_1(\infty;\lambda)}{\phi_2(\infty;\lambda)}
  = \frac{q_N(\lambda)}{p_N(\lambda)}
  .
\end{equation}
To this end, let us define
\begin{equation}
  \label{eq:mCH-w-even-odd}
  w_{2k-1}(\lambda) = \frac{q_{k-1}(\lambda)}{p_k(\lambda)}
  ,\qquad
  w_{2k}(\lambda) = \frac{q_k(\lambda)}{p_k(\lambda)}
  .
\end{equation}
\begin{subequations}
Then \eqref{eq:mCH-transition} gives
\begin{equation*}
  \begin{split}
    w_{2k} - w_{2k-1}
    &
    = \frac{q_k}{p_k} - \frac{q_{k-1}}{p_k}
    = \frac{q_k - q_{k-1}}{p_k}
    = \frac{h_k p_{k-1}}{p_k}
    \\ &
    = h_k \left( 1 + \frac{p_{k-1} - p_k}{p_k} \right)
    = h_k \left( 1 + \frac{\lambda g_k q_{k-1}}{p_k} \right)
    \\ &
    = h_k + \lambda h_k g_k \, \frac{q_{k-1}}{p_k}
    = h_k + \lambda h_k g_k \, w_{2k-1}
    ,
  \end{split}
\end{equation*}
so that
\begin{equation}
  \label{eq:mCH-w-from-odd-to-even}
  w_{2k} = h_k + (1 + \lambda h_k g_k) \, w_{2k-1}
  ,
\end{equation}
and
\begin{equation*}
  \frac{1}{w_{2k+1}} - \frac{1}{w_{2k}}
  = \frac{p_{k+1}}{q_k}
  - \frac{p_{k}}{q_k}
  = \frac{p_{k+1} - p_{k}}{q_k}
  = \frac{-\lambda g_{k+1} q_{k}}{q_k}
  = - \lambda g_{k+1}
  ,
\end{equation*}
so that
\begin{equation}
  \label{eq:mCH-w-from-even-to-odd}
  w_{2k+1} = \frac{1}{- \lambda g_{k+1} + \dfrac{1}{w_{2k}}}
  .
\end{equation}
\end{subequations}
Here we can sense a Stieltjes-type continued fraction for
$W(\lambda) = w_{2N}(\lambda)$ emerging, but the presence of
the $\lambda$-term in \eqref{eq:mCH-w-from-odd-to-even}
is a slight complication.
To illustrate what's going on, let us simply do the computations in the case $N=2$.
First,
\begin{equation*}
  \begin{aligned}
    w_1(\lambda) &= \frac{q_0}{p_1} = \frac{0}{1} = 0
    ,\\
    w_2(\lambda) &= h_1 + (1 + \lambda g_1 h_1) \, w_0 = h_1
    ,\\
    w_3(\lambda) &= \frac{1}{-\lambda g_2 + \dfrac{1}{w_2}}
    = \frac{1}{-\lambda g_2 + \dfrac{1}{h_1}}
    ,
  \end{aligned}
\end{equation*}
and then
\begin{equation*}
  \begin{split}
    W(\lambda)
    = w_4(\lambda)
    &
    = h_2 + (1 + \lambda g_2 h_2) \, w_3
    = h_2 + \frac{1 + \lambda g_2 h_2}{-\lambda g_2 + \dfrac{1}{h_1}}
    \\ &
    = h_2 + \frac{(1 + \lambda g_2 h_2) \, h_1}{-\lambda g_2 h_1 + 1}
    = \frac{h_1 + h_2}{-\lambda g_2 h_1 + 1}
    \\ &
    = \frac{1}{- \lambda \, \frac{g_2 h_1}{h_1 + h_2} + \dfrac{1}{h_1 + h_2}}
    ,
  \end{split}
\end{equation*}
which has the form
\begin{equation*}
  \frac{1}{- \lambda \, c_1 + \dfrac{1}{c_2}}
\end{equation*}
with positive coefficients $c_1 = g_2 h_1 / (h_1+h_2) > 0$ and $c_2 = h_1 + h_2 > 0$.
Now it is known from the theory of Stieltjes continued fractions
that rational functions of the form
\begin{equation*}
  f(\lambda) =
  \cfrac{1}{-\lambda \, c_1 +
    \cfrac{1}{c_2 +
      \cfrac{1}{- \lambda \, c_3 +
        \cfrac{1}{c_4 +
          \cfrac{1}{\raisebox{1.5ex}{$\ddots$} +
          \cfrac{1}{- \lambda \, c_{2n-1} +
            \cfrac{1}{c_{2n}}
          }}}}}}
\end{equation*}
with all $c_k>0$ are in one-to-one correspondence with discrete positive measures on~$\R_+$ of the form
\begin{equation}
  \label{eq:mCH-discrete-measure-on-R+}
  \alpha(\lambda) = \sum_{k=1}^n a_k \, \delta(\lambda - \lambda_k)
  ,
\end{equation}
with $0 < \lambda_1 < \dots < \lambda_n$ and all $a_k>0$,
via (minus) the Stieltjes transform
\begin{equation*}
  f(\lambda) = - \int \frac{d\alpha(z)}{\lambda - z}
  = - \sum_{k=1}^n \frac{a_k}{\lambda - \lambda_k}
  .
\end{equation*}
Thus we have $W(\lambda) = w_4(\lambda) = a_1 / (\lambda_1 - \lambda)$ where $a_1$ and $\lambda_1$ are positive,
and in particular the spectrum is positive and simple.
Of course it is overkill to use the theory in this small case, since we could just have computed
\begin{equation*}
  W(\lambda) = \frac{q_2(\lambda)}{p_2(\lambda)} = \frac{h_1+h_2}{- \lambda g_2 h_1 + 1}
  = \frac{\frac{h_1+h_2}{g_2 h_1}}{\frac{1}{g_2 h_1} - \lambda}
  = \frac{a_1}{\lambda_1 - \lambda}
\end{equation*}
right away, but the point is to illustrate the general pattern, so consider next the case $N=3$,
where we would continue the computation with
\begin{equation*}
  \begin{split}
    w_5(\lambda)
    &
    = \frac{1}{-\lambda g_3 + \dfrac{1}{w_4}}
    = \frac{1}{-\lambda g_3 + \dfrac{1}{\dfrac{1}{- \lambda \, \frac{g_2 h_1}{h_1 + h_2} + \dfrac{1}{h_1 + h_2}}
      }}
    \\ &
    = \frac{1}{-\lambda \Bigl( g_3 + \frac{g_2 h_1}{h_1 + h_2} \Bigr) + \dfrac{1}{h_1 + h_2}}
  \end{split}
\end{equation*}
and
\begin{equation*}
  \begin{split}
    W(\lambda)
    = w_6(\lambda)
    &
    = h_3 + (1 + \lambda g_3 h_3) \, w_5
    \\ &
    = h_3 + \frac{(1 + \lambda g_3 h_3) (h_1+h_2)}{-\lambda \bigl( g_3 (h_1+h_2) + g_2 h_1 \bigr) + 1}
    \\ &
    = c_0 + \frac{1}{- \lambda \, c_1 + \dfrac{1}{c_2}}
    ,
  \end{split}
\end{equation*}
where
\begin{equation*}
  \begin{aligned}
    c_0 &= \frac{h_1 g_2 h_3}{h_1 g_2 + h_1 g_3 + h_2 g_3} > 0
    ,\\
    c_2 &= h_1 + h_2 + h_3 - c_0 = \dotsb > 0
    ,\\
    c_1 &= \frac{h_1 g_2 + h_1 g_3 + h_2 g_3}{c_2} > 0
    .
  \end{aligned}
\end{equation*}
Thus, the Weyl function once more matches the Stieltjes form,
except that there is now also an additive constant $c_0$,
so that we have $W(\lambda) = c_0 + a_1/(\lambda_1 - \lambda)$
for some positive $c_0$, $a_1$ and~$\lambda_1$.
In any case, we conclude again that the spectrum is positive.
(Which is still easy to show directly since there is just one eigenvalue,
but as $N$ increases we will have higher-degree polynomials $p_N(\lambda)$
determining the spectrum, and the coefficients in the Stieltjes continued fractions
will be increasingly horrendous expressions in $\{ g_k, h_k \}$.)

Now the idea should be clear: to prove that the spectrum is positive and simple,
we prove inductively that all $w_n(\lambda)$, and in particular the last one $W(\lambda) = w_{2N}(\lambda)$,
are of the Stieltjes form with only
positive coefficients $c_k^{(n)}$ in their continued fractions
(plus an extra term $c_0^{(n)}>0$ when $n = 4r+2$),
and thus correspond to discrete spectral measures $\alpha^{(n)}$
of the form~\eqref{eq:mCH-discrete-measure-on-R+}.

We have already seen that the statement holds to begin with.
Assume, for the inductive step, that $w_{2k-2}$ has the claimed form.
In the passage from $w_{2k-2}$ to~$w_{2k-1}$
the support of the measure changes, since the denominator changes from $p_{k-1}$ to~$p_{k}$.
In this case, it is easy to see from~\eqref{eq:mCH-w-from-even-to-odd}
that positive coefficients in the continued fraction for~$w_{2k-2}$
imply positive coefficients in the continued fraction for~$w_{2k-1}$,
so that $w_{2k-1}$ has the claimed form too.
Indeed, if $k$ is even, so that $c_0^{(2k-2)} > 0$ by inductive hypothesis,
then~\eqref{eq:mCH-w-from-even-to-odd} immediately gives a
continued fraction of the required form for~$w_{2k-1}$, with $c_1^{(2k-1)} = g_{k}$
as its leading coefficient.
And if $k$ is odd, so that $c_0^{(2k-2)} = 0$,
then the same ``$1/(1/w)=w$ phenomenon'' that we saw in the step from $w_{4}$ to~$w_{5}$ above
implies that the leading coefficient will be $c_1^{(2k-1)} = g_{k} + c_1^{(2k-2)} > g_{k}$.

Next, when going from $w_{2k-1}$ to~$w_{2k}$,
the denominator is $p_k$ in both cases, so the support of the measure is unchanged,
but we need to show that the new measure is still positive,
and here we look directly at the measures rather than at the coefficients in the continued fractions.
Since we just showed that $w_{2k-1}$ has a continued fraction with positive coefficients,
we know that there is a measure $\alpha^{(2k-1)}$ such that
$w_{2k-1}(\lambda) = \int (z-\lambda)^{-1} d\alpha^{(2k-1)}(z)$,
and then \eqref{eq:mCH-w-from-odd-to-even} gives
\begin{equation*}
  \begin{split}
    w_{2k}(\lambda)
    &
    = h_k + (1 + \lambda \, h_k g_k) \, w_{2k-1}(\lambda)
    \\ &
    = h_k + \int \frac{1 + \lambda \, h_k g_k}{z-\lambda} \, d\alpha^{(2k-1)}(z)
    \\ &
    = h_k + \int \left( \frac{1 + z h_k g_k}{z-\lambda} - h_k g_k  \right) \, d\alpha^{(2k-1)}(z)
    \\ &
    = h_k \left( 1 - g_k \int d\alpha^{(2k-1)}(z) \right)
    \\ & \quad
    + \int \frac{1 + z h_k g_k}{z-\lambda}  \, d\alpha^{(2k-1)}(z)
    \\ &
    =: c_0^{(2k)} + \int \frac{d\alpha^{(2k)}(z)}{z-\lambda}
    ,
  \end{split}
\end{equation*}
where the new measure
\begin{equation*}
  \alpha^{(2k)}(z) = (1 + z h_k g_k) \, \alpha^{(2k-1)}(z)
\end{equation*}
is positive and supported on the same set in~$\R_+$ as the old measure~$\alpha^{(2k-1)}(z)$,
and where
\begin{equation*}
  c_0^{(2k)} = h_k \left( 1 - g_k \int d\alpha^{(2k-1)}(z) \right)
\end{equation*}
is zero when $k$ is even and positive when $k$ is odd,
since the integral
\begin{equation*}
  \int d\alpha^{(2k-1)}(z)
  = \lim_{\lambda \to \infty} \bigl( -\lambda \, w_{2k-1}(\lambda) \bigr)
  = \frac{1}{c_1^{(2k-1)}}
\end{equation*}
is equal to or less than~$1/g_k$ depending on the parity of~$k$.
Thus $w_{2k}$ has the claimed form as well, and the inductive step is complete.

This concludes the proof that the Weyl function is the
(shifted) Stieltjes transform of a positive discrete measure $\alpha$ with support inside~$\R_+$,
and in particular that the spectrum is positive and simple.
We saw in~\eqref{eq:mCH-pk-explicit}
that the degree of $p_N(\lambda)$ is $\lfloor N/2 \rfloor$,
so we may summarize the above by saying that
\begin{equation}
  W(\lambda) = c + \int \frac{d\alpha(z)}{z-\lambda}
  = c + \sum_{k=1}^{\floor{N/2}} \frac{a_k}{\lambda_k - \lambda}
\end{equation}
with the spectral measure
\begin{equation}
  \alpha(\lambda) = \sum_{k=1}^{\floor{N/2}} a_k \, \delta(\lambda-\lambda_k)
  ,
\end{equation}
where $0 < \lambda_1 < \dots < \lambda_{\floor{N/2}}$,
where $a_k > 0$ for
$1 \le k \le \floor{N/2}$,
and where $c>0$ when $N$ is odd
and $c=0$ when $N$ is even.

Let us now return to our goal of solving the peakon ODEs \eqref{eq:mCH-distrcompatibility}.
Trivially, all $m_k$ are constant, but it remains to integrate the ODEs for the variables $x_k(t)$,
for given values of the constants $m_1$, \ldots,~$m_N$.
From the time-dependence \eqref{eq:mCH-q-p-evol}
of $q_N$ and~$p_N$ induced by the $t$-equation in the Lax pair,
we readily find that $\dot W = \frac{2}{\lambda} (W - L)$,
so that
\begin{equation}
  \dot a_k = \frac{2 a_k}{\lambda_k}
  ,\qquad
  \dot c = 0
  ,
\end{equation}
and hence
\begin{equation}
  a_k(t) = a_k(0) \, e^{2t/\lambda_k}
  ,\qquad
  c(t) = c(0)
  .
\end{equation}
Thus we know the time evolution of the spectral data encoded in the Weyl function~$W(\lambda)$,
and we can find $x_1(t),\dots,x_N(t)$ by solving the inverse spectral problem of
recovering the quantities
$g_k = m_k \, e^{-x_k}$
and
$h_k = m_k \, e^{x_k}$
(see~\eqref{eq:mCH-gk-hk})
from $W(\lambda)$,
for given values of the constants $m_1$, \ldots,~$m_N$.
The coefficients $c_k$ in the continued fraction for $W(\lambda)$ could in principle
be recovered in a similar way as in Section~\ref{sec:CH},
but they depend in a complicated way on the sought quantities $g_k$ and~$h_k$,
which we still would need to solve for.
Instead, we will present a more direct path to $\{ g_k, h_k \}$ using ideas from multi-point approximation theory.

The inverse problem in question is connected in a natural way to an \emph{interpolation} problem,
where rational functions are required to fit given values at various points,
rather than the \emph{approximation} problems that we have seen before,
where rational functions must match given power series up to a certain order.
Let us iterate \eqref{eq:mCH-transition} from $N-k$ to $N$ and divide by $p_N(\lambda)$
in order to write the resulting expression in terms
of the Weyl function~\eqref{eq:mCH-Weyl}:
\begin{equation}
  \label{eq:mCH-Papprox1}
  \begin{pmatrix} W(\lambda) \\ 1 \end{pmatrix}
  = T_N(\lambda) \, T_{N-1}(\lambda) \dotsm T_{N-k+1}(\lambda)
  \begin{pmatrix} \dfrac{q_{N-k}(\lambda)}{p_{N}(\lambda)} \\[1em] \dfrac{p_{N-k}(\lambda)}{p_{N}(\lambda)} \end{pmatrix}
  .
\end{equation}
Let
\begin{equation}
  \label{eq:mCH-matrix-C}
  \widehat T_{j}(\lambda)
  = \operatorname{adj} T_{N+1-j}(\lambda)
  = \begin{pmatrix} 1 & -h_{N+1-j} \\ \lambda g_{N+1-j} & 1 \end{pmatrix}
\end{equation}
and multiply \eqref{eq:mCH-Papprox1} from the left by
\begin{equation}
  \label{eq:mCH-matrix-Shat}
  \widehat S_k(\lambda) := \widehat T_{k}(\lambda) \dotsm \widehat T_1(\lambda)
\end{equation}
to obtain
\begin{equation*}
  \begin{split}
    & \widehat S_k(\lambda)
    \begin{pmatrix} W(\lambda) \\ 1 \end{pmatrix}
    \\ &
    = |T_N(\lambda)| \, |T_{N-1}(\lambda)| \dotsm |T_{N-k+1}(\lambda)|
    \begin{pmatrix} \dfrac{q_{N-k}(\lambda)}{p_{N}(\lambda)} \\[1em] \dfrac{p_{N-k}(\lambda)}{p_{N}(\lambda)} \end{pmatrix}
    ,
  \end{split}
\end{equation*}
where
\begin{equation*}
  |T_j(\lambda)|
  = \begin{vmatrix} 1 & h_j \\ -\lambda g_j & 1 \end{vmatrix}
  = 1 + \lambda g_j h_j
  = 1 + \lambda m_j^2
  .
\end{equation*}
Since we have proved that the roots of $p_{N}(\lambda)$ are all positive,
we may evaluate this at the negative numbers
$\lambda = -1 / m_{N+1-i}^2$
without risk of dividing by zero,
to obtain
\begin{gather} \label{eq:mCH-Papprox2}
  \left[
    \widehat S_k(\lambda)
    \,
    \begin{pmatrix} W(\lambda) \\ 1 \end{pmatrix}
  \right]_{\lambda = -1 / m_{N+1-i}^2}
  =0
  ,\quad
  1 \le i \le k
  .
\end{gather}
Let us denote the entries in the matrix $\widehat S_k(\lambda)$ by
\begin{equation}
  \widehat S_k(\lambda) = \begin{pmatrix} b_k(\lambda) & c_k(\lambda) \\ d_k(\lambda) & e_k(\lambda) \end{pmatrix}
  .
\end{equation}
Equation~\eqref{eq:mCH-Papprox2}, together with an easy calculation of the polynomial degrees,
shows that these entries solve the following interpolation problem:
\begin{subequations}\label{eq:mCH-Papprox}
  \begin{align}
    \label{eq:mCH-Papprox-bc}
    &
    \Bigl[ b_k(\lambda) W(\lambda) + c_k(\lambda) \Bigr]_{\lambda = - m_{N+1-i}^{-2}} = 0
    , \quad
    1 \le i \le k
    , \\ &
    \deg b_k = \left\lfloor \frac{k}{2} \right\rfloor
    , \quad
    \deg c_k = \left\lfloor \frac{k-1}{2} \right\rfloor
    , \quad
    b_k(0) = 1
    , \\ &
    \label{eq:mCH-Papprox-de}
    \Bigl[ d_k(\lambda) W(\lambda) + e_k(\lambda) \Bigr]_{\lambda = - m_{N+1-i}^{-2}} = 0
    ,\quad
    1 \le i \le k
    , \\ &
    \deg d_k = \left\lfloor \frac{k+1}{2} \right\rfloor
    , \quad
    \deg e_k = \left\lfloor \frac{k}{2} \right\rfloor
    ,
    \notag
    \\ &
    d_k(0) = 0
    , \quad
    e_k(0) = 1
    .
  \end{align}
\end{subequations}
Provided that the numbers $m_{N+1-k}$, \ldots,~$m_N$ are distinct,
the interpolation problem~\eqref{eq:mCH-Papprox} has a unique solution.
Indeed, the conditions \eqref{eq:mCH-Papprox-bc}
and~\eqref{eq:mCH-Papprox-de}
directly amount to systems of linear equations
(both of size $k \times k$)
for the unknown coefficients in the
polynomials $b_k$, $c_k$, $d_k$ and $e_k$, and the only thing that needs to
be proved is that these systems are nonsingular,
which can be done by explicit evaluation of the determinants in question,
which are of Cauchy--Stieltjes--Vandermonde type.
Then Cramer's rule provides determinantal formulas for the sought coefficients;
we omit these formulas here, since they are somewhat unwieldy.

Thus, if the numbers $m_1$, \dots,~$m_N$ are all distinct, we can reconstruct
the matrix $\widehat S_{k}(\lambda)$ for each $k \in \{1, \dots, N \}$.
In order to use this information for obtaining the numbers $\{ g_i, h_i \}_{i=1}^N$,
we just need to figure out how the entries in $\widehat S_{k}(\lambda)$ depend on them.
This can be done like for $p_k(\lambda)$ and $q_k(\lambda)$
in~\eqref{eq:mCH-pk-qk-explicit} above,
since the matrix product $\widehat S_{k} = \widehat T_k \dotsm \widehat T_1$
is of the same form as $S_k = T_k \dotsm T_1$ except with
$-g_{N+1-i}$ and $-h_{N+1-i}$ instead of $g_i$ and~$h_i$.
From this we find, for example, that if $k$ is odd, then
$g_{N+1-k}$ is given by the highest coefficient of the bottom left entry of~$\widehat S_k$
divided by the highest coefficient of the top left entry of~$\widehat S_{k-1}$,
and if $k$ is even, it's the the same except that we use the entries in the second column instead.
In this way we obtain all~$g_i$, and hence also $x_i = \ln(m_i/g_i)$
(as well as $h_i = m_i^2 / g_i$).
All factors of the form $m_i-m_j$ cancel out
in the resulting formulas for $\{ x_k \}_{k=1}^N$,
so these formulas extend by continuity to cover the general case
where some $m_k$ may coincide;
see the examples below for a sample of what they look like.

The matrices $\widehat T_k$ have the following conceptual interpretation.
Consider the boundary value problem \eqref{eq:mCH-xLaxBVP-} again,
but this time using the right regularization $(\alpha, \beta) = (0,1)$ in~\eqref{eq:mCH-alpha-beta}
(which we could express by saying that we are seeking solutions that are continuous from the right).
If we use hats to indicate this, the problem is
\begin{equation} \label{eq:mCH-xLaxBVP+}
  \widehat\Phi_x = \begin{pmatrix} 0 & h \\ -\lambda g & 0 \end{pmatrix} \widehat\Psi
  , \quad
  \widehat\phi_1(-\infty) = \widehat\phi_2(+\infty) = 0
  ,
\end{equation}
where the matrix product is interpreted according to
\begin{equation}
  \label{eq:mCH-right-regularization}
  \widehat\Phi(x) \, \delta(x-x_k) := \widehat\Phi(x_k^+) \, \delta(x-x_k)
  .
\end{equation}
Let us write $\widehat\Phi(x;\lambda)$ for the solution of the following
``backwards'' initial value problem, where we start at $x=\infty$
and go from right to left:
\begin{equation}\label{eq:mCH-xLax+}
  \widehat\Phi_x =
  \begin{pmatrix} 0 & h \\ -\lambda g & 0 \end{pmatrix}
  \widehat\Phi
  , \qquad
  \widehat\Phi(\infty) =
  \begin{pmatrix} 1 \\ 0 \end{pmatrix}
  .
\end{equation}
In the same manner as we defined $q_k(\lambda)$ and~$p_k(\lambda)$ in~\eqref{eq:mCH-def-q-p}, let
\begin{subequations}
\begin{equation}
  \begin{pmatrix} \widehat q_0(\lambda) \\ \widehat p_0(\lambda) \end{pmatrix}
  =
  \begin{pmatrix} 1 \\ 0 \end{pmatrix}
\end{equation}
and
\begin{equation}
  \label{eq:mCH-def-qhat-phat}
  \begin{pmatrix} \widehat q_k(\lambda) \\ \widehat p_k(\lambda) \end{pmatrix}
  =
  \widehat\Phi(x_{N+1-k}^-;\lambda)
  ,\qquad
  1 \le k \le N
  .
\end{equation}
\end{subequations}
Then the jump condition
\begin{equation}
  \widehat\Phi(x_k^+;\lambda) - \widehat\Phi(x_k^-;\lambda)
  = \begin{pmatrix} 0 & h_k \\ -\lambda g_k & 0 \end{pmatrix} \widehat\Phi(x_k^+;\lambda)
\end{equation}
(where \eqref{eq:mCH-right-regularization} has been used)
is equivalent to
\begin{equation*}
  \widehat\Phi(x_k^-;\lambda)
  = \begin{pmatrix} 1 & -h_k \\ \lambda g_k & 1 \end{pmatrix} \widehat\Phi(x_k^+;\lambda)
  ,
\end{equation*}
which upon changing $k$ to $N+1-k$ becomes
\begin{equation}
  \label{eq:mCH-transition-backwards}
  \begin{pmatrix} \widehat q_k(\lambda) \\ \widehat p_k(\lambda) \end{pmatrix}
  = \widehat T_k(\lambda) \begin{pmatrix} \widehat q_{k-1}(\lambda) \\ \widehat p_{k-1}(\lambda) \end{pmatrix}
  ,\qquad
  1 \le k \le N
  ,
\end{equation}
with the same matrix $\widehat T_k(\lambda)$ as in~\eqref{eq:mCH-matrix-C} above
(which explains our choice of notation there).
The spectrum of~\eqref{eq:mCH-xLaxBVP+}
is given by the roots of the polynomial $\widehat q_N(\lambda)$,
which from the explicit expressions can be seen to be identical
with~$p_N(\lambda)$,
so the left-continuous and right-continuous boundary value problems
~\eqref{eq:mCH-xLaxBVP-} and~\eqref{eq:mCH-xLaxBVP+}
have the same spectrum.
Note that even though the two admissible regularizations \eqref{eq:mCH-alpha-beta}
of the the original Lax pair
produce the same compatibility condition, they are both naturally involved
in setting up these two boundary value problems.

We conclude with two examples illustrating the complete solution for $N=2$ and $N=4$.

\begin{example}[The two-peakon solution]
  Even though this example is almost trivial, since the solution just consists of two parallel straight lines,
  one nevertheless learns about some general features of peakon solutions of the mCH equation.
  For $N=2$, the general solution formulas reduce to
  \begin{equation*}
    x_1(t) = \ln\left( \frac{a_1}{\lambda_1 m_1 (1 + \lambda_1 m_2^2)} \right)
    ,\quad
    x_2(t) = \ln\left( \frac{a_1 m_2}{1 + \lambda_1 m_2^2} \right)
    .
  \end{equation*}
  Recall that $a_1 = a_1(t) = a_1(0) \, e^{2t/\lambda_1}$,
  as there is only one eigenvalue, which implies that $x_1$ and~$x_2$
  have the same constant velocity, namely $2/\lambda_1$;
  this is of course also clear directly from the ODEs~\eqref{eq:mCH-twopeakonL}.
\end{example}

\begin{example}[The four-peakon solution]
  When $N=4$, there are two eigenvalues
  $0 < \lambda_1 < \lambda_2$,
  and two positive residues $a_1$ and~$a_2$,
  so the number of parameters in the spectral data
  matches the number of unknown functions $\{ x_k(t) \}_{k=1}^4$.
  The solution formulas take the following form, again with
  $a_k = a_k(t) = a_k(0) \, e^{2t/\lambda_k}$:
  \begin{equation}
    \begin{aligned}
      x_1(t) &= \ln \left( \frac{1}{m_1}\cdot\frac{A}{\lambda_1\lambda_2 \, B} \right)
      ,\\
      x_2(t) &= \ln \left( m_2 \cdot \frac{AC}{BD} \right)
      ,\\
      x_3(t) &= \ln \left( \frac{1}{m_3} \cdot \frac{CE}{DF} \right)
      ,\\
      x_4(t) &= \ln\left(m_4\cdot\frac{E}{F}\right)
      ,
    \end{aligned}
  \end{equation}
  where
  \begin{equation}
    \begin{aligned}
      A &= a_1a_2(\lambda_2-\lambda_1)^2
      ,\\
      B &= a_1\lambda_1(1+\lambda_2m_2^2)(1+\lambda_2m_3^2)(1+\lambda_2m_4^2)
      \\ & \quad
      + a_2\lambda_2(1+\lambda_1m_2^2)(1+\lambda_1m_3^2)(1+\lambda_1m_4^2)
      ,\\
      C &= a_1(1+\lambda_2m_3^2)(1+\lambda_2m_4^2)
      \\ & \quad
      + a_2(1+\lambda_1m_3^2)(1+\lambda_1m_4^2)
      ,\\
      D &= a_1\lambda_1(1+\lambda_2m_3^2)(1+\lambda_2m_4^2)
      \\ & \quad
      + a_2\lambda_2(1+\lambda_1m_3^2)(1+\lambda_1m_4^2)
      ,\\
      E &= a_1(1+\lambda_2m_4^2) + a_2(1+\lambda_1m_4^2)
      ,\\
      F &= (1+\lambda_1m_4^2)(1+\lambda_2m_4^2)
      .
    \end{aligned}
  \end{equation}
  It may happen, even in the pure peakon case,
  that the ordering condition $x_1(t) < x_2(t) < x_3(t) < x_4(t)$
  does not hold for all~$t$, but if it does (and sufficient conditions
  to guarantee this can be formulated),
  then the two eigenvalues uniquely determine two asymptotic velocities;
  as $t \to -\infty$,
  $x_1(t)$ and~$x_2(t)$ travel in parallel lines with same asymptotic velocity $2/\lambda_1$,
  while $x_3(t)$ and~$x_4(t)$ share the same asymptotic velocity $2/\lambda_2$,
  and as $t \to \infty$, it is the other way around.
\end{example}

\begin{remark}
  In general, when $N=2K$ is even, there are $K$ eigenvalues~$0<\lambda_1 < \dots < \lambda_K$,
  and provided that the solution is globally defined,
  the peakons pair up with $x_{2i-1}(t)$ and $x_{2i}(t)$ having the asymptotic
  velocity $2/\lambda_i$ as $t \to -\infty$ and $2/\lambda_{K+1-i}$ as $t \to \infty$.
  If $N=2K+1$ is odd, then the spectral data (and the solution formulas)
  also include the constant~$c$,
  and there will asymptotically be $K$ pairs with velocities $2/\lambda_i$,
  but also a lonesome peakon ``at the slow end'' with asymptotic velocity zero;
  that is,
  $x_{2K+1}(t)$ tends to a constant as $t \to -\infty$,
  with the other peakons pairing up as $(x_1,x_2)$, \ldots, $(x_{2K-1},x_{2K})$,
  while $x_1(t)$ tends to a constant as $t \to \infty$,
  with the other peakons pairing up as $(x_2,x_3)$, \ldots, $(x_{2K},x_{2K+1})$.
\end{remark}

\begin{remark}
  The ODE for $\dot x_k$ in the weak regularization
  differs from that in the Lax regularization
  by having an additional term $\frac23 m_k^2$.
  Moreover, these ODEs depend only on the differences $x_i - x_j$.
  Thus, in the particular case where all the constants~$m_k$ are equal,
  say $m_1 = \dots = m_N = \mu$,
  we can solve the weak peakon ODEs explicitly by letting
  $x_k(t) = f_k(t) + \frac23 \mu^2 t$ (for $1 \le k \le N$),
  where
  $x_k = f_k(t)$ (for $1 \le k \le N$)
  is given by the known solution formulas for the Lax peakon ODEs with the same~$m_k$.
\end{remark}

\section{Additional comments and pointers to the literature} \label{sec:guide}

This final section of our selective tour through the peakon world
contains various comments and remarks that were unsuitable for the main text.
Some are historical in nature,
some concern questions related to peakon equations but not directly to peakons,
some are guides to further reading,
and so on.
As we have already mentioned, the subject is vast,
and this is in no way intended to be a complete review,
but we hope that this section may at least provide some useful additional perspectives.

\subsection{The Camassa--Holm equation}
\label{sec:guide-CH}

The CH equation was put forward as a model of strongly dispersive shallow water waves
by Camassa and Holm~\cite{camassa-holm:1993:CH-orginal-paper} in 1993,
and further studied in a longer paper with Hyman~\cite{camassa-holm-hyman:1994:CH-new-integrable}
the following year.
Before that, in 1981, Fuchssteiner and
Fokas~\cite{fuchssteiner-fokas:1981:symplectic-structures-backlund-transformations-hereditary-symmetries,
  fuchssteiner:1981:lie-algebra-structure-nonlinear-evolution-equations-infinite-dimensional-abelian-symmetry-groups}
mentioned, somewhat indirectly, a family of integrable equations containing
the CH equation as a special case
(see comments by Fokas~\cite[p.~146]{fokas:1995:a-class-of-physically-important-integrable-equations}).
What is perhaps less known is that certain isospectral deformations of
the $x$-member of the Lax pair \eqref{eq:intro-CH-lax-x}, also in the
context of the string problem, were discussed already around 1979 by
Sabatier~\cite{sabatier:1980:needs}.
However, the isospectral deformations considered by him did not cover the CH case.
In particular, he considered one  specific deformation corresponding to
a linear dependence on~$\lambda$ in the time flow,
rather than linear in $1/\lambda$ as it appears in the actual CH flow~\eqref{eq:intro-CH-lax-t}.
Via the compatibility conditions, this choice leads to the nonlinear PDE
\begin{equation*}
  -\tfrac12 m_t = \partial_x (\partial_x^2-1) \, m^{-1/2}
  ,
\end{equation*}
which shows that this type of deformation is not very well suited for
the case of $m$ being a discrete measure.
The CH equation is included in a significantly extended
class of admissible isospectral deformations of an inhomogeneous string,
with general Robin boundary conditions, studied more recently by Szmigielski and
collaborators~\cite{colville-gomez-szmigielski:2016:isospectral-deformations-string,gorski-szmigielski:2018:isospectral-flows-inhomogeneous-string}.
This idea can also be extended to other interesting boundary
value problems. For example, one of the isospectral deformations of
the longitudinal vibrations of an elastic bar was shown by Chang and
Szmigielski~\cite{chang-szmigielski:2020:2mCH-overlapping-peakons-elastic-bar-isospectral}
to be a two-component modified CH equation.
Likewise, it was discovered recently by Beals and
Szmigielski~\cite{beals-szmigielski:2021p:2CH-euler-bernoulli-beam-noncommutative-continued-fractions}
that deforming the Euler--Bernoulli beam leads to a two-component
system akin to the CH equation.

Regarding the CH equation,
the need for a thorough understanding of the $N$-peakon solutions
governed by the ODEs~\eqref{eq:intro-CH-peakon-ode-explicit} was
clear already in the initial papers.
By taking the convolution of the first Lax equation \eqref{eq:intro-CH-lax-x}
with $\tfrac12 e^{-\abs{x}/2}$ and then inserting
the expression \eqref{eq:CH-peakon-m} for~$m$,
Camassa, Holm and Hyman~\cite{camassa-holm-hyman:1994:CH-new-integrable}
obtained a Lax matrix $L = \bigl( m_j \, e^{-\abs{x_i-x_j}} \bigr)_{i,j=1}^N$ whose
characteristic polynomial is time-invariant,
and whose coefficients therefore provide $N$ constants of motion for the peakon ODEs.
These constants of motion
are of degree $1,\dots,N$ in the variables~$m_k$,
and are easily shown to be functionally independent.
In the picture described in Section~\ref{sec:CH},
they appear as the coefficients of the time-independent polynomial
$A(\lambda) = A_N(\lambda)$
determined by \eqref{eq:CH-jump-relation-psi} with $(A_0,B_0)=(1,0)$;
see \eqref{eq:CH-time-dependence-A-B}.
Calogero and Françoise \cite{calogero-francoise:1996:cf-integrable}
proved that the constants of motion are also in involution, thereby verifying the
Liouville integrability of the CH peakon ODEs, and actually of the
entire family of Hamiltonian systems generated by
\begin{equation}
  \label{eq:guide-CF-hamiltonian}
  \begin{split}
    & H(x_1,\dots,x_N,m_1,\dots,m_N)
    \\ &
    = \frac12 \sum_{i,j=1}^N m_i m_j \,
    \Bigl( \mu \cos \nu (x_i-x_j)
    + \mu' \sin \nu \abs{x_i-x_j} \Bigr)
    ,
  \end{split}
\end{equation}
containing the CH peakon Hamiltonian~\eqref{eq:intro-CH-hamiltonian}
as the special case $(\mu,\mu',\nu) = (1,i,i)$.
Ragnisco and Bruschi~\cite{ragnisco-bruschi:1996:peakons-toda}
gave another proof of this result using the $r$-matrix formalism,
noticed that the $r$-matrix in the peakon case was the same as for the
finite nonperiodic Toda lattice previously studied by
Moser~\cite{moser:1975:finitely-many-point-masses-on-the-line, moser:1975:three-integrable},
and showed using Flaschka-type coordinates that
the CH peakon ODEs can be viewed as one of the commuting flows of the Toda hierarchy.
As we have described in Section~\ref{sec:CH},
these ODEs were then solved explicitly for arbitrary~$N$,
including a detailed analysis of peakon--antipeakon collisions,
in a series of papers by Beals, Sattinger and Szmigielski
\cite{beals-sattinger-szmigielski:1998:acoustic-scattering-KdV-hierarchy, beals-sattinger-szmigielski:1999:stieltjes, beals-sattinger-szmigielski:2000:moment},
by exploiting the connection to the inhomogeneous string problem
$\phi_{yy} = -\lambda \, g(y) \, \phi$
and to Stieltjes continued fractions.
It is fair to say that continued fractions figured prominently already in
Moser's work on the Toda lattice,
even though he did not fully use the theory developed by Stieltjes,
despite referring to the Gantmacher--Krein book
where the connection between Stieltjes's work and the inverse problem for
the discrete string problem is presented~\cite[Supplement II]{gantmacher-krein:2002:oscillation-matrices}.
The general inverse problem for an inhomogeneous string,
with a positive mass distribution~$g(y)>0$,
was studied in great detail by Krein already in the 1950s~\cite{krein:1952:inverse-string,kac-krein:1974:spectral-functions};
see the book by Dym and McKean~\cite{dym-mckean:1976:gaussian-processes}
for a full account of this work.
The precise relation between CH peakons and the finite Toda lattice,
and thus to Moser's work, was later examined using spectral methods
and the string interpretation of the mixed system of peakons and antipeakons~\cite{beals-sattinger-szmigielski:2001:peakons-strings-finite-Toda}.
Perhaps the most comprehensive Lie-algebraic picture of the place of
the CH peakon dynamics within the class of Toda-like systems is provided by the work of
Faybusovich and Gekhtman~\cite{faybusovich-gekhtman:2001:inverse-moment-problem-elementary-coadjoint-orbits}.
Here it is also appropriate to mention that
Camassa \cite{camassa:2000:CH-characteristic-variables, camassa:2003:CH-characteristics-initial-value-problem}
showed how to solve the CH $N$-peakon ODEs using factorization techniques
known from generalizations of Toda equations,
and that Li~\cite{li:2009:CH-long-time-behaviour-for-class-of-low-regularity-solutions-infinitely-many-peakons}
used the connection to the Toda lattice to investigate CH solutions consisting of a train
of countably many peakons.

One cannot help but notice that the expression $\frac12  e^{-\abs{x-y}}$
in the CH $N$-peakon Hamiltonian \eqref{eq:intro-CH-hamiltonian}
is the Green's function of the operator $1-\partial_x^2$
with vanishing boundary conditions as $x \to \pm \infty$.
The Calogero--Françoise Hamiltonian \eqref{eq:guide-CF-hamiltonian} is obtained
by replacing this expression with the general Green's function for the same operator,
and thus it can be interpreted as describing CH solutions taking the form
$u(x,t) = A(t) \, e^x + B(t) \, e^{-x}$ on \emph{all} intervals $x_k(t) < x < x_{k+1}(t)$,
including the ``outside'' intervals $x < x_1(t)$ and/or $x_N(t) < x$.
The equations of motion can be integrated in terms of Riemann theta functions.
See also Kardell~\cite{kardell:2015:CH-novikov-peakon-creation}
for elementary examples of such ``unbounded peakon solutions''.
Interestingly, the Calogero--Françoise system also shows up
when studying the \emph{periodic} CH peakon
problem~\cite{beals-sattinger-szmigielski:2002:CF-flows-periodic-peakons-2-peakon-2002,
  beals-sattinger-szmigielski:2005:CF-flows-periodic-peakons-N-peakon-2005},
and the same system also admits an intriguing geometric interpretation
in terms of Higgs fields~\cite{rayan-stanley-szmigielski:2020:CF-higgs-fields}.

The periodic peakon problem has also been studied more recently by
Eckhardt and Kostenko~\cite{eckhardt-kostenko:2020:inverse-spectral-problem-periodic-conservative-CH-multipeakons},
and in fact the whole subject of forward and inverse spectral problems related to the
CH equation has been greatly enriched in the last decade by their work
together with Teschl and other collaborators
\cite{eckhardt-teschl:2013:isospectral-problem-dispersionless-CH,
  eckhardt-kostenko:2016:inverse-spectral-problem-indefinite-strings,
  eckhardt-kostenko:2018:classical-moment-problem-generalized-indefinite-strings,
  eckhardt-kostenko-nicolussi:2020:trace-formulas-continuous-dependence-spectra-periodic-conservative-CH-flow,
  eckhardt:2017:IST-for-conservative-CH-flow-with-decaying-initial-data,
  eckhardt:2022:continued-fraction-expansions-herglotz-nevanlinna-functions-generalized-indefinite-strings-stieltjes-type}.
In the work perhaps most relevant for this article,
they present a very compelling interpretation of the mechanism
of the peakon--antipeakon collisions~\cite{eckhardt-kostenko:2014:CH-isospectral-problem-global-conservative-multipeakons}.
As was pointed out in Section~\ref{sec:CH}, at the time of the
collision the discrete signed string degenerates, changing the number
of point masses by either one or two.
As argued by Beals et~al.~\cite{beals-sattinger-szmigielski:2000:moment},
the peakon profile $u(x,t)$ can be continued past the collision time using the time-invariance
of the Sobolev $H^1(\R)$-norm and the meromorphic nature of the
positions $x_k$ and amplitudes $m_k$ as functions of~$t$,
but the overall picture of the collision from the point of view of
the string boundary value problem had been unclear until it was
addressed by Eckhardt and Kostenko.
Using ideas from the works of Bressan and Constantin~\cite{bressan-constantin:2007:global-conservative-CH}
and
Holden and Raynaud~\cite{holden-raynaud:2007:CH-global-conservative-Lagrangian},
they enlarge the original spectral problem
\begin{equation*}
  \bigl( \partial_x^2 - \tfrac14 \bigr) \, \psi = -\tfrac12 \lambda m \psi
\end{equation*}
to
\begin{equation*}
  \bigl( \partial_x^2 - \tfrac14 \bigr) \, \psi = -\tfrac12 \lambda m \psi - \tfrac14 \lambda^2 v \psi
  ,
\end{equation*}
where $v(x,t)$ is a measure which gets switched on exactly at the times of collisions,
where it absorbs some energy,
rendering a consistent global conservative solution for all~$t$.

Another line of research, studying the analytic aspects of the CH equation
from the point of view of PDE theory, rather than just the dynamical system
governing the peakon solutions,
was taken up by Constantin and Escher
\cite{constantin:1997:CH-periodic-cauchy-problem,
  constantin-escher:1998:CH-global-existence-blowup,
  constantin-escher:1998:CH-global-weak,
  constantin-escher:1998:wave-breaking-for-nonlinear-nonlocal-shallow-water-equations}
and by McKean~\cite{mckean:1998:CH-breakdown}.
This subject has grown enormously, and here we can only mention a very small selection of articles,
with a bias towards work related to peakons.
A useful survey of the work on wellposedness and other analytic aspects up to 2004
was given by Molinet~\cite{molinet:2004:CH-wellposedness-survey}.
In 2007, Bressan and Constantin introduced the distinction between
conservative~\cite{bressan-constantin:2007:global-conservative-CH}
and dissipative~\cite{bressan-constantin:2007:global-dissipative-CH}
global weak solutions to the initial value problem with $u(x,0)=u_0(x)$;
to put it shortly, weak solutions are not uniquely determined by the PDE alone,
but uniqueness can be recovered by imposing additional requirements on the function~$u$.
The basic difference is that the $H^1$-norm
\begin{equation*}
  E(t) = \int_\R \bigl( u(x,t)^2 + u_x(x,t)^2 \bigr) \, dx
\end{equation*}
is preserved for almost all~$t \ge 0$ for conservative solutions,
while it is nonincreasing for $t \ge 0$ for dissipative solutions.
Much work on clarifying the role of peakons in this context has been done by Holden, Raynaud and Grunert,
who also introduced the intermediate concept of $\alpha$-dissipative solutions
\cite{holden-raynaud:2007:CH-global-conservative-Lagrangian,
  holden-raynaud:2007:CH-global-conservative-multipeakon,
  holden-raynaud:2008:CH-periodic-conservative,
  holden-raynaud:2008:CH-global-dissipative-multipeakon,
  holden-raynaud:2009:CH-dissipative-solutions,
  grunert-holden-raynaud:2015:alpha-dissipative-2CH,
  grunert-holden:2016:CH-peakon-antipeakon-alpha-dissipative}.
The non-uniqueness of weak solutions makes the numerical analysis of the CH equation quite
subtle; for a comprehensive list of references covering the multitude of numerical methods
that have been suggested, see the recent paper by
Galtung and Grunert~\cite{galtung-grunert:2021:CH-numerical-study-of-variational-discretizations}.
The stability of peakons, i.e., the question of whether a solution starting out close to
a peakon solution remains close to it,
has been studied by for example Constantin and Strauss
\cite{constantin-strauss:2000:CH-stability-peakons,
  constantin-strauss:2002:CH-stability-solitons},
Lenells \cite{lenells:2004:variational-approach-stability-periodic-peakons:JNMP,
  lenells:2004:stability-periodic-peakons:IMRN},
El Dika and Molinet \cite{eldika-molinet:2009:CH-stability-multi-peakon-antipeakon,
  eldika-molinet:2009:CH-stability-multipeakon}
and recently by Natali and Pelinovsky~\cite{natali-pelinovsky:2020:CH-instability-of-H1-stable-peakons},
whose article is also a good source of up-to-date references to analytic works on the CH equation in general.

The full CH equation \eqref{eq:intro-CH-kappa-expanded},
with a dispersive term $2 \kappa u_x$ where $\kappa>0$,
admits smooth multisoliton solutions vanishing at infinity
\cite{parker:2004:CH-direct-method-1,
  parker:2005:CH-direct-method-2,
  parker:2005:CH-direct-method-3,
  matsuno:2005:CH-multisolitons-parametric-representation,
  holm-ivanov:2010:CH-smooth-and-peaked-solitons,
  holm-ivanov:2011:CH-smooth-peaked-solitons-applications,
  ivanov-lyons-orr:2017:CH-solitons-dressing-method}.
It is possible to recover the determinantal formulas for the $N$-peakon solutions of the dispersionless
CH equation~\eqref{eq:intro-CH-expanded}
from the formulas for these smooth $N$-soliton solutions
by letting $\kappa \to 0$;
see Parker and Matsuno~\cite{parker-matsuno:2006:CH-solitons-peakon-limit}
for the case $N=2$,
and Matsuno~\cite{matsuno:2007:CH-multisolitons-peakon-limit} for general~$N$.
Corresponding statements ought to hold true also for the other peakon PDEs
that we have treated in this article,
but for those equations
it is still an open problem to actually carry out this limiting procedure
for arbitrary~$N$,
since the technical details are much more complicated than for the CH equation.

As a sample of the many other aspects of the CH equation that have been studied, we may also briefly mention
water wave theory
\cite{fokas-liu:1996:asymptotic-integrability-of-water-waves,
  dullin-gottwald-holm:2001:integrable-shallow-water-linear-nonlinear-dispersion,
  dullin-gottwald-holm:2003:CH-KdV5-other-asymptotically-equivalent-shallow-water,
  dullin-gottwald-holm:2004:asymptotically-equivalent-shallow-water,
  johnson:2002:CH-KdV-water-waves,
  johnson:2003:classical-water-waves,
  ionescukruse:2007:variational-derivation-CH-shallow-water-equation,
  constantin-lannes:2009:hydrodynamical-CH-DP},
geometric approaches
\cite{alber-camassa-holm-marsden:1994:geometry-peakons-billiard-solutions,
  misiolek:1998:CH-as-geodesic-flow-on-Bott-Virasoro-group,
  constantin:2000:CH-geometric-approach,
  misiolek:2002:CH-periodic-classical-solutions},
travelling wave solutions~\cite{lenells:2005:CH-traveling-waves, geyer-villadelprat:2015:CH-wavelength-of-smooth-periodic-traveling-waves},
algebro-geometric solutions
\cite{constantin-mckean:1999:CH-shallow-water-equation-on-the-circle,
  gesztesy-holden:2003:CH-hierarchy-alggeo-solutions,
  qiao:2003:CH-hierarchy-alggeo-symplectic-submanifold,
  kalla-klein:2012:CH-construction-alggeo-solutions-numerical-evaluation},
inverse scattering and other integrability aspects
\cite{gilson-pickering:1995:factorization-painleve-analysis-nonlinear-thirdorder-PDE-including-CH,
  constantin:1997:CH-hamiltonian-structure:Expositiones,
  constantin:1997:CH-periodic-spectral-problem,
  constantin:1998:CH-inverse-spectral-problem,
  constantin:1998:CH-periodic-finite-gap-solutions-quasiperiodicity,
  schiff:1998:CH-loop-group-approach,
  fisher-schiff:1999:CH-conserved-quantities-initial-value-problem,
  constantin-lenells:2003:CH-inverse-scattering-approach:PhysLettA,
  constantin-lenells:2003:CH-inverse-scattering-approach:JNMP,
  lenells:2005:CH-conservation-laws,
  li:2008:CH-factorization-problem-on-hilbert-schmidt-group},
and the study of initial--boundary value problems and asymptotics via Riemann--Hilbert problems
\cite{boutetdemonvel-shepelsky:2005:CH-half-line,
  boutetdemonvel-shepelsky:2006:CH-riemann-hilbert-line,
  boutetdemonvel-shepelsky:2008:CH-riemann-hilbert-half-line}.

\subsection{The Degasperis--Procesi equation}
\label{sec:guide-DP}

As we mentioned in the introduction,
the DP equation \eqref{eq:intro-DP} was discovered around 1998 by Degasperis and Procesi
when searching for PDEs of a certain form satisfying an asymptotic integrability
condition~\cite{degasperis-procesi:1999:asymptotic-integrability};
later it has also been derived as an approximate model for
shallow water waves~\cite{johnson:2003:classical-water-waves,
  dullin-gottwald-holm:2004:asymptotically-equivalent-shallow-water,
  constantin-lannes:2009:hydrodynamical-CH-DP}.
The study of this equation began in earnest in 2002, when
Degasperis, Holm and Hone formulated a Lax pair, conservation laws
and a bi-Hamiltonian formulation,
pointed out that the equation has peakon solutions,
and solved the two-peakon case explicitly~\cite{degasperis-holm-hone:2002:new-integrable-equation-DP}.
Soon after that, we set out to study the general $N$-peakon case using inverse spectral
methods~\cite{lundmark-szmigielski:2003:DPshort}.
Since the Lax pair \eqref{eq:DP-lax} contains a third-order differential operator,
the DP case lies outside of the theory of self-adjoint operators,
and is therefore much more difficult to analyze,
but eventually, in 2005, the peakon problem for the DP equation was completely solved,
at least in the pure peakon case~\cite{lundmark-szmigielski:2005:DPlong}.
As we have described in Section~\ref{sec:DP},
the role previously played by the ordinary string
\begin{equation*}
  \phi_{yy} = -\lambda \, g(y) \, \phi
\end{equation*}
is now played by the cubic string
\begin{equation*}
  \phi_{yyy} = -\lambda \, g(y) \, \phi
  ,
\end{equation*}
self-adjointness is replaced by the Gantmacher--Krein theory of oscillatory kernels,
Padé approximation of the single Weyl function $W(\lambda)$ becomes Hermite--Padé approximation
of the pair of Weyl functions $(W(\lambda),Z(\lambda))$
connected by the relation~\eqref{eq:DP-Z-W-symmetry-relation},
and the orthogonal polynomials known from Stieltjes's theory are replaced by
Cauchy biorthogonal polynomials
(see Remark~\ref{rem:DP-CBOP},
and also recent further developments by
Fidalgo, Lagomasino, Peralta and Szmigielski~\cite{lopezlagomasino-medinaperalta-szmigielski:2019:mixed-type-hermite-pade-approximation-inspired-by-DP, fidalgo-lopezlagomasino-medinaperalta:2020:asymptotics-cauchy-biorthogonal-polynomials, medinaperalta:2021:matrix-cauchy-biorthogonal-polynomials}).
Moreover, generalizing the connection between CH peakons and the Toda lattice,
the DP peakons are related to a finite Toda lattice of CKP type;
see Chang, Hu and Li~\cite{chang-hu-li:2018:DP-peakons-Toda-lattice-CKP}.
(See also other related works by Chang and collaborators
\cite{chang-hu-li:2018:moment-modification-multipeakons,
  chang-he-hu-li:2018:partial-skew-orthogonal-polynomials-integrable-lattices-Pfaffian,
  chang-hu-szmigielski:2016:two-component-mCH-Kac-vanMoerbeke,
  chang-hu-szmigielski-zhedanov:2020:isospectral-frobenius-stickelberger-thiele-polynomials,
  chang-chen-hu:2014:generalized-nonisospectral-CH-peakons}.)

Like the CH equation, the DP equation has been the subject of much research from the PDE community,
and the literature is too large to survey here.
Many results have been proved first for the DP equation itself,
either on the form \eqref{eq:intro-DP-expanded} or with an additional term $2 \kappa u_x$ as
in the CH equation~\eqref{eq:intro-CH-kappa-expanded},
and later extended to modified versions of the DP equation,
or to classes of equations containing it as a special case,
such as the ``$b$-family'' \eqref{eq:intro-b-family}.
Early results about periodic and non-periodic strong and weak solutions
were published by Yin
\cite{yin:2003:DP-cauchy-problem:Illinois,
  yin:2003:DP-global-existence-for-new-periodic-integrable-equation,
  yin:2004:DP-global-solutions:Indiana,
  yin:2004:DP-global-weak-solutions-for-new-periodic-integrable-equation-with-peakons};
these weak solutions were required to lie at least in the Sobolev space $H^1$ (with respect to~$x$),
which is enough to handle peakons,
but in fact the DP equation admits much less regular weak solutions
that need not even be continuous.
This is due to the fact that \eqref{eq:intro-b-family} can be rewritten first as
\begin{equation}
  \label{eq:b-family1}
  \begin{split}
    0&=m_t+m_x u+bmu_x
    \\
    &= (u-u_{xx})_t + (b+1)uu_x - b u_x u_{xx} - uu_{xxx}
    \\
    &= (1-\partial_x^2)[u_t+(\tfrac12 u^2)_x] + b (\tfrac12 u^2)_x + (3-b) (\tfrac12 u_x^2)_x
  \end{split}
\end{equation}
and then (say for solutions on the real line vanishing at infinity) as
\begin{equation}
  \label{eq:b-family2}
  0 = u_t + \partial_x \bigl[ \tfrac12 u^2 + \tfrac12 e^{-\abs{x}} * (\tfrac{b}{2} u^2 + \tfrac{3-b}{2} u_x^2) \bigr]
  ,
\end{equation}
where the term $u_x^2$ is absent precisely in the DP case $b=3$.
To obtain uniqueness of such weak solutions, the PDE is supplemented by a so-called entropy condition.
Entropy solutions were studied by Coclite and Carlsen
\cite{coclite-karlsen:2006:DPwellposedness,
  coclite-karlsen:2006:DP-semigroup-of-solutions,
  coclite-karlsen:2007:DPuniqueness,
  coclite-karlsen:2008:DP-bounded-solutions,
  coclite-karlsen:2015:periodic-DP-wellposedness-and-asymptotics},
whose work was influential in connection with the discovery of shockpeakons~\cite{lundmark:2007:DP-shockpeakons},
and at about the same time by Liu and Wang~\cite{liu-wang:2006:DP-local-wellposedness-new-integrable-equation}.
The formation of shockpeakons at peakon--antipeakon collisions
was investigated by Szmigielski and
Zhou~\cite{szmigielski-zhou:2013:DP-peakon-antipeakon, szmigielski-zhou:2013:DP-colliding-peakons-shock-formation}.
An explicit formula for a periodic shockpeakon solution was given by
Escher, Liu and Yin~\cite{escher-liu-yin:2007:periodic-DP-shock-waves-blowup-phenomena},
who used the terminology \emph{``strong'' weak solutions} for the continuous ($H^1$) weak solutions
considered earlier.
We may also mention here
the work by Constantin, Ivanov and Lenells on the inverse scattering transform for the DP equation
\cite{constantin-ivanov-lenells:2010:DP-inverse-scattering-transform},
as well as various articles on explicit solutions (travelling waves, solitons, etc.)
\cite{lenells:2005:DP-traveling-waves,
  vakhnenko-parkes:2004:DP-periodic-and-solitarywave-solutions,
  vakhnenko-parkes:2004:connection-DP-vakhnenko,
  matsuno:2005:DP-N-soliton-solution,
  matsuno:2005:DP-multisolitons-peakon-limit,
  chen-tang:2006:DP-new-type-of-bounded-waves,
  vakhnenko-parkes:2006:generalized-DP-solutions,
  yu-tian-wang:2006:DP-bifurcation-and-peakon,
  zhang-qiao:2007:DP-cuspons-smooth-solitons-under-inhomogeneous-boundary-condition,
  yin-tian-fan:2010:DP-limiting-behavior-of-smooth-periodic-waves,
  zhang-qiao:2010:DP-cusp-solitons,
  stalin-senthilvelan:2012:DP-multiloop-solitons,
  feng-maruno-ohta:2013:DP-tau-functions,
  hou-zhao-fan-qiao:2013:DP-hierarchy-alggeo-solutions,
  constantin-ivanov:2017:DP-dressing-method,
  li-wang-kuang:2020:multisolitons-of-DP-and-its-shortwave-limit-darboux-transformation-approach,
  mao-wang:2020:DP-backlund-transformations},
stability of peakons
\cite{lin-liu:2009:stability-of-DP-peakons,
  kabakouala:2015:stability-in-energy-space-of-DP-N-peakons,
  kabakouala:2016:remark-on-stability-of-DP-peakons,
  molinet:2019:rigidity-result-for-holm-staley-b-family-with-application-to-asymptotic-stability-of-DP-peakon,
  khorbatly-molinet:2020:DP-orbital-stability-antipeakon-peakon-profile,
  khorbatly:2022:DP-asymptotic-stability-of-antipeakon-peakon-profile},
general integrability aspects
\cite{mikhailov-novikov:2002:perturbative,
  hone-wang:2003:prolongation-algebras,
  qiao:2004:integrable-hierarchy-3x3-constrained-systems-parametric-solutions,
  ivanov:2005:integrability-class-nonlinear-dispersive-wave-equations,
  kolev:2009:DP-some-geometric-investigations,
  escher-kolev:2011:DP-as-a-nonmetric-euler-equation,
  tiglay-vizman:2011:generalized-euler-poincare-equations-on-lie-groups,
  kang-liu-olver-qu:2017:Liouville-correspondences-integrable-hierarchies,
  kang-liu-olver-qu:2020:Liouville-correspondences-multicomponent-integrable-hierarchies},
Riemann--Hilbert methods
\cite{boutetdemonvel-shepelsky:2013:DP-riemann-hilbert-approach, lenells:2013:DP-half-line},
numerical methods
\cite{hoel:2007:DP-numerical-scheme-multishockpeakons,
  coclite-karlsen-risebro:2008:DP-numerical-schemes-discontinuous-solutions,
  xia:2014:DP-fourier-spectral-methods-discontinuous-solutions,
  gao-zhang-zhang:2019:DP-adaptive-moving-knots-meshless-method,
  zhang-wang-yang:2019:DP-structure-preserving-methods,
  guo-yang-zhang-wang-song:2019:splitting-method-DP-optimized-WENO-scheme-fourier-pseudospectral-method,
  guo-zhang-yang-wang-song:2019:DP-high-order-operator-splitting-method},
and further analytic developments
\cite{liu-yin:2006:DP-global-existence-and-blowup,
  liu-yin:2007:DP-blowup-phenomena,
  escher-yin:2007:DP-IVBP,
  escher-yin:2008:DP-IBVP:Banach,
  henry:2008:DP-persistence-properties,
  christov-hakkaev:2009:periodic-b-family-cauchy-problem-and-DP-nonuniform-continuity,
  coclite-karlsen-kwon:2009:IBVP-conservation-laws-with-source-terms-and-application-to-DP,
  chen:2011:on-solutions-to-DP,
  fu-liu:2011:periodic-DP-nonuniform-dependence-on-initial-data,
  gui-liu:2011:DP-cauchy-problem,
  himonas-holliman:2011:DP-wellposedness,
  tian-chen-liu-gao:2011:low-regularity-solutions-of-periodic-general-DP,
  guo:2012:DP-wavebreaking-phenomena-decay-properties-limit-behaviour,
  guo-jin:2013:DP-persistence-momentum-support,
  himonas-holliman-grayshan:2014:DP-norm-inflation-illposedness,
  chen-guo:2015:DP-asymptotic-profile,
  chen-guo-liu-qu:2016:DP-novikov-mCH-blowup,
  brandolese:2016:DP-liouville-theorem,
  wu:2018:finite-time-singularities-for-class-of-DP-equations,
  feola-giuliani-pasquali:2019:integrability-DP-control-of-sobolev-norms-birkhoff-resonances,
  li-liu-wu:2020:DP-spectral-stability-of-smooth-solitary-waves,
  pei:2020:DP-exponential-decay-symmetry-solitary-waves}.

\subsection{The Novikov equation}
\label{sec:guide-Novikov}

The Novikov equation \eqref{eq:intro-Novikov} was discovered by Vladimir Novikov \cite{novikov:2009:generalizations-of-CH}
in a classification of cubically nonlinear PDEs admitting infinitely many symmetries.
Hone and Wang \cite{hone-wang:2008:cubic-nonlinearity} found a Lax pair
and a bi-Hamiltonian structure for the PDE,
and studied the two-peakon dynamics.
It is worth noting \cite[p.~3]{hone-wang:2008:cubic-nonlinearity}
that it was actually the mCH equation \eqref{eq:intro-mCH}
that prompted Hone and Wang to ask Novikov to search for other Camassa--Holm type equations
with cubic nonlinearities.
They also provided a Lax pair for the $N$-peakon ODEs~\eqref{eq:intro-Novikov-peakon-ode},
of the form
\begin{equation*}
  \frac{dL}{dt} = [M, L]
  ,
\end{equation*}
where
\begin{equation*}
  \begin{aligned}
    L &= SPEP
    ,\\
    S_{ij} & = \sgn(x_i-x_j)
    = \sgn(i-j)
    \quad\text{(if $x_1 < \dots < x_N$)}
    ,\\
    P &= \operatorname{diag}(m_1,\dots, m_N)
    ,\\
    E_{ij} &= \exp(-\abs{x_i-x_j})
    .
  \end{aligned}
\end{equation*}
However, as they pointed out, this Lax pair does not produce sufficiently many constants of motion
to prove Liouville integrability of the $N$-peakon ODEs;
the coefficients in the characteristic polynomial of~$L = SPEP$
are expressions of degree $4n$ in the variables~$m_k$,
with the expected invariants of degree $4n-2$ missing.
The mystery was resolved in our paper with Hone~\cite{hone-lundmark-szmigielski:2009:novikov},
and the discrepancy turned out to be connected to the problem of the Lax pair being ill-defined
in the peakon sector.
When the Lax pair is defined rigorously as a distributional Lax pair, then the Lax matrix
which is derived from the Lax pair by evaluation at the points of the support of the measure $m$
reads $L = TPEP$, where
\begin{equation}
  \label{eq:guide-matrix-T}
  T_{ij} = 1 + \sgn(i-j)
  .
\end{equation}
Note that this matrix $T = I+S$, a totally nonnegative lower triangular matrix,
is quite different from the skew-symmetric matrix~$S$ in the previous formula $L=SPEP$.
The characteristic polynomial of the new Lax matrix $L = TPEP$
indeed provides the required constants of motion.
Previously it had been observed for small values of~$N$ that the $k$th constant of motion,
as obtained from the eigenvalue problem for the dual cubic string (see Section~\ref{sec:Novikov}),
was given by the sum of \emph{all} $k \times k$ minors (principal and non-principal) of the symmetric
$N \times N$ matrix $PEP$.
On the other hand, the coefficients in the characteristic polynomial of $TPEP$ are of course sums
of $k \times k$ principal minors of $TPEP$.
When trying to reconcile these result, on Canada's national holiday, July 1, 2008,
we stumbled upon the following curious combinatorial fact (the ``Canada Day Theorem'')
\cite{hone-lundmark-szmigielski:2009:novikov,gomez-lundmark-szmigielski:2013:CDT}:
\begin{quote}
  Let $T$ be the $N \times N$ matrix defined by \eqref{eq:guide-matrix-T}.
  For any symmetric $N \times N$ matrix~$X$
  and for any $k = 1, \dots, N$,
  the sum of the $k \times k$ \emph{principal} minors of~$TX$
  equals the sum of \emph{all} $k \times k$ minors of~$X$.
\end{quote}
Moreover, in the pure peakon case where all $m_k$ are positive, the Lax matrix $L = TPEP$ is an
\emph{oscillatory} matrix in the sense of Gantmacher and Krein.
This injection of positivity into the problem shows again that pure peakons belong to the class of
oscillatory systems,
defined by Gantmacher and Krein as an overarching concept for
mechanical vibrational systems like strings, rods, beams, shafts
and other types of elastic objects.
The focus of this theory was on the so-called \emph{oscillatory properties}
of eigenvalue problems known from the theory of small oscillations,
such as the spectrum being positive and simple and
the $j$th eigenfunction having $j$ nodes (with the lowest one corresponding to $j=0$);
the complete list is in the Gantmacher--Krein book \cite[p.~2]{gantmacher-krein:2002:oscillation-matrices}.
They identified a class of kernels, the \emph{oscillatory kernels},
which automatically lead to eigenvalue problems possessing these oscillatory properties.
It is important to emphasize that the mechanical system are almost exclusively described
by symmetric kernels. However, one of the surprising results of the analysis was that
oscillatory kernels do not have to be symmetric.  
The connection to vibrational problems is perhaps obvious for CH peakons,
which are closely connected to the self-adjoint string problem,
but other peakon equations are in general non-self-adjoint, yet many of them are oscillatory.

For mixed peakon--antipeakon solutions of Novikov's equation, the situation is more complicated;
see Remark~\ref{rem:Novikov-peakon-antipeakon}.

Weak solutions of Novikov's equation are usually defined
by rewriting the PDE \eqref{eq:intro-Novikov} first as
\begin{equation}
  \begin{split}
    0 &=
    m_t + \bigl( (u m)_x + 2 u_x m \bigr) \, u
    \\ &
    = u_t - u_{xxt} + 4 u^2 u_x -  u^2 u_{xxx} - 3 u u_x u_{xx}
    \\ &
    = (1 - \partial_x^2)(u_t + u^2 u_x) + \partial_x \bigl( u^3 + \tfrac32 u u_x^2 \bigr) + \tfrac12 u_x^3
  \end{split}
\end{equation}
and then as the nonlocal transport equation
\begin{equation}
  \label{eq:guide-Novikov-weak}
  \begin{split}
    0 &=
    u_t + u^2 u_x
    \\ &
    + \partial_x (1-\partial_x^2)^{-1} \bigl( u^3 + \tfrac32 u u_x^2 \bigr)
    \\ &
    + (1-\partial_x^2)^{-1} \bigl( \tfrac12 u_x^3 \bigr)
    ,
  \end{split}
\end{equation}
where (for solutions on the real line) the operator $(1-\partial_x^2)^{-1}$ is realized
as convolution with $\tfrac12 e^{-\abs{x}}$.
Note that only $u$ and~$u_x$ appear in \eqref{eq:guide-Novikov-weak},
not the second derivative $u_{xx}$,
so this formulation has no problems handling peakon solutions,
which have a weak first derivative $u_x$.
Chen, Chen and Liu~\cite{chen-chen-liu:2018:novikov-global-conservative-weak-existence-uniqueness}
have studied conservative weak solutions,
where an important role is played by the quantities
\begin{equation}
  E(t) = \int_{\mathbb{R}} (u^2 + u_x^2) \, dx
\end{equation}
and
\begin{equation}
  F(t) = \int_{\mathbb{R}} (u^4 + 2 u^2 u_x^2 - \tfrac13 u_x^4) \, dx
  ,
\end{equation}
which are conserved for smooth solutions.
The natural function space to use in this context is
$H^1(\R) \cap W^{1,4}(\R)$,
since
$u(\cdot, t) \in H^1(\R) = W^{1,2}(\R)$
means precisely that $E(t)$ is finite,
while
$u(\cdot, t) \in W^{1,4}(\R)$
means that $u^4$ and $u_x^4$ are integrable,
and hence so also $u^2 u_x^2$ by Cauchy--Schwarz, so that $F(t)$ is finite.
The precise definition of a conservative weak solution is too technical to describe here,
but Chen et~al. show that if $u_0 \in H^1 \cap W^{1,4}$ is absolutely continuous, then there is a unique global
conservative weak solution to the initial value problem with $u(x,0)=u_0(x)$,
with the property that $E(t) = E(0)$ and $F(t) \ge F(0)$ for all $t \ge 0$, with
$F(t) = F(0)$ for almost all~$t \ge 0$;
the higher-order ``energy'' in $F(t)$ may become concentrated
(from the integral of $-\tfrac13 u_x^4$), for example at peakon--antipeakon collisions,
but it immediately returns to its previous value again.

The literature surrounding the Novikov equation is not yet quite as overwhelming as for the CH and DP equations,
but there is no shortage of articles about PDE-analytic questions,
in addition to the one just mentioned
\cite{ni-zhou:2011:novikov-wellposedness-and-persistence-properties,
  tiglay:2011:novikov-periodic-cauchy-problem,
  wu-yin:2011:novikov-global-weak-solutions,
  jiang-ni:2012:novikov-blowup,
  himonas-holliman:2012:novikov-cauchy-problem,
  yan-li-zhang:2012:novikov-cauchy-problem:JDiffEqu,
  wu-yin:2012:novikov-wellposedness-global-existence,
  grayshan:2013:novikov-data-to-solution-map,
  himonas-holmes:2013:novikov-holder-continuity,
  lai:2013:novikov-global-weak-solutions,
  yan-li-zhang:2013:novikov-cauchy-problem:NoDEA,
  lai-li-wu:2013:novikov-global-solutions,
  wu-yin:2013:novikov-note-on-cauchy-problem,
  chen-guo-liu-qu:2016:DP-novikov-mCH-blowup,
  wu-guo:2016:periodic-novikov-global-wellposedness,
  guo:2017:novikov-asymptotic-profile-momentum-support,
  cai-chen-chen-shen:2018:novikov-lipschitz-metric,
  himonas-holliman-kenig:2018:novikov-2-peakon-illposedness,
  zhou-yang-mu:2018:novikov-global-dissipative-solutions,
  wu:2018:novikov-global-analytic-solutions-and-traveling-wave-solutions,
  coclite-diruvo:2019:note-on-convergence-novikov,
  ma-cao-guo:2019:large-time-behavior-momentum-support-novikov-type-equation,
  shen:2019:novikov-weak-solution-optimal-control,
  li-li-zhu:2020:novikov-nonuniform-dependence-besov-spaces}.
Stability of peakons has been considered by several researchers
\cite{liu-liu-qu:2014:novikov-stability-of-peakons,
  wang-tian:2018p:novikov-stability-periodic-peakons,
  palacios:2020:novikov-peakons-asymptotic-stability,
  palacios:2021:novikov-train-of-peakons-orbital-and-asymptotic-stability,
  chen-lian-wang-xu:2021:novikov-rigidity-property-asymptotic-stability-of-peakons,
  chen-pelinovsky:2021:novikov-W-1-infty-instability},
and likewise solitons
\cite{matsuno:2013:novikov-multisolitons-peakon-limit,
  li:2014:novikov-exact-cuspon-and-compactons,
  pan-yi:2015:novikov-soliton-solutions,
  pan-li:2016:novikov-further-results-on-smooth-and-nonsmooth-solitons,
  wu-li-li:2019:novikov-solitons-negative-flow,
  mao:2021:novikov-backlund-transformation-and-applications,
  zheng-xiao-ouyang:2021:novikov-smooth-soliton-periodic-cuspon}
and integrability aspects
\cite{stalin-senthilvelan:2011:novikov-prolongation-structure,
  bozhkov-freire-ibragimov:2014:novikov-group-analysis,
  kardell:2015:CH-novikov-peakon-creation,
  boutetdemonvel-shepelsky-zielinski:2016:novikov-riemann-hilbert-approach,
  rasin-schiff:2019:simple-looking-relative-of-novikov-hirotasatsuma-sawadakotera}.
However, the numerical analysis community has not yet jumped on the bandwagon;
we are only aware of two (rather similar) studies
\cite{chen-yang-li-zhu:2017:novikov-conservative-finite-difference-scheme,
  chen-zhu-yang:2017:generalized-novikov-conservative-finite-difference-scheme}.

\subsection{The Geng--Xue equation}
\label{sec:guide-GX}

As already mentioned in Sections~\ref{sec:intro} and~\ref{sec:GX},
the two-component Geng--Xue equation \eqref{eq:intro-GX} was
obtained by modifying the $3 \times 3$ matrix Lax pair for Novikov's equation,
where the quantity $m = u - u_{xx}$ appears in two of the entries in the $x$-equation~\eqref{eq:Novikov-lax-x}.
Geng and Xue~\cite{geng-xue:2009:GX-peakon-equation-cubic-nonlinearity}
changed $m$ to $n = v - v_{xx}$ in one of these two entries, and also
changed some $u$ to~$v$ and some $m$ to~$n$ in the more complicated $t$-equation~\eqref{eq:Novikov-lax-t},
to obtain the Lax pair~\eqref{eq:GX-laxI}, for which the GX equation is the compatibility condition.
Since the GX equation is symmetric with respect to the interchange of $u$ and~$v$,
it is also the compatibility condition of the ``twin'' Lax pair ~\eqref{eq:GX-laxII}.
As we saw in Section~\ref{sec:GX}, both Lax pairs need to be used in the inverse
spectral approach to computing explicit peakon solutions
\cite{lundmark-szmigielski:2016:GX-inverse-problem,
  lundmark-szmigielski:2017:GX-dynamics-interlacing,
  shuaib-lundmark:2019:GX-noninterlacing}.
The relevant approximation problems are again of mixed Hermite--Padé type and the resulting
Cauchy biorthogonal polynomials involve two spectral measures which are independent of each other,
which distinguishes the Geng--Xue equation from the DP and Novikov equations,
where the second spectral measure is identical to the first one (for Novikov)
or related to it in a very simple way (for DP).

Formally the GX equation reduces to the Novikov equation when $v=u$ (and to the DP equation when $v=1$),
but one needs to be careful when it comes to weak solutions.
In expanded form, the system reads
\begin{equation}
  \begin{aligned}
    0 &= u_t - u_{xxt} + (4 u u_x - 3 u_x u_{xx} - u u_{xxx}) \, v
    , \\
    0 &= v_t - v_{xxt} + (4 v v_x - 3 v_x v_{xx} - v v_{xxx}) \, u
    .
  \end{aligned}
\end{equation}
In order to define a general concept of weak solution that would
encompass peakons in the same manner as for Novikov's equation above,
we would like to write these equations as nonlocal equations for $u_t$
and $v_t$ with no explicit appearance of $u_{xx}$ or $v_{xx}$.
For the first equation, that would require expressing the thrice differentiated terms
$3 u_x u_{xx} v + u u_{xxx} v$ as a linear combination
\begin{equation*}
  \begin{split}
    &
    a (u^2 v)_{xxx}
    + b (u^2 v_x)_{xx}
    + c (uu_x v)_{xx}
    \\ & \quad
    + d (uu_x v_x)_x
    + e (u_x^2 v)_x
    + f u_x^2 v_x
    .
  \end{split}
\end{equation*}
This leads to
\begin{equation*}
  \begin{split}
    &
    3 u_x u_{xx} v + u u_{xxx} v
    \\ & =
    (a+b) \, u^2 \, v_{xxx}
    + (6a+4b+c+d) \, u u_x v_{xx}
    \\ & \quad
    + (6a+2b+2c+d) \, u u_{xx} v_x
    \\ & \quad
    + (6a+2b+2c+d+e+f) \, u_x^2v_x
    \\ & \quad
    + (2a+c) \, u u_{xxx}v
    + (6a+3c+2e) \, u_x u_{xx} v
    ,
  \end{split}
\end{equation*}
so that $2a+c=1$ and $6a+3c+2e=3$, while the remaining coefficients are zero.
Unfortunately this means that
$0 = (6a+4b+c+d) - (6a+2b+2c+d) = 2b-c$ and hence $1 = 1 + 0 = (2a+c) + (2b-c) = 2a+2b = 2(a+b) = 0$, a contradiction,
so the linear system is inconsistent and our task is impossible.
On the other hand, if we write the system as
\begin{equation}
  \begin{aligned}
    0 &= m_t + v \, (4 - \partial_x^2) \, \partial_x \bigl( \tfrac12 u^2 \bigr)
    , \\
    0 &= n_t + u \, (4 - \partial_x^2) \, \partial_x \bigl( \tfrac12 v^2 \bigr)
  \end{aligned}
\end{equation}
and consider only \emph{non-overlapping} peakons,
meaning that no peakon in~$u$ is located at the same site as a peakon in~$v$,
then the expression $(4 - \partial_x^2) \, \partial_x \bigl( \tfrac12 u^2 \bigr)$ in the first equation
will give rise to singular distributions (Dirac deltas and derivatives thereof)
at the sites of the peakons in~$u$,
but since the function $v$ is infinitely differentiable at all those points,
the product
$v \cdot (4 - \partial_x^2) \, \partial_x \bigl( \tfrac12 u^2 \bigr)$
is well-defined,
and similarly in the second equation, of course.
In fact, as long as there is no overlapping, the same reasoning shows that
the GX equation even admits \emph{shockpeakon} solutions in this distributional
sense~\cite{lundmark-szmigielski:2017:GX-dynamics-interlacing}.
While our argument here refers specifically to peakons, it uses only the standard definition
of the product between a smooth function and a distribution,
so one may speculate that there could be some general definition of weak solution
that would allow one component to be ``worse than usual'' at points where the other one is ``good enough'',
in order to accommodate at least non-overlapping peakon and shockpeakon solutions.
Further research is needed to clarify this,
but clearly this requirement of peakons being non-overlapping
is incompatible with letting $v=u$,
so at present it is not clear to us whether it is justifiable to use
Novikov peakons as a source of counterexamples for the GX equation,
as has been done in the literature~\cite{himonas-mantzavinos:2016:GX-initial-value-problem}.
Speaking of literature, we are aware of a few analytic studies
\cite{mi-mu-tao:2013:GX-cauchy-problem,
  tang-liu:2015:GX-cauchy-problem,
  himonas-mantzavinos:2016:GX-initial-value-problem,
  barostichi-himonas-petronilho:2016:ovsyannikov-theorem,
  chen-qiao-zhou:2019:GX-persistence-properties-wavebreaking-criteria,
  wang-chong-wu:2021:GX-cauchy-problem}
as well as some papers about integrability aspects
\cite{li-liu:2013:GX-bihamiltonian,
  li-niu:2014:GX-reciprocal,
  li-wang:2017:GX-new-liouville-transformation,
  li-liu:2022p:GX-smooth-multisoliton-solutions}.
There is a also a bewildering array of other multi-component peakon equations
generalizing the Novikov and/or GX equations, which we will not attempt to survey here,
although we may mention the work by Zhao and Qu~\cite{zhao-qu:2021:two-component-novikov-type-system-peakons-H1}
who have classified all two-component Novikov-type cubic equations which admit peakons in
the standard weak sense and in addition conserve the integral $\int (u^2 + u_x^2 + v^2 + v_x^2) \, dx$
(the GX equation is not one of them).

\subsection{The modified Camassa--Holm equation}
\label{sec:guide-mCH}

The modified CH equation \eqref{eq:mCH-again},
\begin{equation*}
  m_t + \bigl( (u^2-u_x^2) \, m \bigr)_x = 0
  ,\qquad
  m = u-u_{xx}
  ,
\end{equation*}
originally arose from the methods developed by Fuchssteiner and Fokas~\cite{fuchssteiner-fokas:1981:symplectic-structures-backlund-transformations-hereditary-symmetries}
for producing new integrable PDEs from previously known ones.
They derived a family containing the CH equation by taking
the KdV equation $u_t = u_{xxx} + 6 u u_x$ as their starting point,
and the analogous procedure applied to the modified KdV equation $u_t = u_{xxx} + 6 u^2 u_x$
gives rise to the family
\begin{equation}
  \label{eq:guide-mCH-fokas-version}
  \begin{split}
    &
    u_t + u_x + \nu u_{xxt} + \gamma u_{xxx} + \alpha u u_x + \tfrac13 \nu \alpha ( u u_{xxx} + 2 u_x u_{xx} )
    \\ &
    + 3 \mu \alpha^2 u^2 u_x + \nu \mu \alpha^2 ( u^2 u_{xxx} + u_x^3 + 4 u u_x u_{xx} )
    \\ &
    + \nu^2 \mu \alpha^2 (u_x^2 u_{xxx} + 2 u_x u_{xx}^2) = 0
    ,
  \end{split}
\end{equation}
of which equation \eqref{eq:mCH-again} is a special case
(see Marinakis~\cite{marinakis:2009:mCH-comment-on-qiao});
hence the name ``modified CH equation''.
As far as we know, the explicit form \eqref{eq:guide-mCH-fokas-version}
was first published in a 1995 article by Fokas~\cite[eq.~(3.9)]{fokas:1995:KdV-and-beyond},
together with a sketch of how it can be derived from water wave theory,
and the same year in another paper of his~\cite[eq.~(7)]{fokas:1995:a-class-of-physically-important-integrable-equations}.
The following year, the family~\eqref{eq:guide-mCH-fokas-version} was mentioned by Fuchssteiner~\cite[eq.~(3.5)]{fuchssteiner:1996:symmetry-toolbox-tricks-generalizations-of-CH},
and rediscovered in the form~\eqref{eq:mCH-again} by Olver and Rosenau~\cite[eq.~(25)]{olver-rosenau:1996:trihamiltonian-duality-solitons-compact-support},
again as a ``dual counterpart of the mKdV equation'' in the same sense as the CH equation arises
from the KdV equation through their formalism.
A zero-curvature representation was derived by Schiff~\cite{schiff:1996:CH-dual-hierarchies-zero-curvature-formulations} soon thereafter.
Later, in 2006, the mCH equation was rediscovered by Qiao
\cite{qiao:2006:mCH-new-integrable-cuspons-WMpeakons, qiao:2007:mCH-new-integrable-hierarchy-parametric-cuspons-peakons-MWpeakons}
in the form~\eqref{eq:mCH-again},
starting from the two-dimensional Euler equations of fluid dynamics;
he also discussed some new types of non-smooth soliton solutions (cuspons and ``W/M-shaped'' solitons,
but not peakons, curiously enough).
The equation also appeared in Novikov's 2009 classification of integrable CH-type equations
\cite[eq.~(32)]{novikov:2009:generalizations-of-CH}.
The initials of Fokas/Fuchssteiner, Olver, Rosenau and Qiao
explain the name ``FORQ equation'', which is also commonly used for~\eqref{eq:mCH-again}.

Regarding smooth multisoliton solutions of the mCH/FORQ equation, see
Ivanov and Lyons~\cite{ivanov-lyons:2012:qiao-hierarchy-dark-solitons},
Matsuno \cite{matsuno:2013:modified-cubic-CH-backlund-multisolitons, matsuno:2014:modified-cubic-CH-multisolitons},
Bies, Górka and Reyes~\cite{bies-gorka-reyes:2012:mCH-geometry-and-local-analysis},
Xia, Zhou and Qiao~\cite{xia-zhou-qiao:2016:CH-mCH-darboux-transformations-multisolitons},
Hu, Yun and Wu~\cite{hu-yin-wu:2016:mCH-bilinear-equations-new-multisoliton-solutions},
as well as the more recent works
by Boutet de Monvel, Karpenko and Shepelsky~\cite{boutetdemonvel-karpenko-shepelsky:2020:mCH-nonzero-boundary-conditions-riemann-hilbert-approach},
Wang, Liu and Mao~\cite{wang-liu-mao:2020:mcH-backlund-transformation-and-nonlinear-superposition-formula}
and Mao and Kuang~\cite{mao-kuang:2021:mCH-solitons-via-dressing-method},
all three of which contain good up-to-date lists of references covering many other aspects of this equation.

We should perhaps warn the reader that there are several other PDEs that are also referred to
as ``the modified CH equation'', which may lead to some confusion when browsing the literature.
For example,
there is the case $b=2$ of the ``modified $b$-family''
\begin{equation}
  u_t - u_{xxt} = u u_{xxx} + b u_x u_{xx} - (b+1) u^2 u_x
\end{equation}
introduced by Wazwaz~\cite{wazwaz:2006:modified-CH-DP-b-family-solitary-waves},
where $uu_x$ has been replaced with $u^2 u_x$, like in the mKdV equation.
There is also the equation
\begin{equation}
  m_t + m_x u + 2mu_x = 0
  ,\quad
  m=(1-\partial_x^2)^k u
  ,
\end{equation}
where $k \ge 2$ is an integer,
studied by McLachlan and Zhang~\cite{mclachlan-zhang:2009:wellposedness-of-modified-CH-equations}.
And last but not least, we mention the equation
\begin{equation}
  G \gamma_t = \gamma_x^2 + \gamma \gamma_{xx} + \lambda \gamma_x
  - G \partial_x \left( \gamma G^{-1} \left( \gamma_x + \frac{\gamma^2}{2\lambda} - \frac{\lambda}{2} \right) \right)
  ,
\end{equation}
where $G = \partial_x^2 - 1$,
proposed by Górka and Reyes \cite[eq.~(5.6)]{gorka-reyes:2011:modified-CH-but-not-FORQ}
as a natural ``modified'' counterpart to the CH equation,
based on a transformation from the CH equation
analogous to the classical Miura map between the KdV and mKdV equations.
(There is also Miura-type map between the CH and mCH/FORQ equations,
found by Kang, Liu, Olver and Qu~\cite{kang-liu-olver-qu:2016:liouville-correspondence-between-mKdV-hierarchy-and-its-dual-mCH-hierarchy}.)

\subsection{Other equations with peakon (or peakon-like) solutions}
\label{sec:guide-other}

Non-smooth solitons were studied already
in the early 80s by Ichikawa, Konno, Wadati, Sanuki and Shimizu
\cite{ichikawa-konno-wadati-sanuki:1980:spiky-soliton-circular-polarized-alfven-wave,
  wadati-ichikawa-shimizu:1980:cusp-soliton-new-integrable-nonlinear-evolution-equation,
  konno-ichikawa-wadati:1981:loop-soliton-propagating-along-stretched-rope,
  ichikawa-konno-wadati:1981:nonlinear-transverse-oscillation-of-elastic-beams-under-tension,
  ichikawa-konno-wadati:1983:new-integrable-nonlinear-evolution-equations-exotic-solitons}.
These singular solutions were obtained by direct integration,
usually in the form of a travelling wave ansatz,
followed by a variety of limiting cases, which, at least in some cases,
produced solutions with sharp edges~\cite[Figure 5]{ichikawa-konno-wadati:1981:nonlinear-transverse-oscillation-of-elastic-beams-under-tension}).
In their own words~\cite{ichikawa-konno-wadati:1981:nonlinear-transverse-oscillation-of-elastic-beams-under-tension}:
``through a series of our investigations we have revealed existence of new species of solitons''.
We would like to point out that Wadati, Ichikawa and Shimizu~\cite[Sect.~2]{wadati-ichikawa-shimizu:1980:cusp-soliton-new-integrable-nonlinear-evolution-equation}
identify the Lax pair for their new integrable equation
\begin{equation}
  q_t - 2 \left( \frac{1}{\sqrt{1+q}} \right)_{xxx} = 0
\end{equation}
to be
\begin{equation}
  \begin{aligned}
    \psi_{xx} &= - \lambda^2 (1+q) \psi
    ,\\
    \psi_t &= 2 \lambda^2 \left[ \frac{2}{\sqrt{1+q}} \, \psi_x - \Biggl( \frac{1}{\sqrt{1+q}} \Biggl)_x \psi \right]
    ,
  \end{aligned}
\end{equation}
where the first equation is an inhomogeneous string problem
\begin{equation*}
  \psi_{xx} = -z m \psi
\end{equation*}
with $z = \lambda^2$ and $m = 1+q$.
Clearly, $q$ in this context can be viewed as a perturbation of the homogeneous string
with constant density~$1$.
Thus the work of Wadati, Ichikawa and Shimizu is intrinsically tied to the CH equation.
In fact, it is in the same hierarchy of isospectral deformations of the inhomogeneous string,
although in their treatment the string has infinite length,
while the string connected to the CH equation has a finite length.
The importance of the work of Wadati, Ichikawa and Shimizu is deservedly highlighted
in the paper by Olver and Rosenau~\cite{olver-rosenau:1996:trihamiltonian-duality-solitons-compact-support}.

After Camassa and Holm discovered the CH equation in 1993,
no further PDEs with peakon solutions were known until the
formulation of the DP equation in 2001.
(The peakon solutions of the mCH/FORQ equation were
not considered until later.)
Then the floodgates opened, and nowadays a large number of such equations
(integrable as well as non-integrable) have been found,
many of them by Qiao;
in addition to rediscovering the mCH equation as discussed above,
he proposed an integrable two-component version of it with Song and Qu~\cite{song-qu-qiao:2011:2mCH-new-integrable-twocomponent-cubic-nonlinearity},
and later an avalanche of other multi-component peakon systems together with various collaborators
\cite{qiao-xia:2013:integrable-peakon-systems-with-kink-peakon,
  xia-qiao:2015:two-component-CH-with-peakons,
  xia-qiao-zhou:2015:synthetical-twocomponent-model-peakons,
  xia-zhou-qiao:2015:cubic-3CH-peakons,
  xia-qiao:2016:multicomponent-generalization-CH,
  luo-qiao-lopez:2014:integrable-generalization-associated-CH,
  hu-qiao:2016:multicomponent-peakon-system-arbitrary-polynomial-analyticity-gevrey-regularity-unique-continuation,
  lou-qiao:2017:alice-bob-peakon-systems,
  yan-qiao-zhang:2018:new-twocomponent-b-family-cubic,
  zhou-qiao-mu:2020:continuity-generalized-crosscoupled-CH-waltzing-peakons-higher-order-nonlinearities}.

In 2002 Holm and Staley
\cite{holm-staley:2003:wave-structure-nonlinear-balances-family-of-evolutionary-PDEs,
  holm-staley:2003:nonlinear-balance-exchange-of-stability-solitons-ramps-cliffs-leftons}
introduced the $b$-family \eqref{eq:intro-b-family},
which is integrable if and only if $b=2$ (the CH case) or $b=3$ (the DP case),
but has peakon solutions of the form \eqref{eq:intro-peakons} for all~$b$;
further details can be found in the work of Degasperis, Holm and Hone~\cite{degasperis-holm-hone:2003:integrable-and-nonintegrable-peakons}.

Another interesting class of peakon equations was proposed in 2010 by Lenells, Misiołek and
Tığlay~\cite{lenells-misiolek-tiglay:2010:evolution-equations-tensor-densities-peakons}.
This particular direction of research goes back to
Misiołek's geometric interpretation, in the Euler--Poincaré--Arnold formalism,
of the periodic CH equation as an Euler equation on the dual $\mathfrak{g}^*$
to the Lie algebra
$\mathfrak{g} = \diff(\circleS)$
associated with the Lie group
$G = \Diff(\circleS)$
of orientation-preserving diffeomorphisms of the circle~\cite{misiolek:1998:CH-as-geodesic-flow-on-Bott-Virasoro-group}.
In short, the picture involves a Lie algebra $\mathfrak{g}$ and its dual~$\mathfrak{g^*}$.
The adjoint action $\ad_a(b) = [a,b]$ of $\mathfrak{g}$ on itself
induces a coadjoint action $\ad^*$ on~$\mathfrak{g}^*$;
see \eqref{eq:guide-coadjoint-action} below.
Suppose now that an inner product on $\mathfrak{g}$ is given.
This is equivalent, at least in finite dimensions, to the existence of
an isomorphism  $A \colon \mathfrak{g} \to \mathfrak{g}^*$
generalizing the familiar inertia tensor from the dynamics of a free rigid body.
Then the Euler equation reads
\begin{equation*}
  m_t=-\ad^*_{A^{-1} m} m
  , \qquad
  m \in \mathfrak{g}^*
  .
\end{equation*}
Some well-known examples are the Euler equations of a free rigid body,
where $\mathfrak{g}=\mathfrak{so}(3)$,
and the Euler equations from fluid dynamics, where $\mathfrak{g} = \diff_V(\R^3)$
(volume-preserving diffeomorphisms).
The KdV equation also fits into this setup, with $\mathfrak{g} = \diff(\circleS) \oplus \R$
(the Virasoro algebra, a central extension of $\diff(\circleS)$, with a suitably defined Lie bracket)
and with the $L^2$ inner product on the $\diff(\circleS)$ part~\cite{ovsienko-khesin:1987:KdV-to-Euler}.
And so does the full CH equation \eqref{eq:intro-CH-kappa-expanded},
including the linear dispersion term $2 \kappa u_x$,
with the same Lie algebra~$\mathfrak{g}$ as for KdV but with the
Sobolev $H^1$ inner product on the $\diff(\circleS)$ part~\cite{misiolek:1998:CH-as-geodesic-flow-on-Bott-Virasoro-group}.
For the dispersionless CH equation \eqref{eq:intro-CH-expanded} with $\kappa=0$,
the central extension is not needed, and $\mathfrak{g} = \diff(\circleS)$ suffices;
let us briefly sketch how this works.
The natural geometric way of interpreting elements $u \in \diff(\circleS)$ is to view them as vector fields
$u \, \partial_x$, and the dual space $\diff^* (\circleS)$ as the space of quadratic
differentials $\Omega^{\otimes^2}$, with the diffeomorphism-invariant pairing
\begin{equation*}
  \langle m \, dx^{2}, u \, \partial_x \rangle = \int_{\circleS} m u \, dx
  .
\end{equation*}
Recall that the coadjoint Lie algebra action on the dual is given by
\begin{equation}
  \label{eq:guide-coadjoint-action}
  \langle \ad_a^*\xi, b \rangle
  = - \langle \xi, \ad_a b \rangle
  = - \langle \xi, [a,b] \rangle
  ,
\end{equation}
for $\xi \in \mathfrak{g}^*$ and  $a,b \in \mathfrak{g}$.
The Lie bracket on $\mathfrak{g}=\diff(\circleS)$ is the Lie bracket of vector fields,
\begin{equation*}
  [u \, \partial _x , v \, \partial_x] = (uv_x-u_x v) \, \partial_x
  ,
\end{equation*}
and hence, if we integrate by parts,
\begin{equation*}
  \begin{split}
    \langle \ad_{u \, \partial_x}^* (m \, dx^2), v \, \partial_x \rangle
    &
    = - \langle m \, dx^2, [u \, \partial _x , v \, \partial_x] \rangle
    \\ &
    = - \int_{\circleS} m \, (u v_x - u_x v) \, dx
    \\ &
    = \int_{\circleS}\big( (u m)_x + u_x m \big) \, v \, dx
    ,
  \end{split}
\end{equation*}
so that
$\ad^*_{u \, \partial_x} (m \, dx^2) = \bigl( (um)_x + u_x m \bigr) \, dx^2$.
Note the appearance of the expression on the right-hand side,
which coincides with the negative of the right-hand side of the CH equation $m_t = -((um)_x + u_x m)$.
A priori there is of course no relation between $m$ and~$u$.
However, if we equip the Lie algebra $\diff(\circleS)$ with the $H^1$ inner product
\begin{equation*}
  (u \, \partial_x , v \,\partial_x) = \int_{\circleS} \big( uv + u_xv_x) \, dx
  ,
\end{equation*}
then after one integration by parts the inner product can be written
\begin{equation*}
  (u \, \partial_x , v \, \partial_x)
  = \int_{\circleS} \big( u-u_{xx}\big) \, v \, dx
  = \langle Au \, dx^2 , v \partial_x \rangle
  ,
\end{equation*}
with $A = 1-\partial_x^2$.
Hence, $A$ is our inertia tensor and this shows that the CH equation
(with $\kappa=0$)
is the Euler equation for the group $\Diff(\circleS)$
and a particular pairing $m=Au$ encoded by~$A = 1-\partial_x^2$.
One can then write the CH equation, referring only to the vector field $u \, \partial _x$
and the mapping~$A$, as
\begin{equation} \label{eq:ACH}
  (Au)_t + \bigl( u \, Au \bigr)_x + u_x \, Au = 0
  .
\end{equation}

Let us now see how the picture changes if we choose a different inner product,
\begin{equation*}
  (u \, \partial_x , v \, \partial_x) = \mu(u) \, \mu(v) + \int_{\circleS} u_xv_x \, dx
  ,
\end{equation*}
where
\begin{equation*}
  \mu(f)=\int_{\circleS} f(x) \, dx
\end{equation*}
is the average of~$f$ over $\circleS =  \mathbf{R}/\mathbf{Z}$.
Integration by parts gives
\begin{equation*}
  (u \, \partial_x , v \, \partial_x)
  = \int_{\circleS} \bigl( \mu(u) - u_{xx} \bigr) \, v \, dx
  = \langle Au \, dx^2 , v \, \partial_x \rangle
  ,
\end{equation*}
where this time $Au = \mu(u) - u_{xx}$,
or $A = \mu - \partial_x^2$ for short.
If we substitute this into \eqref{eq:ACH}, we obtain
\begin{equation*}
  \begin{split}
    0 &
    = \bigl( \mu(u) - u_{xx} \bigr)_t
    + \Bigl( u \, \bigl( \mu(u) - u_{xx} \bigr) \Bigr)_x
    + u_x \, \bigl( \mu(u) - u_{xx} \bigr)
    .
  \end{split}
\end{equation*}
Integrating this equation over~$\circleS$ shows that $\mu(u)_t = 0$,
so the final form of this equation,
obtained by Khesin, Lenells and
Misiołek~\cite{khesin-lenells-misiolek:2008:generalized-HS-circle-diffeomorphisms}, is
\begin{equation}
  \label{eq:guide-mu-HS}
  u_{xxt} - 2 \mu(u) \, u_x + 2 u_x u_{xx} + u u_{xxx} = 0
  .
\end{equation}
Moreover, $\mu(u)_t = 0$ means that the non-local term $\mu(u)$ is actually a constant determined
by the initial condition $u(x,0)$,
and the equation has the character of a PDE rather than an integro-differential equation.
The Lax pair for \eqref{eq:guide-mu-HS} is
\begin{align*}
  \psi_{xx} &=\lambda m \psi
  , \\
  \psi_t &= \left( \tfrac{1}{2\lambda} - u \right) \, \psi_x + \tfrac12 u_x \psi
  ,
\end{align*}
where $m = Au = \mu(u) - u_{xx}$,
which interestingly is in principle the same Lax pair as for the Hunter--Saxton (HS)
equation~\cite{hunter-saxton:1991:dynamics-of-director-fields,
  hunter-zheng:1994:completely-integrable-hyperbolic-variational-equation}
\begin{equation*}
  u_{xxt} + 2 u_x u_{xx} + u u_{xxx} = 0
  ,
\end{equation*}
except that for the HS equation $m = -u_{xx}$ instead.
The reason why this Lax pair covers both equations is that the
compatibility conditions actually are
$m_t + (um)_x + u_xm = 0$ and $m_x = -u_{xxx}$,
so that there is some freedom; we may have $m = -u_{xx} + c$,
where $c$ possibly depends on~$t$ but not on~$x$.
For the HS equation one takes $c=0$, while \eqref{eq:guide-mu-HS} corresponds to the choice $c=\mu(u)$.

Lenells, Misiołek and Tığlay~\cite{lenells-misiolek-tiglay:2010:evolution-equations-tensor-densities-peakons}
have generalized this picture by observing that the coadjoint action on quadratic
differentials is just a special case of the Lie algebra action of $\diff(\circleS)$ on densities of
arbitrary weight~$b$.
This more general action reads
\begin{equation*}
  \mathcal{L}_{u\partial_x} (m \, dx^b)
  = \bigl( (um)_x + (b-1) u_x m \bigr) \, dx^b
  ,
\end{equation*}
where $b=2$ gives the CH case above, while $b=3$ gives the DP case.
In other words, the flows generalizing the Euler flow are postulated to be
\begin{equation*}
  m_t + (um)_x + (b-1) u_x m = 0
  .
\end{equation*}
The formal substitution $m=Au$ results in the equation
\begin{equation*}
  (Au)_t + (u \, Au)_x + (b-1) \, u_x ,\ Au = 0
  ,
\end{equation*}
which for $b=3$ specializes to the DP equation if $A = 1 - \partial_x^2$
and to what the authors call the $\mu$-version of the DP equation if $A = \mu - \partial_x^2$.
Like the DP equation, this $\mu$DP equation admits not just peakon solutions but also shockpeakons.

Let us conclude this article with a few more examples of generalizations of peakon equations.
Anco, da Silva and
Freire~\cite{anco-dasilva-freire:2015:wavebreaking-equations-generalizing-CH-and-novikov},
studied a $4$-parameter family of PDEs,
\begin{equation*}
  u_t - u_{txx} + a u^p u_x - b u^{p-1} u_x u_{xx} - c u^p u_{xxx} = 0
  ,
\end{equation*}
and established that this equation admits peakon solutions of the form \eqref{eq:intro-peakons}
for any $N \ge 1$
only when $a = b + c$, $c \neq 0$ and $p\ge 0$;
see also other similar works by
Anco et~al. \cite{anco-recio-gandarias-bruzon:2015:nonlinear-generalization-of-CH-with-peakons,
  anco-recio:2019:accelerating-dynamical-peakons}.
Anco and Mobasheramini \cite{anco-mobasheramini:2017:integrable-U1-invariant-peakon-equations-NLS}
derived a pair of complex-valued integrable peakon equations
(first obtained as two-component systems by Xia, Qiao and Zhou~\cite{xia-qiao:2015:two-component-CH-with-peakons,xia-qiao-zhou:2015:synthetical-twocomponent-model-peakons})
from the nonlinear Schrödinger (NLS) hierarchy.
One of them is a complex counterpart of the mCH equation,
while the other one is similar to the NLS equation itself, and features ``peakon breathers''.
These equations were further generalized by
Anco, Chang and Szmigielski~\cite{anco-chang-szmigielski:2018:conservative-peakons-U1-invariant-NLS-Hirota},
in the form of a family of peakon equations parametrized by the real projective line~$\mathbf{RP}^1$,
\begin{equation}
  m_t + \bigl( \operatorname{Re} (e^{i\theta} Q) \, m \bigr)_x - i \operatorname{Im} ( e^{i\theta} Q ) \, m = 0
  ,
\end{equation}
where $\theta \in [0, \pi)$ and $Q = (u-u_x)(\bar u + \bar u_x)$.
The peakon solutions of this family were computed
using a modification of the inverse spectral problem employed earlier to solve the mCH peakon ODEs.

\section{Acknowledgements}

Jacek Szmigielski's research is supported by the Natural Sciences and Engineering Research Council of Canada (NSERC).

\bibliographystyle{arxiv-bibulous}
\bibliography{peakons-orth-poly-arXiv-v2}

\end{document}